\pdfoutput=1
\documentclass[12pt,a4paper]{article}

\usepackage{ifthen} 
\usepackage{rotating}
\newboolean{pdflatex}
\setboolean{pdflatex}{true} 

\newboolean{articletitles}
\setboolean{articletitles}{true} 

\newboolean{uprightparticles}
\setboolean{uprightparticles}{false} 

\def\paperauthors{LHCb collaboration} 
\def\paperasciititle{Observation of strangeness enhancement with charmed mesons in high-multiplicity pPb collisions at sqrt(sNN) = 8.16 TeV} 
\def\papertitle{Observation of strangeness enhancement with charmed mesons in high-multiplicity $p$Pb collisions at \sqsnn = 8.16 \tev}
\def\paperkeywords{{High Energy Physics}, {LHCb}} 
\def\papercopyright{\the\year\ CERN for the benefit of the LHCb collaboration} 
\def\paperlicence{CC BY 4.0 licence}
\def\paperlicenceurl{https://creativecommons.org/licenses/by/4.0/}
\usepackage[top=1in, bottom=1.25in, left=1in, right=1in]{geometry}

\usepackage{microtype}
\usepackage{lineno}  
\usepackage{xspace} 
\usepackage{caption} 

\usepackage{graphicx}  
\usepackage{color}
\usepackage{colortbl}
\graphicspath{{./figs/}} 

\usepackage{amsmath} 
\usepackage{amssymb}
\usepackage{amsfonts}
\usepackage{upgreek} 

\newcommand*\patchAmsMathEnvironmentForLineno[1]{%
\expandafter\let\csname old#1\expandafter\endcsname\csname #1\endcsname
\expandafter\let\csname oldend#1\expandafter\endcsname\csname
end#1\endcsname
 \renewenvironment{#1}%
   {\linenomath\csname old#1\endcsname}%
   {\csname oldend#1\endcsname\endlinenomath}%
}
\newcommand*\patchBothAmsMathEnvironmentsForLineno[1]{%
  \patchAmsMathEnvironmentForLineno{#1}%
  \patchAmsMathEnvironmentForLineno{#1*}%
}
\AtBeginDocument{%
\patchBothAmsMathEnvironmentsForLineno{equation}%
\patchBothAmsMathEnvironmentsForLineno{align}%
\patchBothAmsMathEnvironmentsForLineno{flalign}%
\patchBothAmsMathEnvironmentsForLineno{alignat}%
\patchBothAmsMathEnvironmentsForLineno{gather}%
\patchBothAmsMathEnvironmentsForLineno{multline}%
\patchBothAmsMathEnvironmentsForLineno{eqnarray}%
}

\usepackage{hyperxmp}

\usepackage[pdftex,
            pdfauthor={\paperauthors},
            pdftitle={\paperasciititle},
            pdfkeywords={\paperkeywords},
            pdfcopyright={Copyright (C) \papercopyright},
            pdflicenseurl={\paperlicenceurl}]{hyperref}

\usepackage[colorinlistoftodos,textsize=scriptsize]{todonotes}

\usepackage[bottom,flushmargin,hang,multiple]{footmisc}

\usepackage[all]{hypcap} 

\usepackage{xspace} 
\usepackage{upgreek}

\def\lhcb   {\mbox{LHCb}\xspace}

\def\MagUp {\mbox{\em Mag\kern -0.05em Up}\xspace}

\ifthenelse{\boolean{uprightparticles}}%
{

 \def\Pmu         {\ensuremath{\upmu}\xspace}

 \def\Ppi         {\ensuremath{\uppi}\xspace}

 \def\Ppsi        {\ensuremath{\uppsi}\xspace}

 \def\PDelta      {\ensuremath{\Delta}\xspace}                 
 \def\PXi         {\ensuremath{\Xi}\xspace}                 
 \def\PLambda     {\ensuremath{\Lambda}\xspace}                 
 \def\PSigma      {\ensuremath{\Sigma}\xspace}                 
 \def\POmega      {\ensuremath{\Omega}\xspace}                 
 \def\PUpsilon    {\ensuremath{\Upsilon}\xspace}
 \let\oldPi\Pi
 \def\PPi         {\ensuremath{\oldPi}\xspace}

 \def\PB      {\ensuremath{\mathrm{B}}\xspace}                 
                  
 \def\PD      {\ensuremath{\mathrm{D}}\xspace}

 \def\PJ      {\ensuremath{\mathrm{J}}\xspace}                 
 \def\PK      {\ensuremath{\mathrm{K}}\xspace}

 \def\Pe      {\ensuremath{\mathrm{e}}\xspace}

 \def\Pi      {\ensuremath{\mathrm{i}}\xspace}

 \def\Pp      {\ensuremath{\mathrm{p}}\xspace}

 \def\Ps      {\ensuremath{\mathrm{s}}\xspace}

 \def\thebaroffset{0.0em}
}
{

 \def\Pmu         {\ensuremath{\mu}\xspace}

 \def\Ppi         {\ensuremath{\pi}\xspace}

 \def\Ppsi        {\ensuremath{\psi}\xspace}                 
                  
 \mathchardef\PDelta="7101
 \mathchardef\PXi="7104
 \mathchardef\PLambda="7103
 \mathchardef\PSigma="7106
 \mathchardef\POmega="710A
 \mathchardef\PUpsilon="7107
 \mathchardef\PPi="7105
                  
 \def\PB      {\ensuremath{B}\xspace}                 
                  
 \def\PD      {\ensuremath{D}\xspace}

 \def\PJ      {\ensuremath{J}\xspace}                 
 \def\PK      {\ensuremath{K}\xspace}

 \def\Pe      {\ensuremath{e}\xspace}

 \def\Pi      {\ensuremath{i}\xspace}

 \def\Pp      {\ensuremath{p}\xspace}

 \def\Ps      {\ensuremath{s}\xspace}

 \def\thebaroffset{0.18em}
}
\newcommand{\offsetoverline}[2][\thebaroffset]{\kern #1\overline{\kern -#1 #2}}%

\makeatletter
\ifcase \@ptsize \relax
  \newcommand{\miniscule}{\@setfontsize\miniscule{4}{5}}
\or
  \newcommand{\miniscule}{\@setfontsize\miniscule{5}{6}}
\or
  \newcommand{\miniscule}{\@setfontsize\miniscule{5}{6}}
\fi
\makeatother

\DeclareRobustCommand{\optbar}[1]{\shortstack{{\miniscule (\rule[.5ex]{1.25em}{.18mm})}
  \\ [-.7ex] $#1$}}

\def\electron   {{\ensuremath{\Pe}}\xspace}

\def\epem       {{\ensuremath{\Pe^+\Pe^-}}\xspace}

\def\mumu       {{\ensuremath{\Pmu^+\Pmu^-}}\xspace}

\def\squark    {{\ensuremath{\Ps}}\xspace}

\def\pion   {{\ensuremath{\Ppi}}\xspace}

\def\pip    {{\ensuremath{\pion^+}}\xspace}
\def\pim    {{\ensuremath{\pion^-}}\xspace}

\def\kaon    {{\ensuremath{\PK}}\xspace}

\def\KorKbar {\kern \thebaroffset\optbar{\kern -\thebaroffset \PK}{}\xspace}

\def\KS      {{\ensuremath{\kaon^0_{\mathrm{S}}}}\xspace}

\def\D       {{\ensuremath{\PD}}\xspace}

\def\DorDbar {\kern \thebaroffset\optbar{\kern -\thebaroffset \PD}\xspace}
\def\Dz      {{\ensuremath{\D^0}}\xspace}

\def\Dp      {{\ensuremath{\D^+}}\xspace}
\def\Dm      {{\ensuremath{\D^-}}\xspace}

\def\DpDm    {\ensuremath{\Dp {\kern -0.16em \Dm}}\xspace}

\def\Ds      {{\ensuremath{\D^+_\squark}}\xspace}

\def\B       {{\ensuremath{\PB}}\xspace}

\def\BorBbar {\kern \thebaroffset\optbar{\kern -\thebaroffset \PB}\xspace}

\def\Bd      {{\ensuremath{\B^0}}\xspace}

\def\BdorBdbar {\kern \thebaroffset\optbar{\kern -\thebaroffset \Bd}\xspace}

\def\Bs      {{\ensuremath{\B^0_\squark}}\xspace}

\def\BsorBsbar {\kern \thebaroffset\optbar{\kern -\thebaroffset \Bs}\xspace}

\def\jpsi     {{\ensuremath{{\PJ\mskip -3mu/\mskip -2mu\Ppsi}}}\xspace}

\def\Y#1S{\ensuremath{\PUpsilon{(#1S)}}\xspace}

\def\proton      {{\ensuremath{\Pp}}\xspace}

\def\LorLbar     {\kern \thebaroffset\optbar{\kern -\thebaroffset \PLambda}\xspace}

\def\BF         {{\ensuremath{\mathcal{B}}}\xspace}
\def\BR         {\BF}

\def\to                 {\ensuremath{\rightarrow}\xspace}

\def\AT#1     {\ensuremath{A_{\mathrm{T}}^{#1}}\xspace}           

\def\C#1      {\ensuremath{\mathcal{C}_{#1}}\xspace}                       
\def\Cp#1     {\ensuremath{\mathcal{C}_{#1}^{'}}\xspace}                    
\def\Ceff#1   {\ensuremath{\mathcal{C}_{#1}^{\mathrm{(eff)}}}\xspace}        
\def\Cpeff#1  {\ensuremath{\mathcal{C}_{#1}^{'\mathrm{(eff)}}}\xspace}       
\def\Ope#1    {\ensuremath{\mathcal{O}_{#1}}\xspace}                       
\def\Opep#1   {\ensuremath{\mathcal{O}_{#1}^{'}}\xspace}                    

\newcommand{\aunit}[1]{\ensuremath{\text{\,#1}}}       

\newcommand{\tev}{\aunit{Te\kern -0.1em V}\xspace}
\newcommand{\gev}{\aunit{Ge\kern -0.1em V}\xspace}
\newcommand{\mev}{\aunit{Me\kern -0.1em V}\xspace}
\newcommand{\kev}{\aunit{ke\kern -0.1em V}\xspace}
\newcommand{\ev}{\aunit{e\kern -0.1em V}\xspace}
 
\newcommand{\mevc}{\ensuremath{\aunit{Me\kern -0.1em V\!/}c}\xspace}
\newcommand{\gevc}{\ensuremath{\aunit{Ge\kern -0.1em V\!/}c}\xspace}
\newcommand{\mevcc}{\ensuremath{\aunit{Me\kern -0.1em V\!/}c^2}\xspace}
\newcommand{\gevcc}{\ensuremath{\aunit{Ge\kern -0.1em V\!/}c^2}\xspace}

\def\mbarn{\aunit{mb}\xspace}

\def\nb {\aunit{nb}\xspace}
\def\invnb {\ensuremath{\nb^{-1}}\xspace}

\newcommand{\chisq}{\ensuremath{\chi^2}\xspace}

\newcommand{\chisqip}{\ensuremath{\chi^2_{\text{IP}}}\xspace}

\def\deriv {\ensuremath{\mathrm{d}}}

\def\gsim{{~\raise.15em\hbox{$>$}\kern-.85em
          \lower.35em\hbox{$\sim$}~}\xspace}
\def\lsim{{~\raise.15em\hbox{$<$}\kern-.85em
          \lower.35em\hbox{$\sim$}~}\xspace}

\def\sqs   {\ensuremath{\protect\sqrt{s}}\xspace}
\def\sqsnn {\ensuremath{\protect\sqrt{s_{\scriptscriptstyle\text{NN}}}}\xspace}
\def\pt         {\ensuremath{p_{\mathrm{T}}}\xspace}

\def\photos     {\mbox{\textsc{Photos}}\xspace}

\def\tell1  {TELL1\xspace}
\def\ukl1   {UKL1\xspace}

\newcommand{\lhcborcid}[1]{\href{https://orcid.org/#1}{\hspace*{0.1em}\raisebox{-0.45ex}{\includegraphics[width=1em]{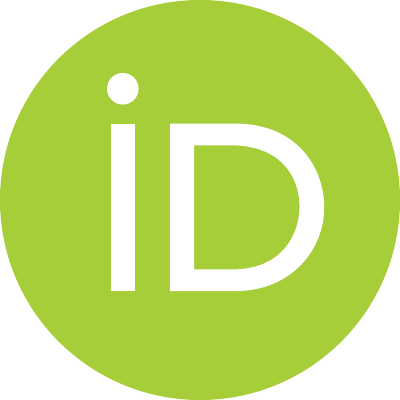}}}}

\usepackage{cite} 
\usepackage{mciteplus}

\usepackage{longtable} 

\begin{document}

\renewcommand{\thefootnote}{\fnsymbol{footnote}}
\setcounter{footnote}{1}


\begin{titlepage}
\pagenumbering{roman}

\vspace*{-1.5cm}
\centerline{\large EUROPEAN ORGANIZATION FOR NUCLEAR RESEARCH (CERN)}
\vspace*{1.5cm}
\noindent
\begin{tabular*}{\linewidth}{lc@{\extracolsep{\fill}}r@{\extracolsep{0pt}}}
\ifthenelse{\boolean{pdflatex}}
{\vspace*{-1.5cm}\mbox{\!\!\!\includegraphics[width=.14\textwidth]{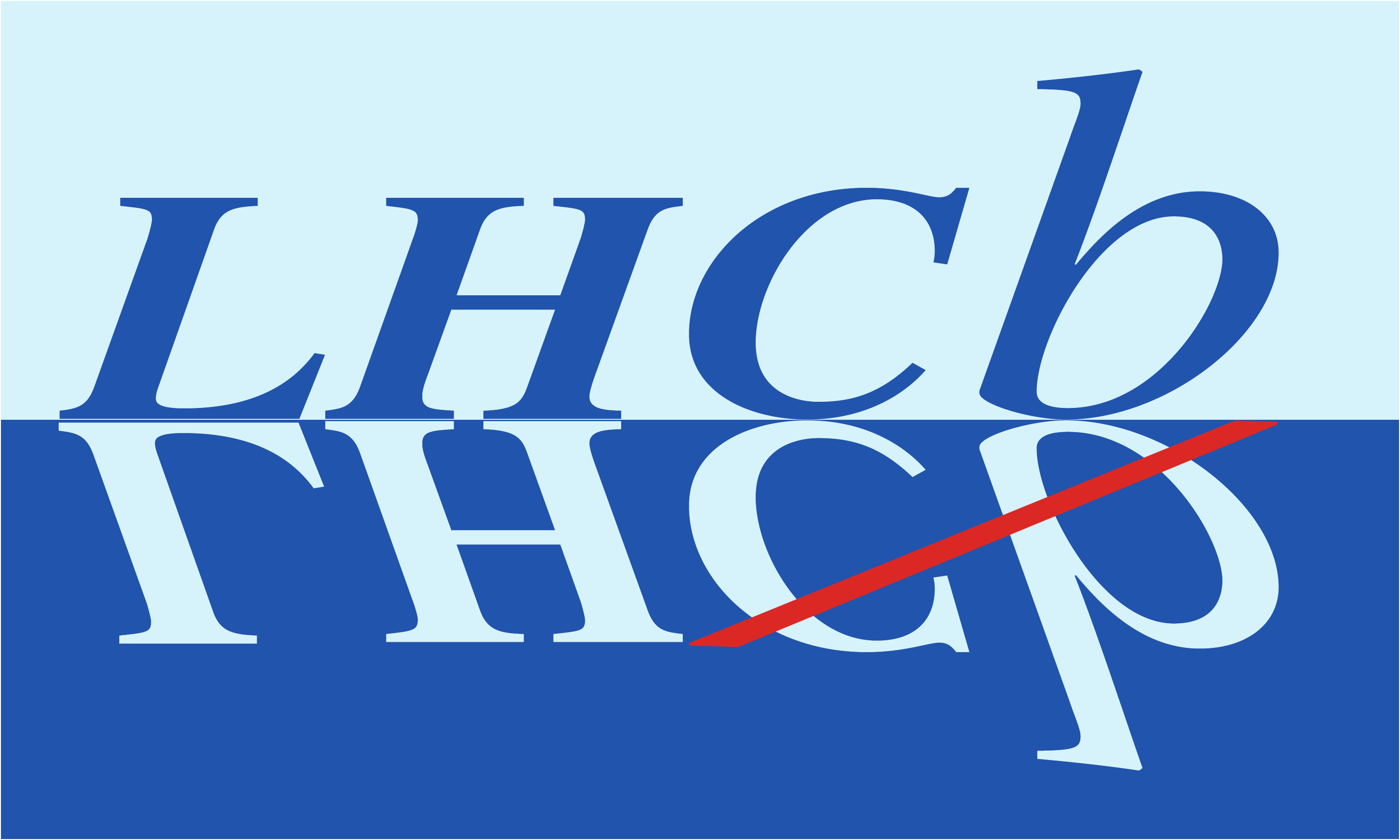}} & &}%
{\vspace*{-1.2cm}\mbox{\!\!\!\includegraphics[width=.12\textwidth]{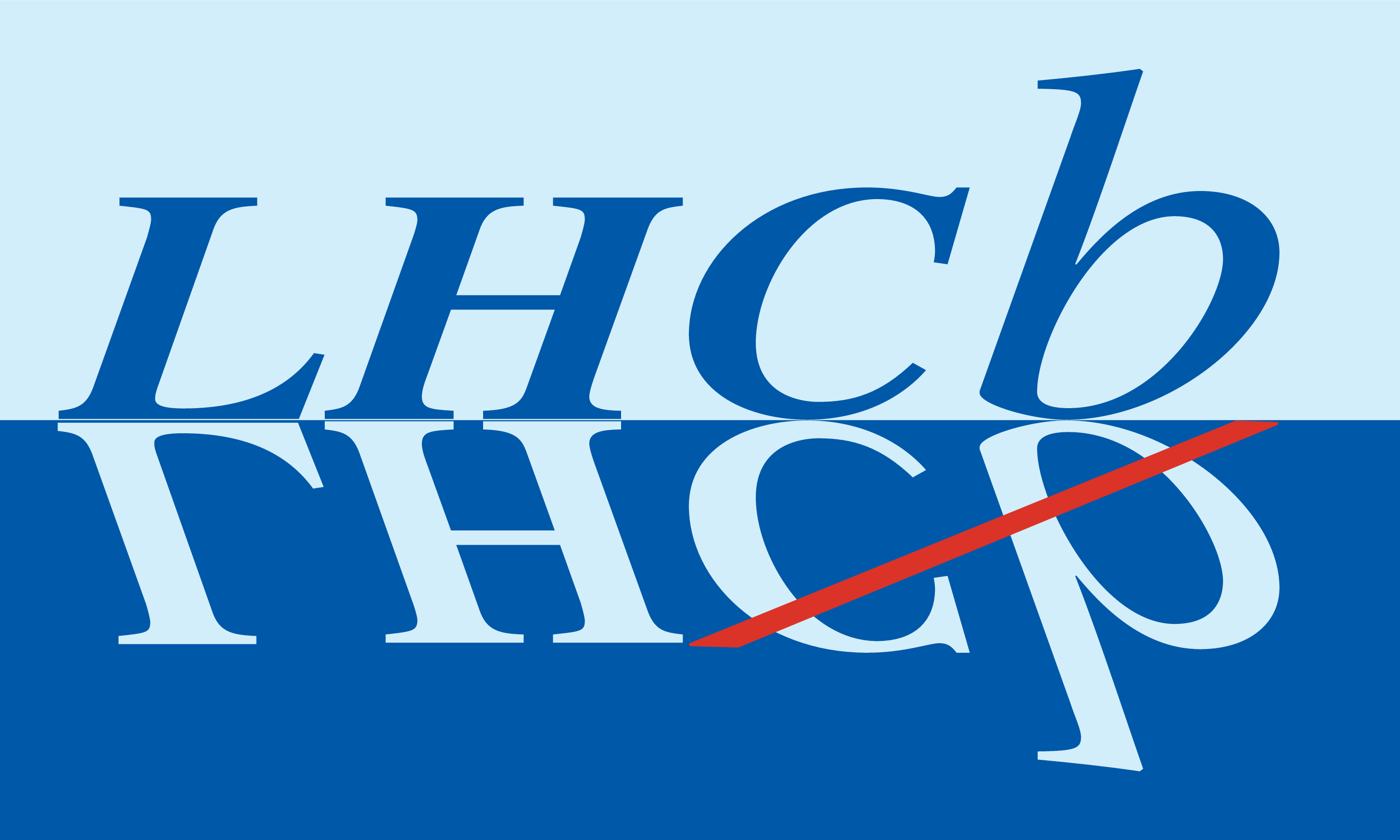}} & &}%
\\
 & & CERN-EP-2023-236 \\  
 & & LHCb-PAPER-2023-021 \\  
 & & \today \\ 
 & & \\
\end{tabular*}

\vspace*{3.0cm}

{\normalfont\bfseries\boldmath\huge
\begin{center}
  \papertitle 
\end{center}
}

\vspace*{2.0cm}

\begin{center}
\paperauthors\footnote{Authors are listed at the end of this paper.}
\end{center}

\vspace{\fill}

\begin{abstract}
  \noindent
  The production of prompt \Ds and \Dp mesons is measured by the LHCb experiment in proton-lead ($p$Pb) collisions in both the forward ($1.5<y^*<4.0$) and backward ($-5.0<y^*<-2.5$) rapidity regions at a nucleon-nucleon center-of-mass energy of $\sqsnn=8.16\tev$. The nuclear modification factors of both \Ds and \Dp mesons are determined as a function of transverse momentum, \pt, and rapidity. In addition, the \Ds to \Dp cross-section ratio is measured as a function of the primary charged particle multiplicity in the event. An enhanced \Ds to \Dp production in high-multiplicity events is observed for the whole measured \pt range, in particular at low \pt and backward rapidity, where the significance exceeds six standard deviations. This constitutes the first observation of strangeness enhancement in charm quark hadronization in high-multiplicity $p$Pb collisions. The results are also qualitatively consistent with the presence of quark coalescence as an additional charm quark hadronization mechanism in high-multiplicity proton-lead collisions.

\end{abstract}

\vspace*{2.0mm}

\begin{center}
  Published in Phys. Rev. D 110 (2024) L031105
\end{center}

\vspace{\fill}

{\footnotesize 
\centerline{\copyright~\papercopyright. \href{\paperlicenceurl}{\paperlicence}.}}
\vspace*{2mm}

\end{titlepage}


\newpage
\setcounter{page}{2}
\mbox{~}
%
%
%
%


\renewcommand{\thefootnote}{\arabic{footnote}}
\setcounter{footnote}{0}

\cleardoublepage


\pagestyle{plain} 
\setcounter{page}{1}
\pagenumbering{arabic}



At hadron colliders, charm quarks are mainly produced by hard parton-parton interactions in the initial stages of the collisions, which are well described by perturbative quantum chromodynamics (pQCD) calculations. These calculations are based on the factorisation theorem, according to which the charmed hadron cross-sections are dependent on the parton distribution functions (PDFs) of the incoming nucleons, the hard parton-parton scattering cross-section, and the fragmentation functions~\cite{Webber:1983if,Andersson:1983ia}. 

In proton–lead collisions, various effects could modify the charmed hadron cross-sections compared to $pp$ collisions. In the initial state, the charmed hadron production can be affected by the modification of the parton distribution functions of bound nucleons (nPDFs)~\cite{Hirai:2007sx,Eskola:2021nhw} compared to those of free nucleons. Furthermore, the increased gluon density at small momentum fraction $\it{x}$ leads to non-perturbative features, even if the coupling constant is weak. The color-glass condensate (CGC) effective theory~\cite{Gelis:2012ri,Fujii:2013yja} provides an appropriate theoretical framework in this regime. A recent measurement from the LHCb experiment has shown a discrepancy with the theoretical calculations based on nPDFs~\cite{LHCb:2022dmh}.
In the final state, the fragmentation functions are typically parameterised based on measurements performed in \epem or \electron \proton collisions, assuming that the hadronization of charm quarks to charmed hadrons is a universal process independent of the colliding system~\cite{Braaten:1994bz}. A recent measurement from the ALICE experiment has shown that charm quark hadronization differs between \epem and \proton\proton collisions~\cite{ALICE:2021dhb,ALICE:2023sgl}. This result suggests the existence of other hadronization mechanisms beyond fragmentation. An alternative mechanism is quark coalescence~\cite{Oh:2009zj,He:2019vgs,Minissale:2020bif,CMS:2018loe}, where charm quarks recombine with other quarks to form charmed hadrons. This mechanism requires that multiple quarks overlap in velocity-position space. As a result, the fraction of charmed hadrons produced by coalescence is expected to be larger when the number of quarks produced in the collision is large, for example in relativistic heavy-ion collisions where quark-gluon plasma (QGP) is formed~\cite{STAR:2005gfr,PHENIX:2004vcz}. This mechanism is also expected to be more prominent at relatively low transverse momentum, \pt, as most quarks or particles are produced in that kinematic region. 

Relativistic heavy-ion collisions are often accompanied by strangeness enhancement, which was originally considered as a signature of QGP~\cite{PhysRevLett.48.1066}. The enhanced strangeness production~\cite{STAR:2011fbd,ALICE:2013xmt} and the coalescence mechanism result in an increased yield of strange charmed mesons relative to non-strange charmed mesons compared to $pp$ collisions~\cite{STAR:2021tte,ALICE:2021kfc}. Additionally, the ALICE collaboration observed the production enhancement of strange light hadrons in both high-multiplicity $pp$~\cite{ALICE:2016fzo} and $p$Pb~\cite{ALICE:2013wgn,ALICE:2015mpp} collisions. Although the origin of the strangeness enhancement in ``small" systems (proton-proton or proton-nucleus collisions) is still under debate~\cite{Kanakubo:2019ogh,Bierlich:2022ned}, it may indicate a common underlying physics mechanism which gradually compensates the strangeness suppression in fragmentation. If the coalescence mechanism contributes to the charm quark hadronization in small systems, the production rates of \Ds mesons ($c\bar{s}$) relative to \Dp mesons ($c\bar{d}$) could also increase with the event multiplicity.

This letter reports LHCb measurements of the prompt $D_{(s)}^+$ ($D_{s}^+$ and $D^+$) differential production cross-sections, of their nuclear modification factors and forward-backward cross-section ratio in $p$Pb collisions at $\sqsnn=8.16$ \tev. Additionally, the cross-section ratio, $\sigma_{\Ds}/\sigma_{\Dp}$, as a function of the primary charged particle multiplicity of the events is reported. 

The LHCb detector is a single-arm forward spectrometer covering the pseudorapidity range $2 < \eta < 5$, described in detail in Refs.~\cite{LHCb-DP-2008-001,LHCb-DP-2014-002}.
The present measurement covers the forward rapidity range of $1.5 < y^{*} < 4.0$ when the proton beam points towards the LHCb arm, and the backward rapidity range of $-5.0 < y^{*} < -2.5$ when the lead beam does. Here, $y^{*}$ is the rapidity in the nucleon-nucleon center-of-mass frame. The centre-of-mass frame does not coincide with the laboratory frame due to the asymmetry of the colliding beam energies, with a constant boost of \mbox{$ y_{\textbf{lab}}-y^* = 0.5 \log (A/Z) = 0.465$} in the direction of the proton beam, where $A = 208$ is the lead nucleus mass number and $Z = 82$ is the lead nucleus atomic number. The corresponding integrated luminosity for the forward (backward) rapidity data sample is $12.18\pm0.32\invnb$ ($18.57\pm0.46\invnb$). 

Simulation is used to model the effects of detector acceptance and selection requirements. The $D_{(s)}^+$ mesons are generated using Pythia 8~\cite{Sjostrand:2007gs,*Sjostrand:2006za} and embedded into minimum-bias (MB) $p$Pb events using the EPOS generator~\cite{Pierog:2013ria}, calibrated with LHC data~\cite{LHCb:2011dpk}. The decays of unstable particles are described by EvtGen~\cite{Lange:2001uf}, in which final-state radiation is generated using \photos~\cite{Golonka:2005pn}. The interaction of the generated particles with the detector, and its response, are implemented using the Geant4 toolkit~\cite{GEANT4:2002zbu} as described in Ref.~\cite{Clemencic:2011zza}. The simulated $D_{(s)}^+$ event multiplicity distribution is weighted to match the background-subtracted distribution that is extracted from data using the \textit{sPlot} method~\cite{Pivk:2004ty}. 


The double-differential cross-section in a given (\pt, $y^*$) interval is defined as
    \begin{equation}\label{eq:cross-section}
        \frac{\deriv^2 \sigma_{p\text{Pb}}}{\deriv \pt \deriv y^*}=
        \frac{N}
        {\mathcal{L}\times\epsilon^{\text{acc}}\times\epsilon^{\text{trig}}\times\epsilon^{\text{PID}}\times\epsilon^{\text{rec\&sel}}
        \times\mathcal{B}\times\Delta\pt \times \Delta y^*}~,
    \end{equation}
where $N$ is the observed number of prompt $D_{(s)}^+$ and $D_{(s)}^-$ mesons, $\mathcal{L}$ the integrated luminosity,
$\BR$  the branching fraction of the corresponding $D_{(s)}^+$ meson decay, $\epsilon^{\text{acc}}$, $\epsilon^{\text{trig}}$, $\epsilon^{\text{PID}}$, $\epsilon^{\text{rec\&sel}}$ are the LHCb acceptance, trigger, particle identification (PID), reconstruction and selection efficiencies, respectively, and $\Delta\pt$ and $\Delta y^*$ are the \pt and $y^*$ interval widths.
The $D_{(s)}^+$ mesons are reconstructed through the \mbox{$\Dp \to K^{-} \pip \pip$} and \mbox{$\Ds \to K^{-} K^{+} \pip$} decay channels, where the mass of the $K^+K^-$ pair is required to be within $20$\mevcc of the known mass of the $\phi(1020)$ meson. 
The corresponding branching fractions are $\BR=(2.24\pm0.13)\%$ for the \mbox{$\Ds \to K^- K^+ \pi^+$} decay~\cite{CLEO:2008hzo}, and $\BR=(9.38\pm0.16\%)$ for the \mbox{$\Dp \to K^- \pi^+ \pi^+$} decay~\cite{ParticleDataGroup:2022pth}.

The selection criteria applied to $D_{(s)}^+$ candidates are similar to those used in the recent \Dz production measurements in $p$Pb collisions at $\sqsnn$ = 8.16 \tev~\cite{LHCb:2022dmh}.

The sample of $D_{(s)}^+$ candidates includes $D_{(s)}^+$ mesons originating from the collision point and from the decay of $b$ hadrons. These categories are referred to as ``prompt" and ``from-\textit{b}", respectively. The inclusive signal yield is determined using an extended unbinned maximum-likelihood fit to the invariant-mass distributions of the $K^-K^+\pi^+$ or $K^-\pi^+\pi^+$ combinations. The invariant mass of the signal is described by the sum of a Crystal Ball function~\cite{Skwarnicki:1986xj} and a Gaussian function, where both functions share a common mean, while the background shape is described by a linear function. The prompt signal yield is determined by fitting the distribution of $\log_{10}(\chisqip)$ of the candidates, where \chisqip is defined as the difference in the vertex-fit \chisq of a given primary vertex (PV) reconstructed with and without the candidate under consideration. Combinatorial background in the $\log_{10}(\chisqip)$ distribution is subtracted using the \textit{sPlot} method with the charm meson invariant mass as discriminating variable. The shapes of the $\log_{10}(\chisqip)$ distributions corresponding to the prompt and from-\textit{b} components are described by Bukin functions~\cite{Bukin:2007}. The parameters of the function describing the from-\textit{b} component are fixed from simulation, and the parameters describing the prompt component are allowed to float. Typical invariant mass and $\log_{10}(\chisqip)$ distributions are shown in the Supplemental Material~\cite{Supplemental:2023}.

The LHCb acceptance, trigger, reconstruction and selection efficiencies are evaluated with $p$Pb simulated samples. The track reconstruction efficiency is calibrated with MB \jpsi\to\mumu and \KS\to\pip\pim samples, using the tag-and-probe approach of Ref.~\cite{LHCb-DP-2013-002}. The PID efficiencies are estimated using a tag-and-probe method~\cite{LHCb-PUB-2016-021,LHCb-DP-2018-001}.

The various sources of systematic uncertainties considered in this measurement are listed in Table~\ref{tab:sys}. The uncertainty from the invariant mass fit is determined by describing signal and background shapes with alternative models~\cite{LHCb:2023kqs}. For the estimation of the uncertainty associated to the $\log_{10}(\chisqip)$ fit, the data are fitted again with different models and after varying any fixed parameters to evaluate the change in signal yield. The uncertainties on the tracking and PID calibration are dominated by the limited size of calibration samples. The uncertainty associated to the simulation multiplicity correction is estimated by weighting simulated events using different multiplicity variables. The larger uncertainty from multiplicity corrections in the backward region primarily stems from a worse agreement between simulation and data in that region. For the trigger efficiency, the difference between the efficiencies derived from simulation and from collision data~\cite{LHCb-PUB-2014-039} are considered as a systematic uncertainty. The uncertainties associated to the luminosity, the branching fractions and the simulated samples size are also included. 

\begin{table}[tbh]
    \centering
    \caption{Systematic uncertainties on the measured double-differential cross-section. Each range indicates the minimum and the maximum value across all kinematic intervals. The uncertainties due to the mass and $\text{log}_{10}(\chisqip)$ fits are uncorrelated across the intervals. The other sources of uncertainty are 100\% correlated between the different intervals.}
    \begin{tabular}{lcc}
        \hline
        Uncertainty source & Forward [\%] & Backward [\%]\\
        \hline
        Mass fit & 0.1 -- 6.1 & 0.1 -- 9.6 \\
        $\text{log}_{10}(\chisqip)$ fit & 0.1 -- 22.2 & 0.1 -- 17.3 \\
        Tracking calibration & 0.9 -- 3.6 & 1.4 -- 9.6  \\
        PID calibration & 1.2 -- 14.0 & 1.4 -- 8.9 \\
        Multiplicity correction & 0.5 -- 3.5 & 4.9 -- 11.3 \\
        Trigger efficiency & 0.0 -- 1.6 & 0.0 -- 1.5 \\
        Luminosity & 2.6 & 2.5 \\
        Branching fraction \Ds & 5.8 & 5.8 \\
        Branching fraction \Dp & 1.7 & 1.7 \\
        \hline
    \end{tabular}\label{tab:sys}
\end{table}

The double-differential cross-sections for prompt \Ds (\Dp) mesons are measured in the \pt range $1<\pt<13\,\gevc$ ($1<\pt<14\,\gevc$) and the rapidity ranges $1.5<y^*<4.0$ and $-5.0<y^*<-2.5$ for the forward and backward rapidity regions, respectively. The results and numerical values are given in the Supplemental Material~\cite{Supplemental:2023}. The total prompt $D_{(s)}^+$ production cross-sections, obtained by integrating the double-differential results in the measured kinematic ranges, are $42.83\pm 0.29 \pm 3.45 \mbarn~(92.36\pm 0.18 \pm 4.96 \mbarn)$ for the forward rapidity region, and $42.96\pm 0.36 \pm 4.91 \mbarn~(84.09\pm 0.17 \pm 8.39 \mbarn)$ for the backward rapidity region, where the first uncertainty is statistical and the second systematic.

The nuclear modification factor $R_{p\mathrm{Pb}}$ is defined as the ratio of differential cross-sections
\begin{equation}
\label{eq:nuclear modification factor}
R_{p \mathrm{Pb}}\left(p_{\mathrm{T}}, y^{*}\right) \equiv \frac{1}{A} \frac{\mathrm{d}^{2} \sigma_{p \mathrm{Pb}}\left(p_{\mathrm{T}}, y^{*}\right) / (\mathrm{d} p_{\mathrm{T}} \mathrm{d} y^{*})}{\mathrm{d}^{2} \sigma_{pp}\left(p_{\mathrm{T}}, y^{*}\right) / (\mathrm{d} p_{\mathrm{T}} \mathrm{d} y^{*})},
\end{equation}
where $A = 208$ is the lead nucleus mass number and $\sigma_{pp}$ is the prompt $D_{(s)}^+$ meson cross-section in $pp$ collisions at $\sqs = 8.16 \tev$. The latter are obtained by an interpolation between LHCb measurements at $\sqs = 5.02 \tev$ and $\sqs = 13 \tev$~\cite{LHCb-PAPER-2016-042,LHCb-PAPER-2015-041}. 
The interpolation is performed within the common kinematic range $1 < \pt < 10 \gevc$ and $2.0 < y < 4.5$, using a power-law function. The difference obtained when using a linear function is assigned as a systematic uncertainty.

\begin{figure}[tb]
    \begin{center}
    \includegraphics[width=1.0\linewidth]{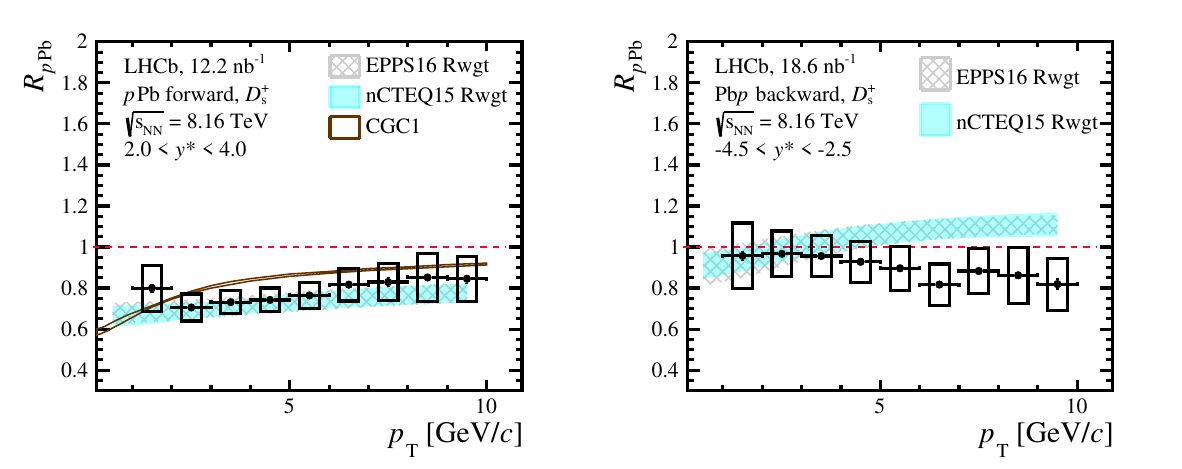}
    \includegraphics[width=1.0\linewidth]{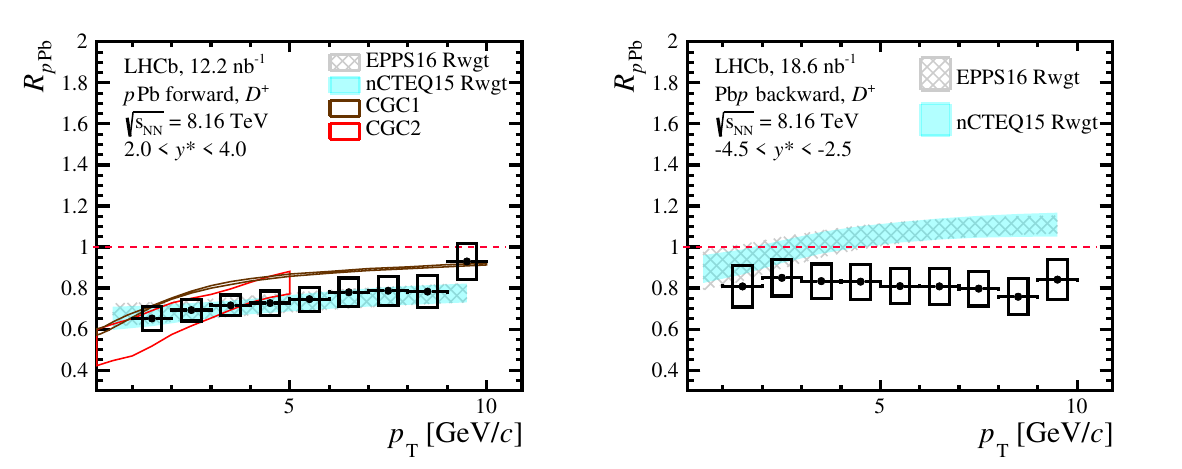}
    \vspace*{-0.5cm}
    \end{center}
    \caption{Nuclear modification factor $R_{p\mathrm{Pb}}$ as a function of \pt for prompt (upper) \Ds and (lower) \Dp mesons. Forward rapidity results are shown on the left and backward rapidity on the right. The vertical error bars show the statistical uncertainties and the boxes show the systematic uncertainties. The theoretical calculations are also shown~\cite{Eskola:2016oht, Kovarik:2015cma, Ducloue:2015gfa, Ducloue:2016ywt, Ma:2018bax}.}
    \label{fig:RpPb_pt}
\end{figure}

The nuclear modification factors for $D_{(s)}^+$ mesons as a function of \pt are displayed in Fig.~\ref{fig:RpPb_pt}, where the results are integrated over the rapidity range $2.0 < y^{*} < 4.0$ for the forward rapidity region and $-4.5 < y^{*} < -2.5$ for the backward region. A significant suppression of $D_{(s)}^+$ production in $p$Pb collisions, with respect to those in $pp$ collisions scaled by the lead mass number, is observed at forward rapidity. Figures showing $R_{p\mathrm{Pb}}$ in different $y^*$ intervals of width $\Delta y^*=0.5$, as well as the numerical values, are given in the Supplemental Material~\cite{Supplemental:2023}. 

The $R_{p\mathrm{Pb}}$ results are compared with nPDF theoretical calculations. 
These calculations use the  HELAC-Onia approach~\cite{Shao:2012iz,Shao:2015vga}, which is based on a data-driven modeling of the scattering at partonic level folded with free proton PDFs~\cite{Lansberg:2016deg}. They are first tuned by fitting the cross-sections measured in $pp$ collisions at the LHC. Then, the modified PDFs of nucleons in the Pb nucleus are introduced to calculate the cross-sections in $p$Pb collisions and to estimate the effect of nPDFs.
Reweighted EPPS16~\cite{Eskola:2016oht} or nCTEQ15~\cite{Kovarik:2015cma} nPDF sets, which incorporate LHC heavy-flavor data~\cite{LHCb-PAPER-2017-015,ALICE:2014xjz,ALICE:2016cpm,ALICE:2016yta} in a Bayesian-reweighting analysis~\cite{Kusina:2017gkz}, are used in these calculations. This procedure leads to considerably reduced uncertainties with respect to calculations using the default nPDFs. The theoretical uncertainties shown in Fig.~\ref{fig:RpPb_pt} are dominated by the nPDF parameterisations and correspond to a 68\% confidence interval.
At forward rapidity, the calculations are in satisfactory agreement with data. At backward rapidity, the data are lower than the calculations, indicating a weaker antishadowing effect or possible final-state effects that depend weakly on charm hadronization.

The nuclear modification factors in the forward rapidity region (small momentum fraction $\it{x}$) are also compared with two calculations based on the CGC effective field theory, CGC1~\cite{Ducloue:2015gfa,Ducloue:2016ywt} and CGC2~\cite{Ma:2018bax}.  
The most significant theoretical uncertainty in CGC2 is the initial saturation scale of the target nucleus. The CGC1 predictions have much smaller uncertainties than the CGC2 predictions, as they include only variations of the charm quark mass and of the factorisation scale, which largely cancel out in the $R_{p\mathrm{Pb}}$ ratio. The CGC1 calculations are consistent with the upper bound of the CGC2 predictions and slightly overshoot the data. The CGC2 predictions show a stronger suppression than HELAC-Onia, especially for $\pt<3\gevc$.

\begin{figure}[tb]
    \begin{center}
    \includegraphics[width=0.49\linewidth]{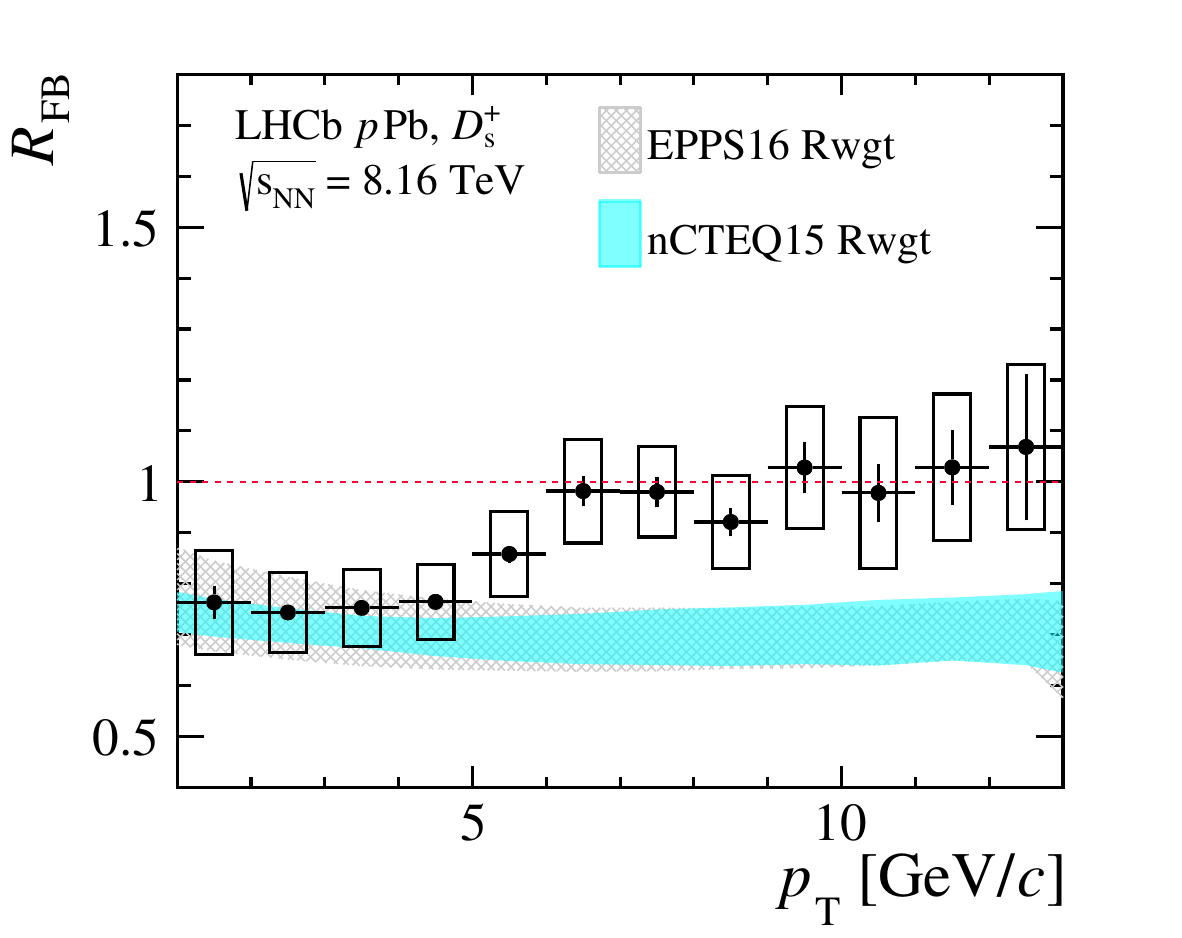}
    \includegraphics[width=0.49\linewidth]{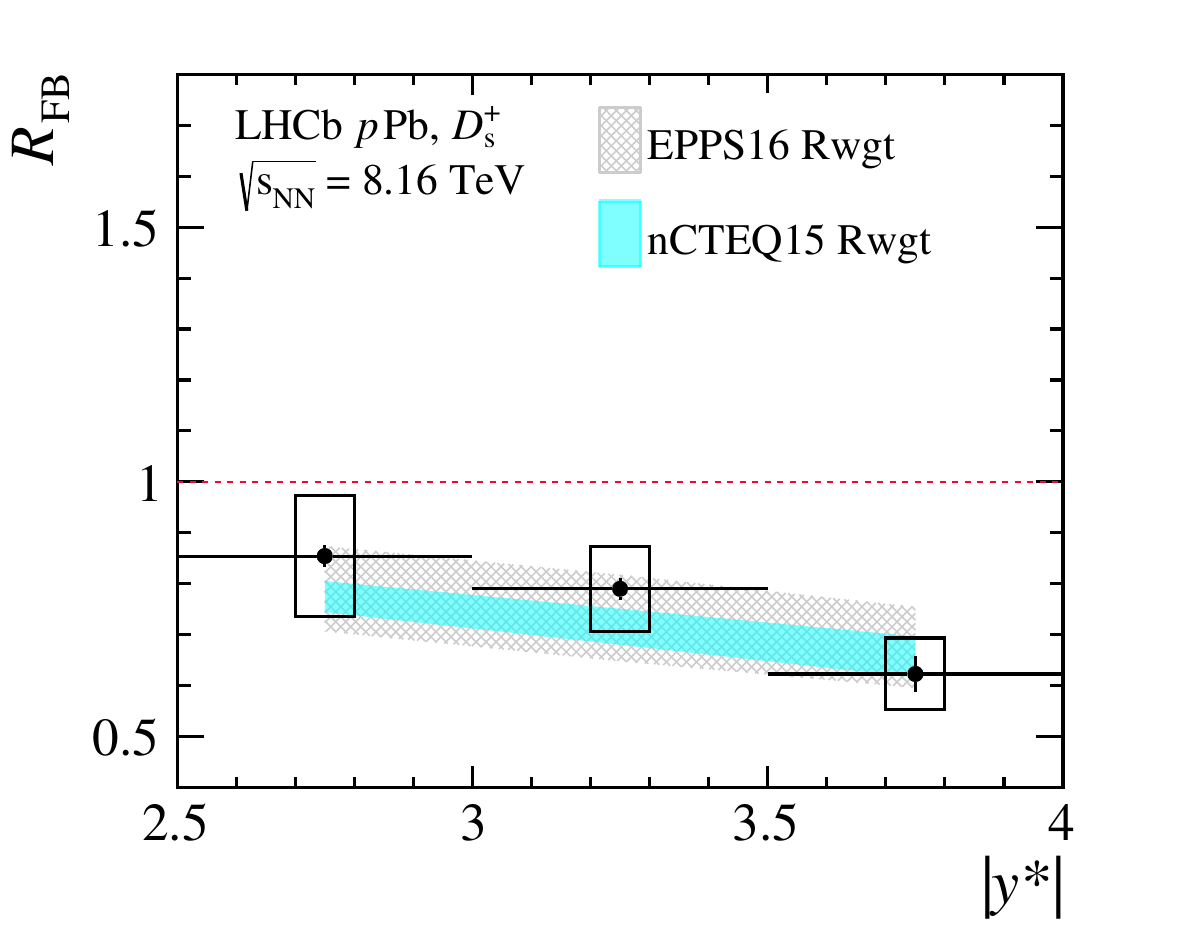}
    \includegraphics[width=0.49\linewidth]{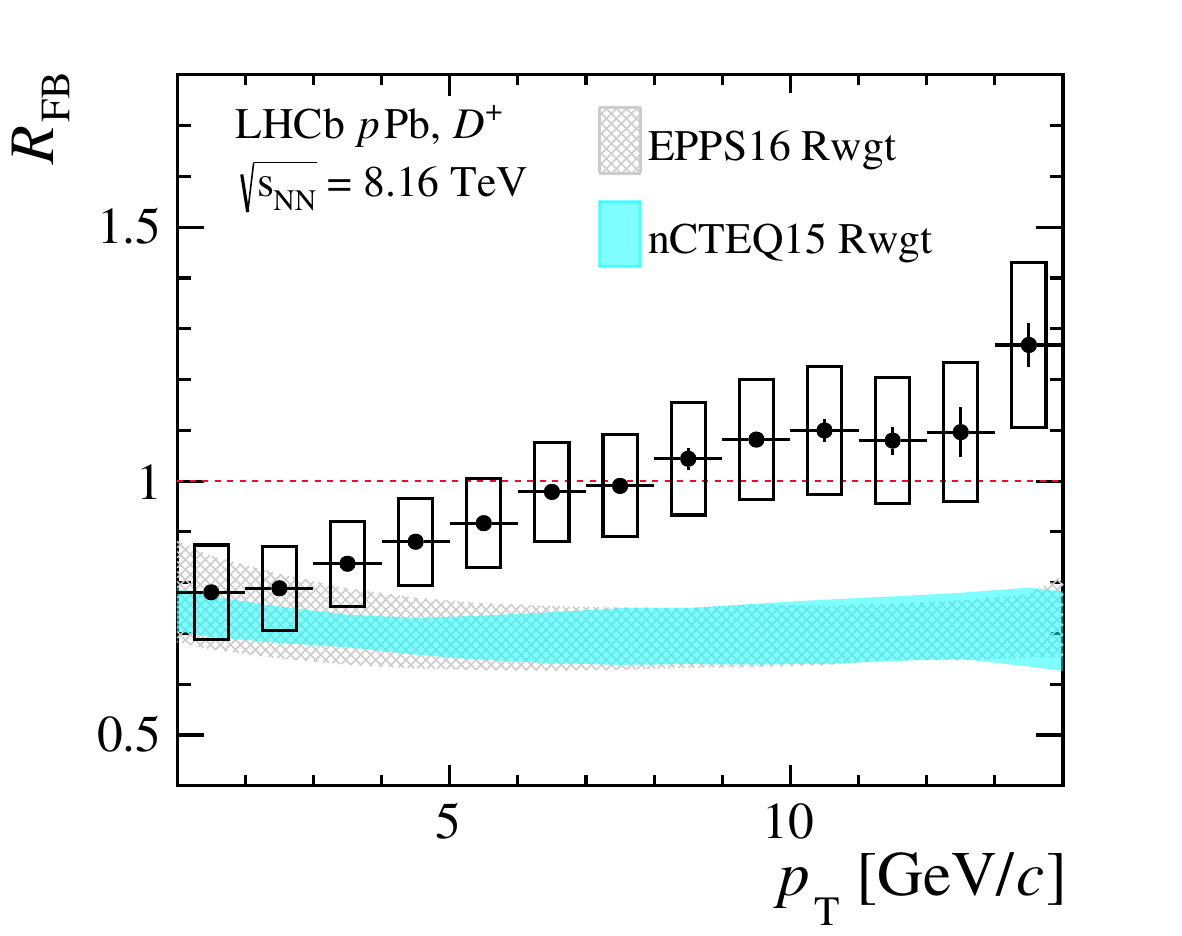}
    \includegraphics[width=0.49\linewidth]{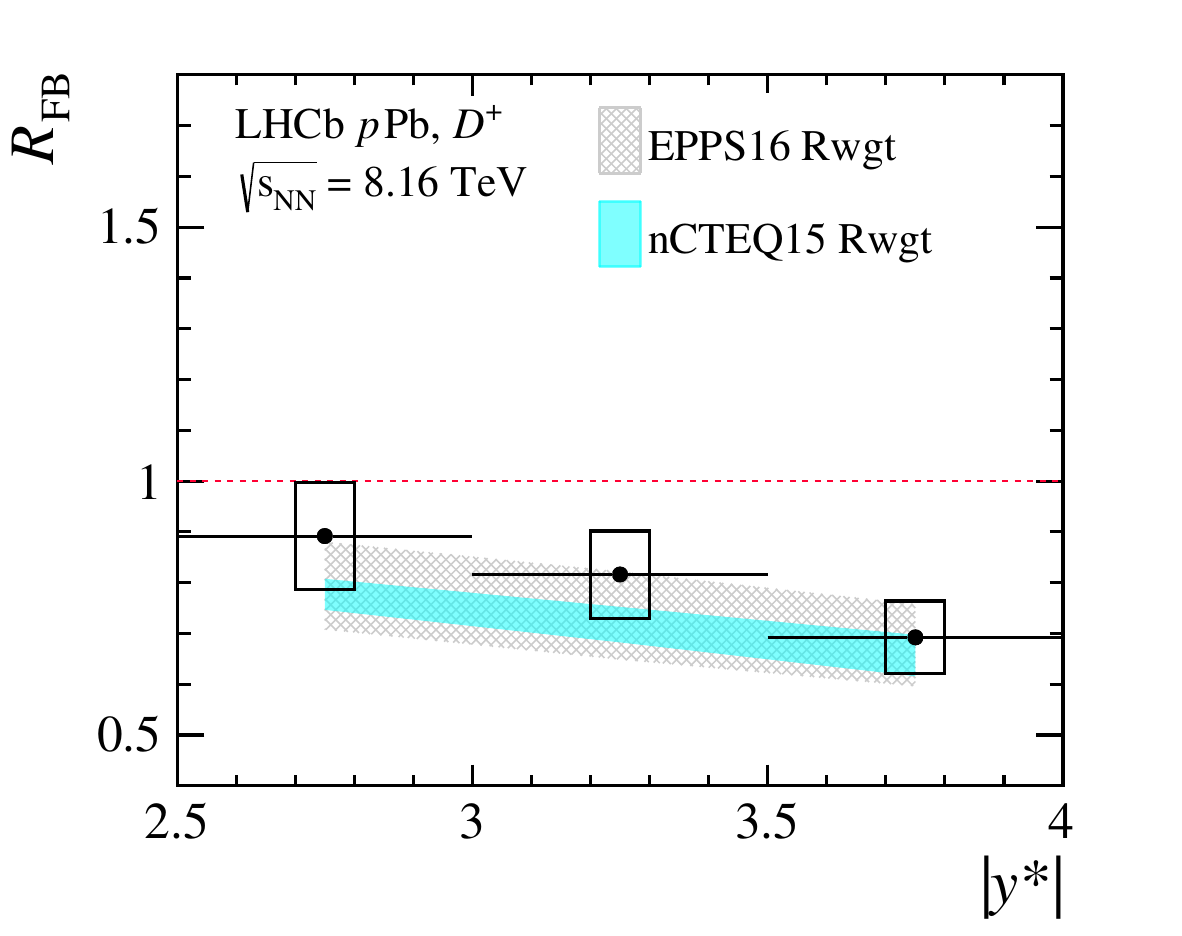}
    \vspace*{-0.5cm}
    \end{center}
    \caption{
    Forward-backward cross-section ratio $R_{\mathrm{FB}}$ for prompt (upper) \Ds and (lower) \Dp mesons as a function of (left) \pt and (right) $y^{*}$. The vertical error bars show the statistical uncertainties and the boxes show the systematic uncertainties. The coloured bands represent the theoretical calculations, incorporating nPDFs EPPS16 (gray)~\cite{Eskola:2016oht} and nCTEQ15 (cyan)~\cite{Kovarik:2015cma}.}
    \label{fig:RFB}
\end{figure}

The forward-backward cross-section ratio $R_{\mathrm{FB}}$ is defined as
\begin{equation}
    R_{\text{FB}}(\pt,|y^*|) = \frac{\deriv^2\sigma_{p\text{Pb}}(\pt,+|y^*|)/(\deriv \pt \deriv y^*)}{\deriv^2\sigma_{\text{Pb}p}(\pt,-|y^*|)/(\deriv \pt \deriv y^*)}~,
\end{equation}
and calculated in the common $|y^*|$ interval of the forward-backward acceptances, namely $2.5 < |y^*| < 4$. The measurements of $R_{\mathrm{FB}}$ are shown as a function of \pt and $|y^{*}|$ in Fig.~\ref{fig:RFB}, along with the nPDF calculations~\cite{Eskola:2016oht,Kovarik:2015cma}. Good agreement with nPDF calculations is found at low \pt, however, the data show a clear rising trend with increasing \pt, reaching unity at the highest \pt values. This is in contrast to the nPDF calculations, which predict $R_{\mathrm{FB}} \sim 0.7$ almost independently of \pt. This discrepancy originates from the observed suppression of high-\pt  ~$D_{(s)}^+$ mesons at backward rapidity.

The cross-section ratio $\sigma_{\Ds}/\sigma_{\Dp}$, which is written as
\begin{equation}\label{eq:cross-section ratio}
        \frac{\sigma_{\Ds}}{\sigma_{\Dp}} = \frac{N_{\Ds}}{N_{\Dp}}
        \times\frac{\mathcal{B}_{\Dp}}{\mathcal{B}_{\Ds}}
        \times\frac{\epsilon^{\text{acc}}_{\Dp}}{\epsilon^{\text{acc}}_{\Ds}}
        \times\frac{\epsilon^{\text{trig}}_{\Dp}}{\epsilon^{\text{trig}}_{\Ds}}
        \times\frac{\epsilon^{\text{PID}}_{\Dp}}{\epsilon^{\text{PID}}_{\Ds}}
        \times\frac{\epsilon^{\text{rec\&sel}}_{\Dp}}{\epsilon^{\text{rec\&sel}}_{\Ds}},
\end{equation}
is more precisely measured thanks to a cancellation of systematic uncertainties. The dependence of $\sigma_{\Ds}/\sigma_{\Dp}$ versus the primary charged particle multiplicity is measured in the $D_{(s)}^+$ kinematic intervals $2<\pt<12$\gevc and $1.8<y^{*}<3.3$ ($-4.3<y^{*}<-2.8$) for forward (backward) rapidity. The primary charged particle multiplicity, denoted as $N_{\text{ch}}$, represents the number of charged particles originating from the collisions, including decay products. In this Letter, it is estimated within the forward-pseudorapidity region ($2 < \eta < 4.8$) by measuring the number of tracks used to reconstruct the primary vertex, denoted as $N^{\text{PV}}_{\text{Tracks}}$. The correlation between the measured $N^{\text{PV}}_{\text{Tracks}}$ and $N_{\text{ch}}$ is obtained from simulation.

\begin{figure}[tb]
  \begin{center}
    \includegraphics[width=1.\linewidth]{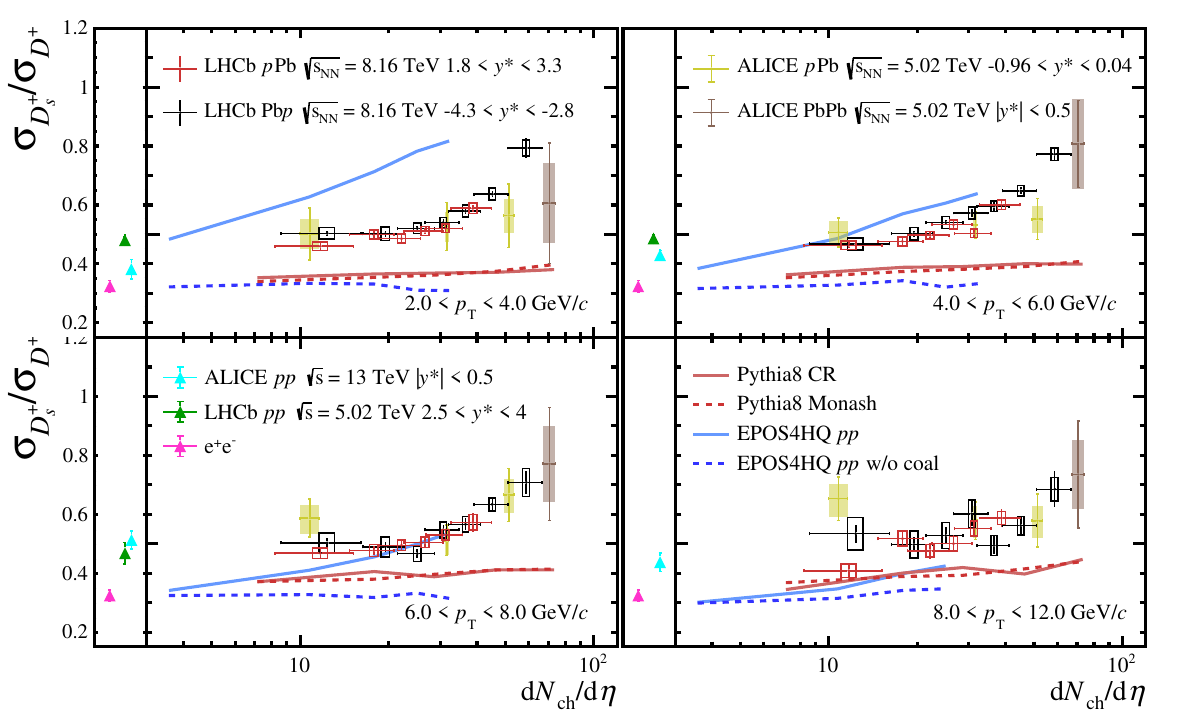}
    \vspace*{-0.5cm}
  \end{center}
  \caption{
    Cross-section ratio $\sigma_{\Ds}/\sigma_{\Dp}$ versus the primary charged particles per unit of pseudorapidity in \epem~\cite{Lisovyi:2015uqa}, $pp$~\cite{LHCb:2016ikn,ALICE:2023sgl}, $p$Pb~\cite{ALICE:2019fhe}, PbPb~\cite{ALICE:2018lyv} collisions in different $D_{(s)}^+$ \pt ranges. The vertical error bars show the statistical uncertainties and the boxes show the systematic uncertainties. The colored bands contain both statistical and systematic uncertainties. The calculations from Pythia 8~\cite{Christiansen:2015yqa, Skands:2014pea}, EPOS4HQ~\cite{Zhao:2023ucp,Zhao:2024ecc} and EPOS4HQ without coalescence mechanism are also shown. These calculations are applicable to $pp$ collisions at $\sqs = 8.16 \tev$ within the rapidity range of $1.8<y^{*}<3.3$.}
  \label{fig:DsDpratio}
\end{figure}

Figure~\ref{fig:DsDpratio} shows the dependence of $\sigma_{\Ds}/\sigma_{\Dp}$ on primary charged particle multiplicity in four different \pt intervals (integrated over rapidity). Plots of $\sigma_{\Ds}/\sigma_{\Dp}$ in different $y^*$ intervals and the derived numerical values are given in the Supplemental Material~\cite{Supplemental:2023}. These measurements show that the $\sigma_{\Ds}/\sigma_{\Dp}$ ratio increases significantly as a function of the primary charged particle multiplicity, especially in the low-\pt and backward rapidity regions. They deviate from a flat distribution, expected if only the fragmentation mechanism is considered, by 6.1 ($2<\pt<4$\gevc), 6.8 ($4<\pt<6$\gevc), 2.7 ($6<\pt<8$\gevc) and 3.2 ($8<\pt<12$\gevc) standard deviations in the forward rapidity region, and by 7.9 ($2<\pt<4$\gevc), 10.5 ($4<\pt<6$\gevc), 4.4 ($6<\pt<8$\gevc) and 1.1 ($8<\pt<12$\gevc) standard deviations at backward rapidity.
As a comparison, the measured $\sigma_{\Ds}/\sigma_{\Dp}$ ratios in \epem~\cite{Lisovyi:2015uqa}, $pp$~\cite{LHCb:2016ikn,ALICE:2023sgl}, $p$Pb~\cite{ALICE:2019fhe} and PbPb~\cite{ALICE:2018lyv} collisions are also shown in the Fig.~\ref{fig:DsDpratio}. There are significant differences in the $\sigma_{\Ds}/\sigma_{\Dp}$ ratios between $pp$ and PbPb collisions. The LHCb measurements reveal a trend where the ratio tends to resemble that of $pp$ collisions in low-multiplicity $p$Pb collisions, while it converges towards the behavior observed in PbPb collisions in high-multiplicity $p$Pb collisions. In $p$Pb collisions, the LHCb data are compatible with the ratio measured by ALICE within uncertainties. The $\sigma_{\Ds}/\sigma_{\Dp}$ pattern is similar in both the forward and backward rapidity regions. This suggests that the $\sigma_{\Ds}/\sigma_{\Dp}$ ratio is independent of rapidity, and the mechanism contributing to this ratio increase is strongly correlated with the charged particle density. Additionally, theoretical calculations are compared using PYTHIA 8 with Monash~\cite{Skands:2014pea} and CR~\cite{Christiansen:2015yqa} tunes, along with EPOS4HQ~\cite{Zhao:2023ucp,Zhao:2024ecc}. EPOS4HQ extends the EPOS4 framework to include heavy quarks and incorporates a coalescence mechanism in hadronization. These calculations are applicable to $pp$ collisions. Theoretical calculations from Pythia 8 underestimate experimental measurements and and do not fully capture the trends dependent on multiplicity. While EPOS4HQ also exhibits some discrepancies with experimental data, it can depict the multiplicity-dependent trends across all \pt intervals by introducing a coalescence mechanism.


In summary, the prompt $D_{(s)}^+$ production cross-sections are measured by the \lhcb experiment in $p$Pb collisions at $\sqsnn=8.16 \tev$, both in the forward and backward rapidity regions. The nuclear modification factors are measured and found to be consistent with the previous results with \Dz mesons~\cite{LHCb:2022dmh}. The results show a strong suppression of the $D_{(s)}^+$ cross-sections at forward rapidity, consistent with the nPDF and CGC effective theory calculations. At backward rapidity, the $R_{p\mathrm{Pb}}$ values of $D_{(s)}^+$ mesons are lower than nPDF calculations at high \pt, indicating a weaker antishadowing effect than predicted by the models or additional hadronization-independent final-state effects. Moreover, the forward-backward cross-section ratio also shows a deviation from the nPDF calculations at high \pt. Combined with the nuclear modification factors, this deviation may arise from the observed suppression of high-\pt  ~$D_{(s)}^+$ mesons at backward rapidity. The production of \Ds mesons is significantly enhanced relative to \Dp mesons in high particle multiplicity proton-lead collision events, in particular for low \pt and backward rapidity. This is the first observation of strangeness enhancement in charm quark hadronization in high-multiplicity small collision systems. The multiplicity-dependent trend is well understood within EPOS4HQ.

\section*{Acknowledgements}
%
%
\noindent We express our gratitude to our colleagues in the CERN
accelerator departments for the excellent performance of the LHC. We
thank the technical and administrative staff at the LHCb
institutes.
We acknowledge support from CERN and from the national agencies:
CAPES, CNPq, FAPERJ and FINEP (Brazil); 
MOST and NSFC (China); 
CNRS/IN2P3 (France); 
BMBF, DFG and MPG (Germany); 
INFN (Italy); 
NWO (Netherlands); 
MNiSW and NCN (Poland); 
MCID/IFA (Romania); 
MICINN (Spain); 
SNSF and SER (Switzerland); 
NASU (Ukraine); 
STFC (United Kingdom); 
DOE NP and NSF (USA).
We acknowledge the computing resources that are provided by CERN, IN2P3
(France), KIT and DESY (Germany), INFN (Italy), SURF (Netherlands),
PIC (Spain), GridPP (United Kingdom), 
CSCS (Switzerland), IFIN-HH (Romania), CBPF (Brazil),
and Polish WLCG (Poland).
We are indebted to the communities behind the multiple open-source
software packages on which we depend.
Individual groups or members have received support from
ARC and ARDC (Australia);
Key Research Program of Frontier Sciences of CAS, CAS PIFI, CAS CCEPP, 
Fundamental Research Funds for the Central Universities, 
and Sci. \& Tech. Program of Guangzhou (China);
Minciencias (Colombia);
EPLANET, Marie Sk\l{}odowska-Curie Actions, ERC and NextGenerationEU (European Union);
A*MIDEX, ANR, IPhU and Labex P2IO, and R\'{e}gion Auvergne-Rh\^{o}ne-Alpes (France);
AvH Foundation (Germany);
ICSC (Italy); 
GVA, XuntaGal, GENCAT, Inditex, InTalent and Prog.~Atracci\'on Talento, CM (Spain);
SRC (Sweden);
the Leverhulme Trust, the Royal Society
 and UKRI (United Kingdom).

\clearpage
\section*{Supplemental material}
\label{sec:Supplementary}
\begin{figure}[h]
  \begin{center}
    \includegraphics[width=0.99\linewidth]{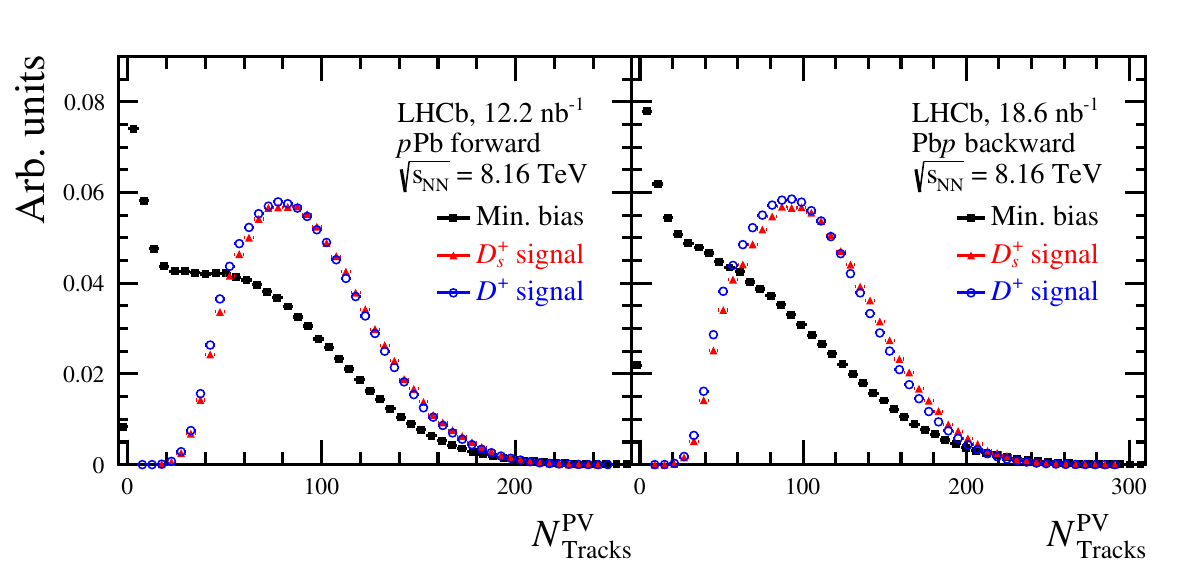}
    \vspace*{-0.5cm}
  \end{center}
  \caption{
    Distribution of the number of charged tracks used to reconstruct the PV for $D_{(s)}^+$ signal and minimum-bias events in (left) forward and (right) backward configurations, each with only one primary vertex. The vertical scale is arbitrary.}
  \label{fig:Minbias}
\end{figure}

The multiplicity variable used in this paper is the number of tracks used to reconstruct the primary vertex (PV), $N^{\text{PV}}_{\text{Tracks}}$. The $N^{\text{PV}}_{\text{Tracks}}$ distribution is affected by the position of the primary vertex along the beam axis. This is due to the asymmetry of the $p$Pb collisions and the pseudorapidity coverage limitations of vertex locator (VELO). To address this effect, a selection is made on the position of the primary vertex along the beam axis to ensure the stable distribution of $N^{\text{PV}}_{\text{Tracks}}$ within this range. The $N^{\text{PV}}_{\text{Tracks}}$ distributions for three categories of events, namely minimum-bias events, \Ds signal events, and \Dp signal events, with the additional requirement of one reconstructed primary vertex for each category, are shown in Fig.~\ref{fig:Minbias}. The multiplicity distributions for \Ds and \Dp signal events are extracted from data; background is removed using the \textit{sPlot} method \cite{Pivk:2004ty}. 

The $\sigma_{\Ds}/\sigma_{\Dp}$ ratios are extracted in different multiplicity classes defined as 10-60, 60-80, 80-100, 100-120, 120-140, 140-200 (10-60, 60-80, 80-100, 100-120, 120-140, 140-180, 180-250) $N^{\text{PV}}_{\text{Tracks}}$ for forward (backward) rapidity region. The normalised multiplicity is defined as $N^{\text{PV}}_{\text{Tracks}}/\langle N^{\text{PV}}_{\text{Tracks}}\rangle_{\text{MB}}$, where $\langle N^{\text{PV}}_{\text{Tracks}}\rangle_{\text{MB}}$ is the average multiplicity for MB events in the corresponding beam configuration. For the forward (backward) rapidity sample $\langle N^{\text{PV}}_{\text{Tracks}}\rangle_{\text{MB}}$ = 60.3 (69.0) with negligible uncertainty. The primary charged particles per unity of pseudorapidity is defined as $\text{d}N_{\text{ch}}/\text{d}\eta$, where $\eta$ range from 2 to 4.8. The primary charged particle multiplicity, denoted as $N_{\text{ch}}$, represents the number of charged particles originating from the collisions, including decay products. It is estimated within the forward-pseudorapidity region ($2 < \eta < 4.8$) by measuring $N^{\text{PV}}_{\text{Tracks}}$. In the forward (backward) rapidity region, the means and standard deviations of $N_{\text{ch}}$ in different $N^{\text{PV}}_{\text{Tracks}}$ intervals are denoted as 32.8 $\pm$ 9.8, 49.9 $\pm$ 8.6, 61.9 $\pm$ 10.0, 74.5 $\pm$ 11.4, 87.5 $\pm$ 12.5, 108.4 $\pm$ 17.0 (34.6 $\pm$ 10.5, 54.6 $\pm$ 8.8, 70.1 $\pm$ 10.2, 86.0 $\pm$ 11.5, 102.3 $\pm$ 12.7, 126.2 $\pm$ 16.7, 164.7 $\pm$ 22.1).

\begin{figure}[ht]
  \begin{center}
    \includegraphics[width=0.49\linewidth]{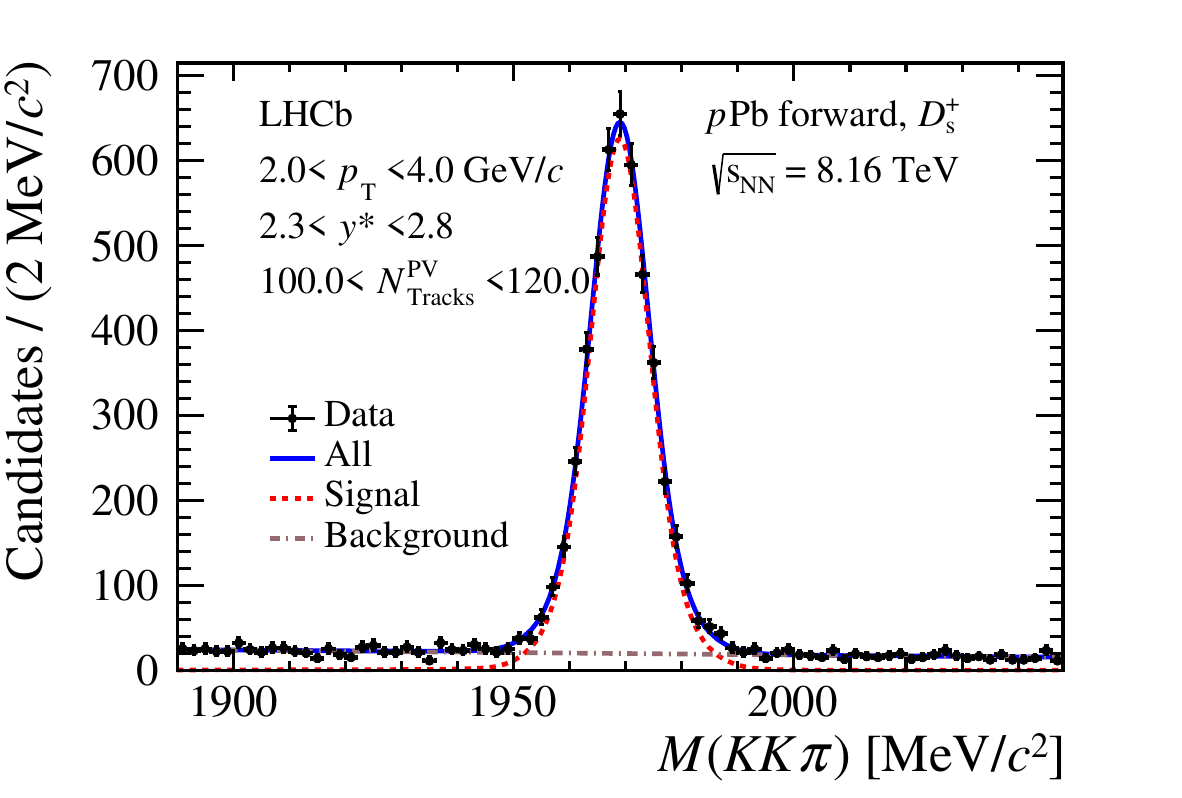}
    \includegraphics[width=0.49\linewidth]{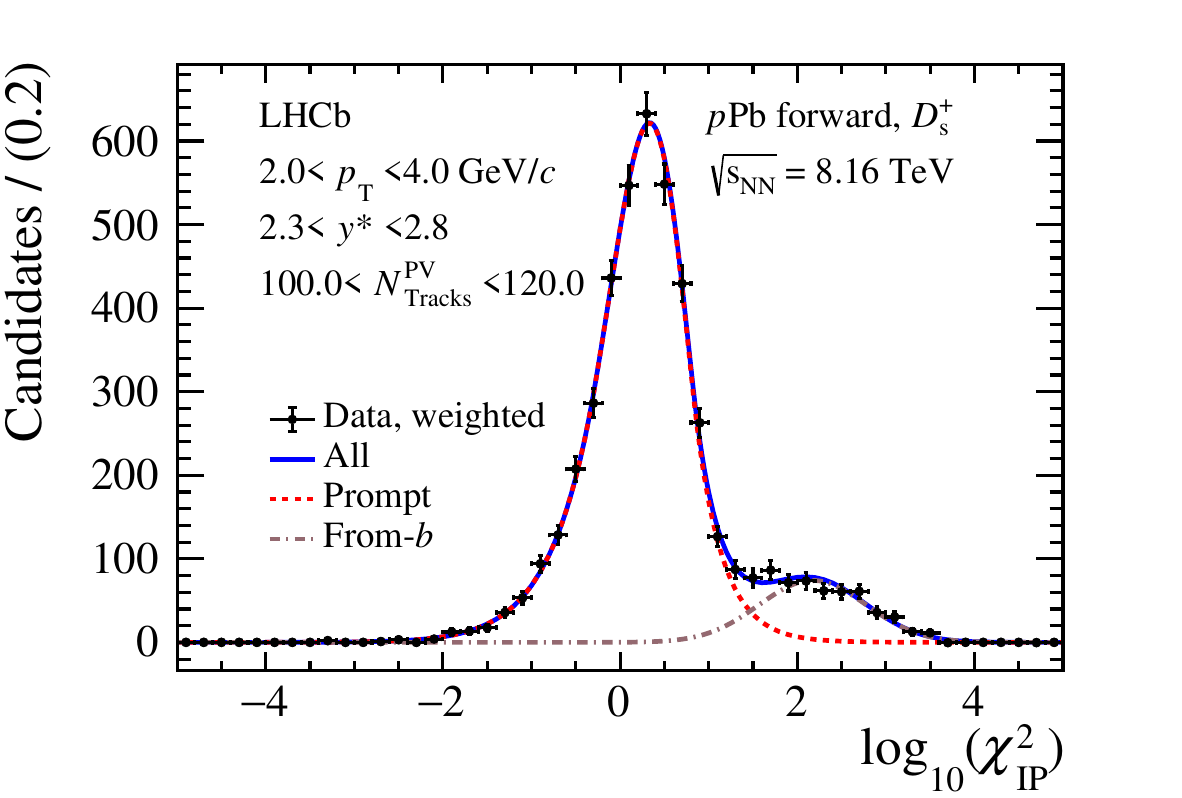}
    \vspace*{-0.5cm}
  \end{center}
  \caption{Distributions of (left) $M(KK\pi)$ and (right) $\log_{10}(\chisqip)$ for inclusive \Ds mesons in the forward data sample in the interval of $2.0 < \pt < 4.0 \gevc$, $2.3 < y^{*} < 2.8$ and $100 < N^{\text{PV}}_{\text{Tracks}} < 120$. The fit results are overlaid. For the $\log_{10}(\chisqip)$ fit, the data are weighted using the \textit{sPlot} method to subtract the background component.}
  \label{fig:Dsmassfit0}
\end{figure}

\begin{figure}[ht]
  \begin{center}
    \includegraphics[width=0.49\linewidth]{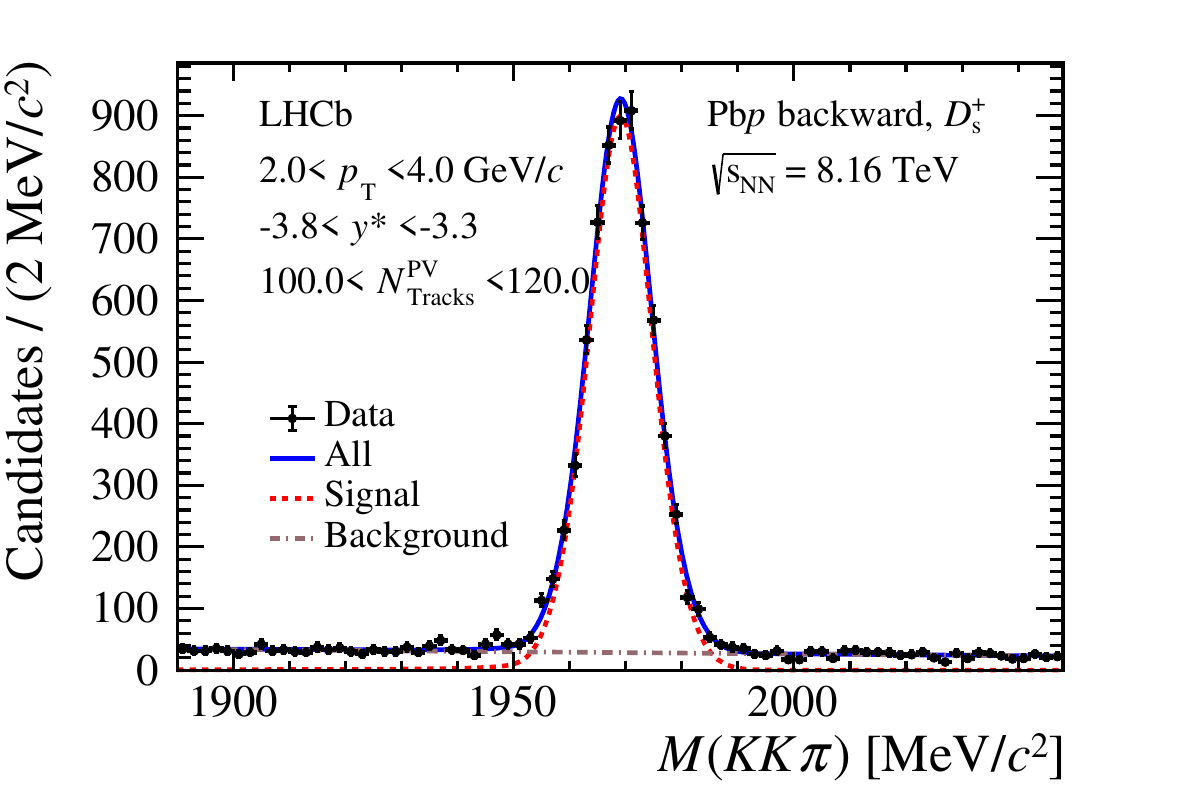}
    \includegraphics[width=0.49\linewidth]{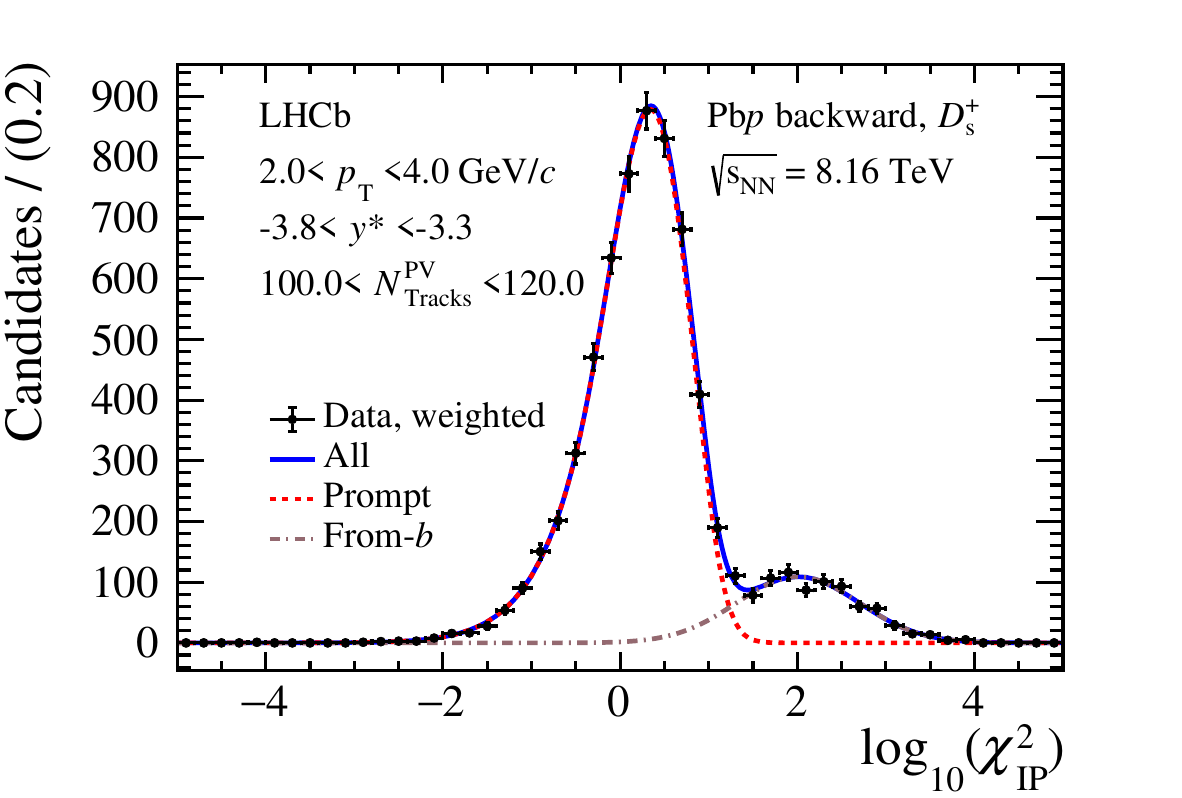}
    \vspace*{-0.5cm}
  \end{center}
  \caption{Distributions of (left) $M(KK\pi)$ and (right) $\log_{10}(\chisqip)$ for inclusive \Ds mesons in the backward data sample in the interval of $2.0<\pt<4.0\gevc$, $-3.8< y^{*} <-3.3$ and $100< N^{\text{PV}}_{\text{Tracks}} <120$. The fit results are overlaid. For the $\log_{10}(\chisqip)$ fit, the data are weighted using the \textit{sPlot} method to subtract the background component.}
  \label{fig:Dsmassfit1}
\end{figure}

\begin{figure}[ht]
  \begin{center}
    \includegraphics[width=0.49\linewidth]{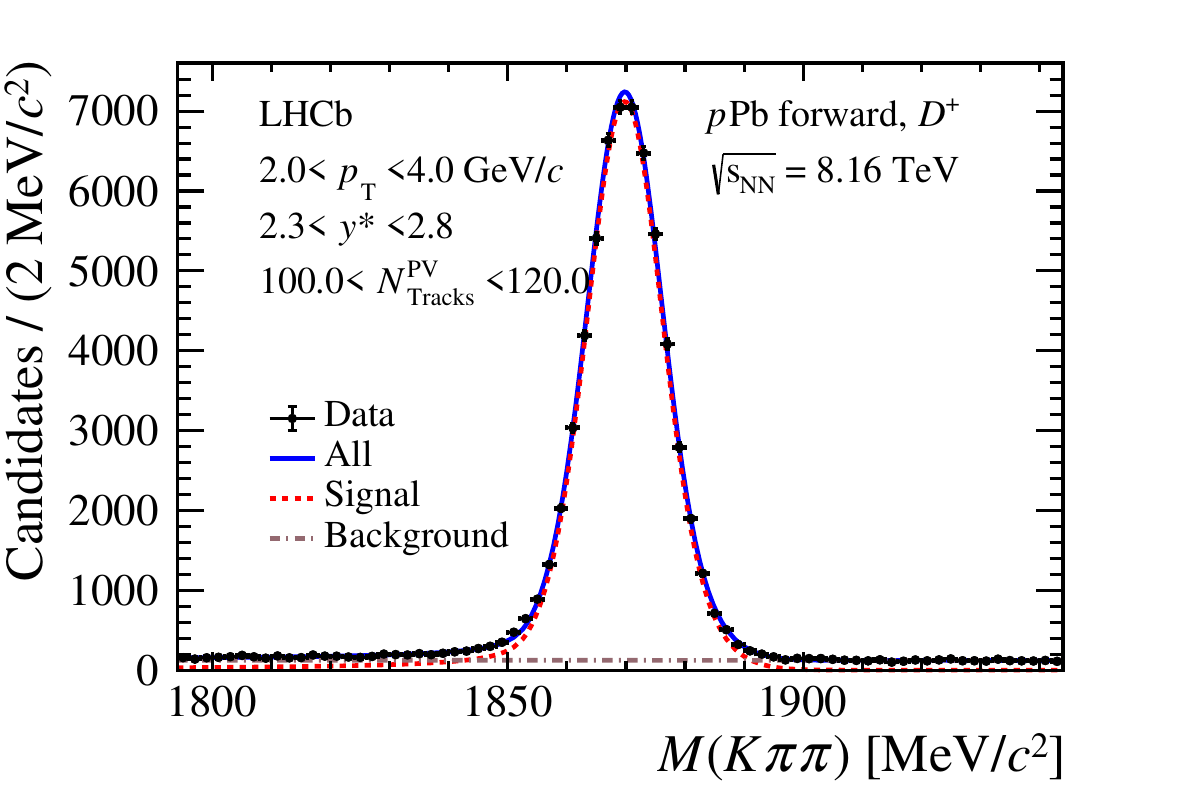}
    \includegraphics[width=0.49\linewidth]{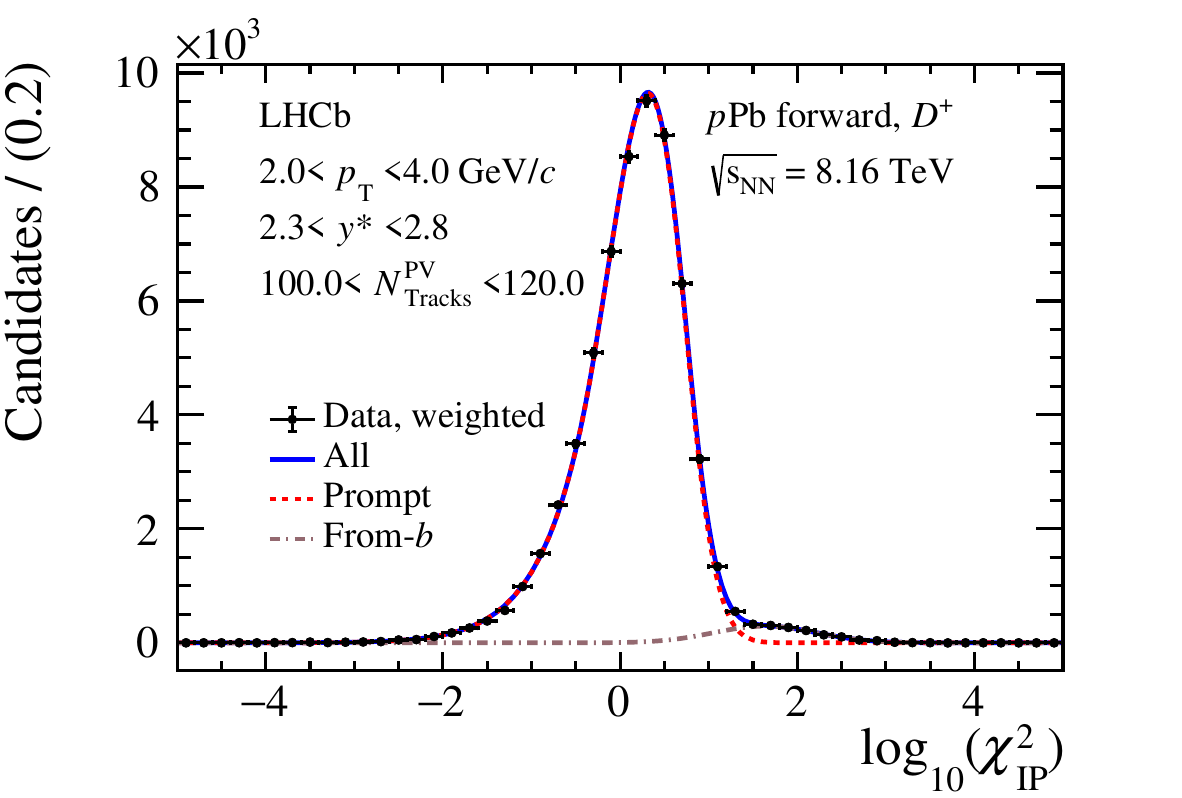}
    \vspace*{-0.5cm}
  \end{center}
  \caption{Distributions of (left) $M(K\pi\pi)$ and (right) $\log_{10}(\chisqip)$ for inclusive \Dp mesons in the forward data sample in the interval of $2.0 < \pt < 4.0 \gevc$, $2.3 < y^{*} < 2.8$ and $100 < N^{\text{PV}}_{\text{Tracks}} < 120$. The fit results are overlaid. For the $\log_{10}(\chisqip)$ fit, the data are weighted using the \textit{sPlot} method to subtract the background component.}
  \label{fig:Dpmassfit0}
\end{figure}

\begin{figure}[ht]
  \begin{center}
    \includegraphics[width=0.49\linewidth]{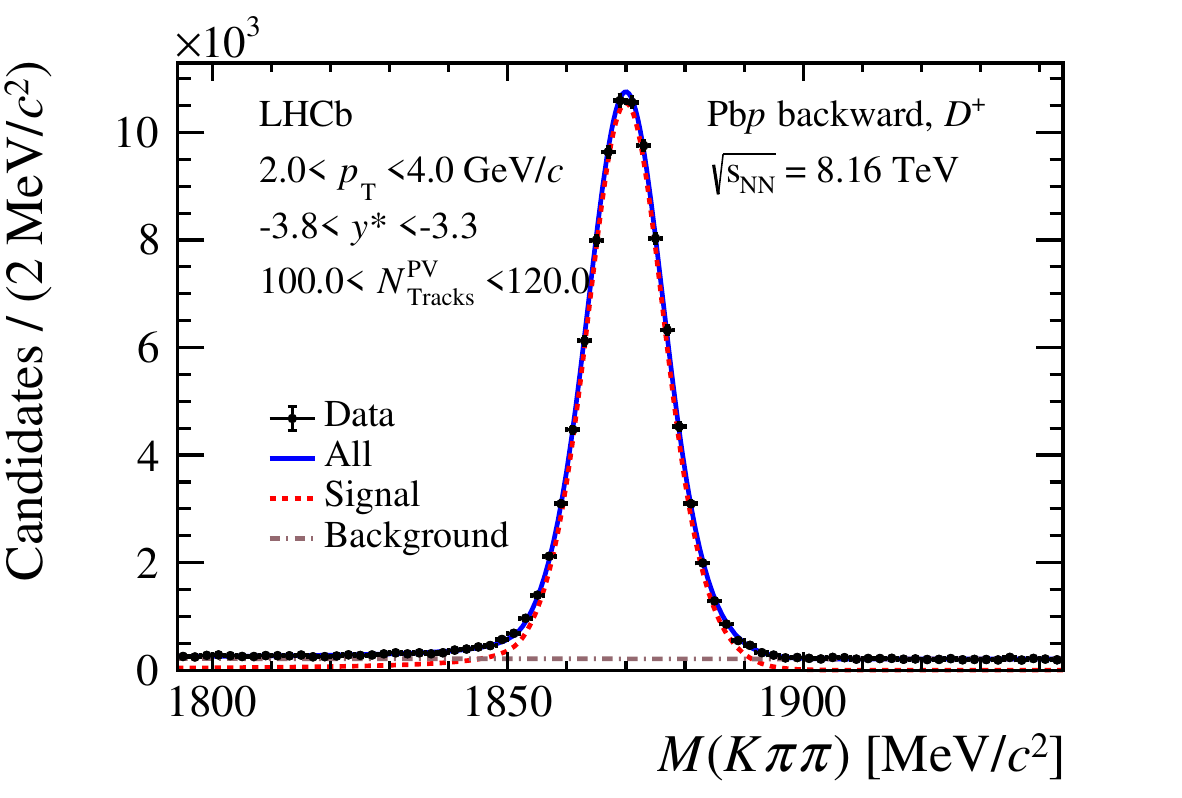}
    \includegraphics[width=0.49\linewidth]{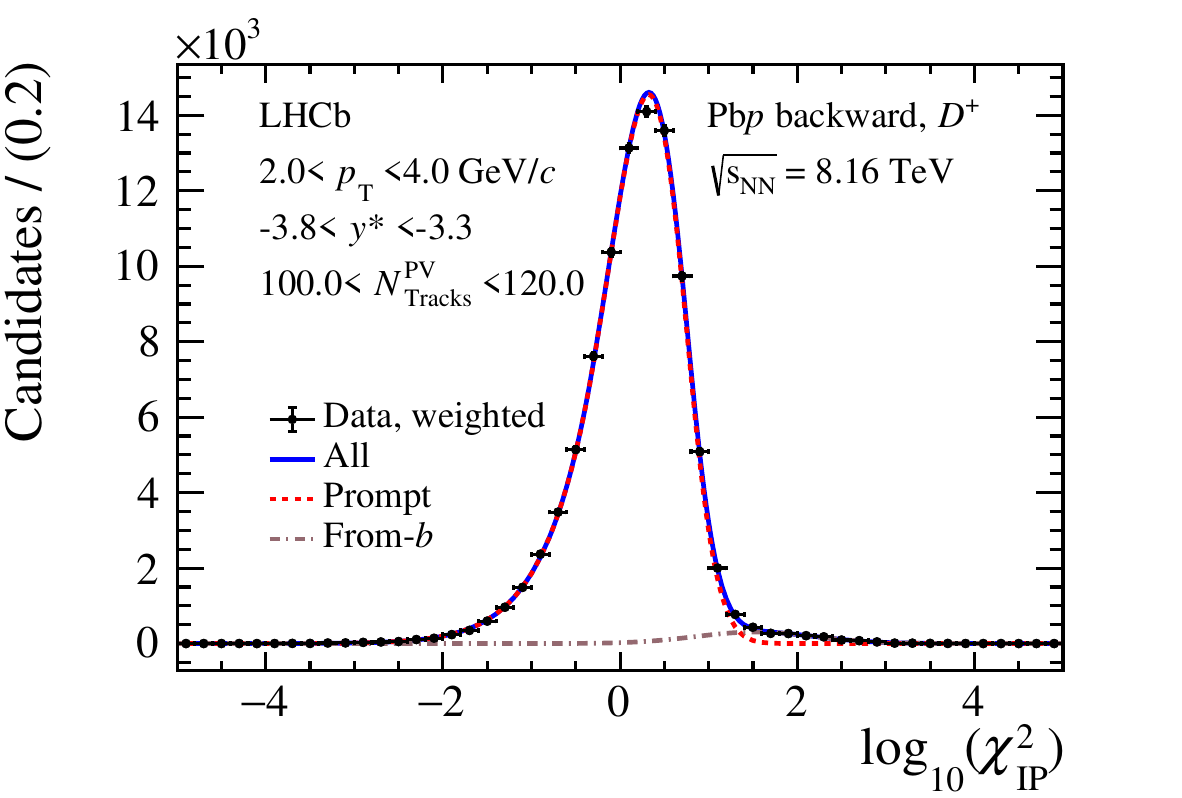}
    \vspace*{-0.5cm}
  \end{center}
  \caption{Distributions of (left) $M(K\pi\pi)$ and (right) $\log_{10}(\chisqip)$ for inclusive \Dp mesons in the backward data sample in the interval of $2.0<\pt< 4.0\gevc$, $-3.8< y^{*}< -3.3$ and $100< N^{\text{PV}}_{\text{Tracks}} <120$. The fit results are overlaid. For the $\log_{10}(\chisqip)$ fit, the data are weighted using the \textit{sPlot} method to subtract the background component.}
  \label{fig:Dpmassfit1}
\end{figure}

The results of the fits to the invariant-mass and $\log_{10}(\chisqip)$ distributions in the forward and backward rapidity intervals are shown in Fig.~\ref{fig:Dsmassfit0}--\ref{fig:Dpmassfit1}.

\begin{figure}[ht]
    \begin{center}
    \includegraphics[width=0.49\linewidth]{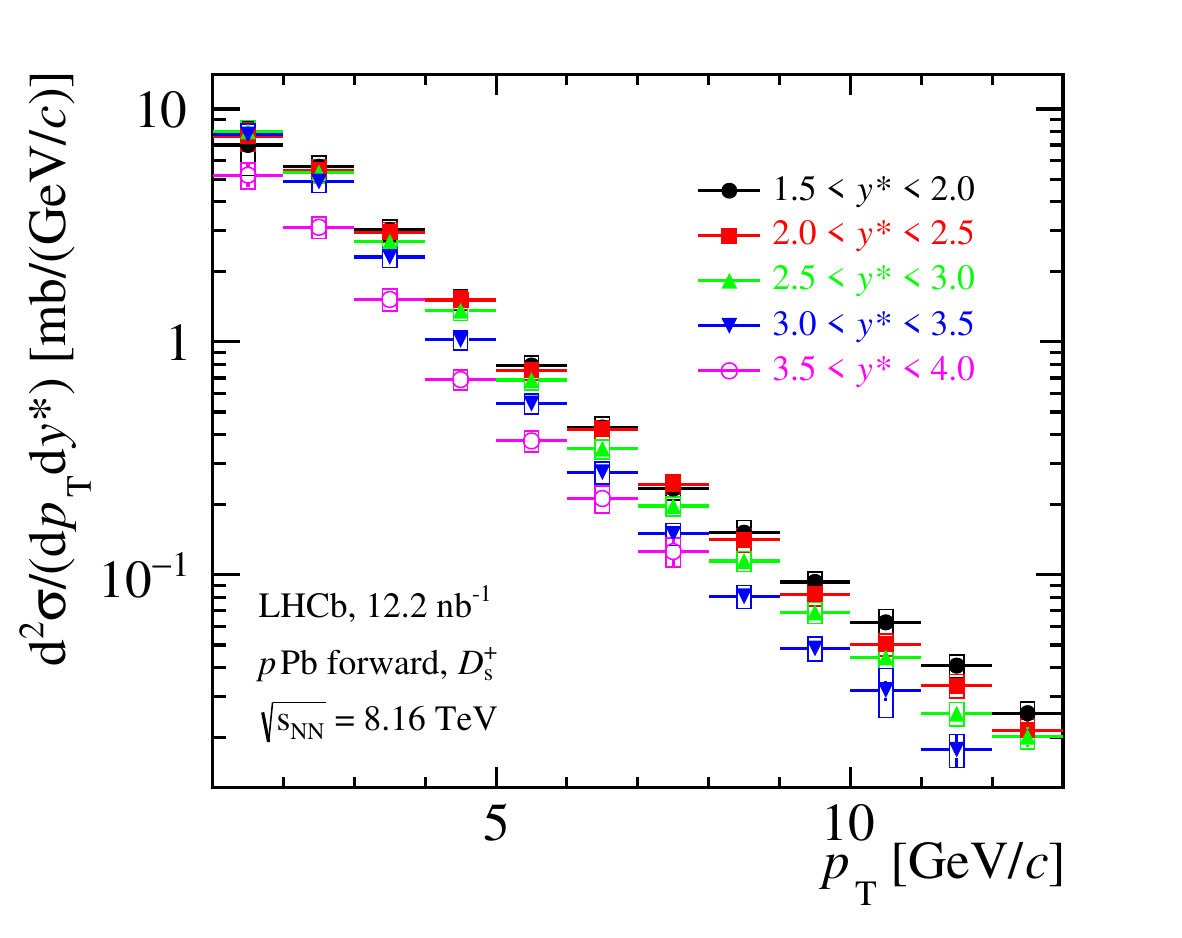}
    \includegraphics[width=0.49\linewidth]{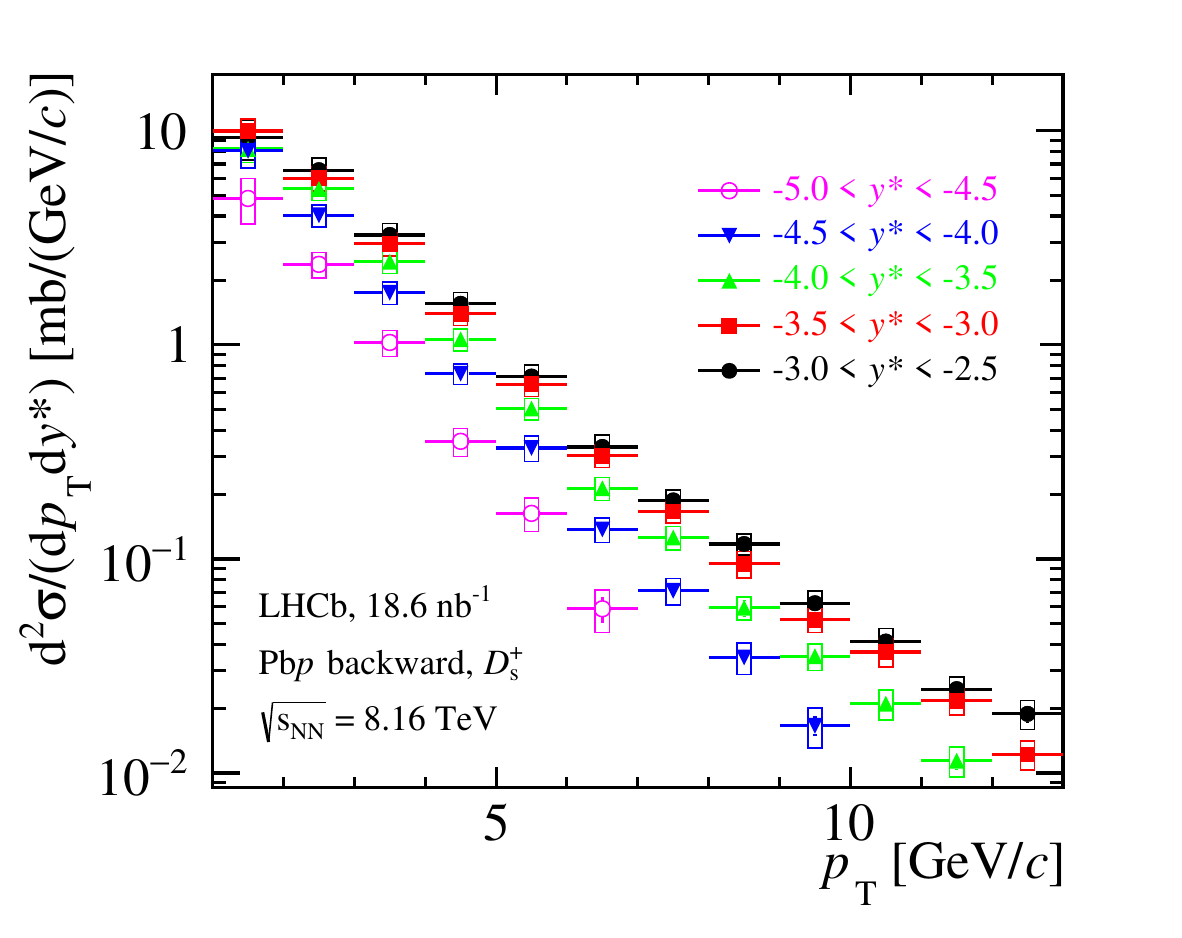}
    \vspace*{-0.5cm}
    \end{center}
    \caption{Double-differential cross-section of prompt $\Ds$ production in $p$Pb collisions at (left) forward and (right) backward rapidities. The vertical error bars show the statistical uncertainties and the boxes show the systematic uncertainties. }
    \label{fig:cross_section_Ds}
\end{figure}

\begin{figure}[ht]
    \begin{center}
    \includegraphics[width=0.49\linewidth]{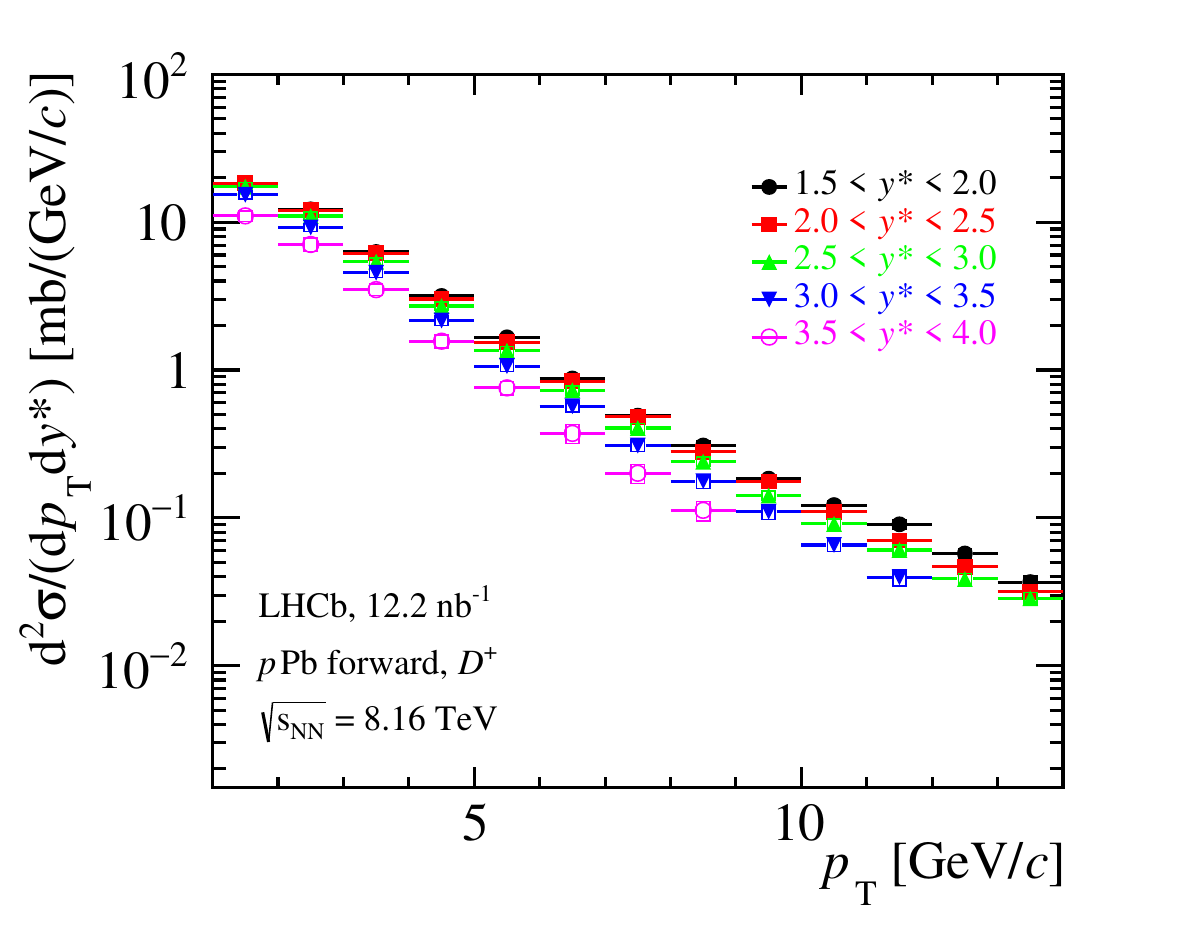}
    \includegraphics[width=0.49\linewidth]{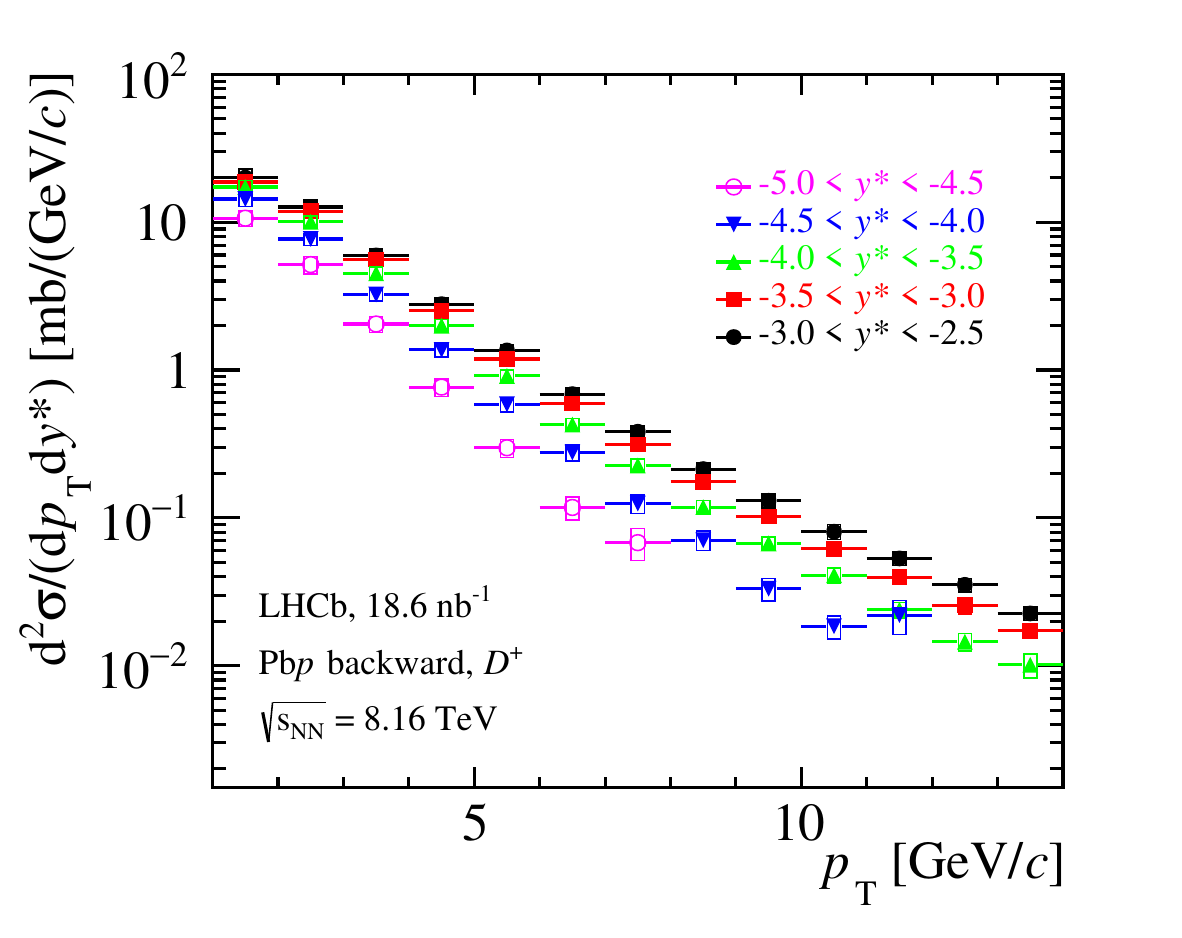}
    \vspace*{-0.5cm}
    \end{center}
    \caption{Double-differential cross-section of prompt $\Dp$ production in $p$Pb collisions at (left) forward and (right) backward rapidities. The vertical error bars show the statistical uncertainties and the boxes show the systematic uncertainties.}
    \label{fig:cross_section_Dp}
\end{figure}

\begin{figure}[tb]
    \begin{center}
    \includegraphics[width=0.49\linewidth]{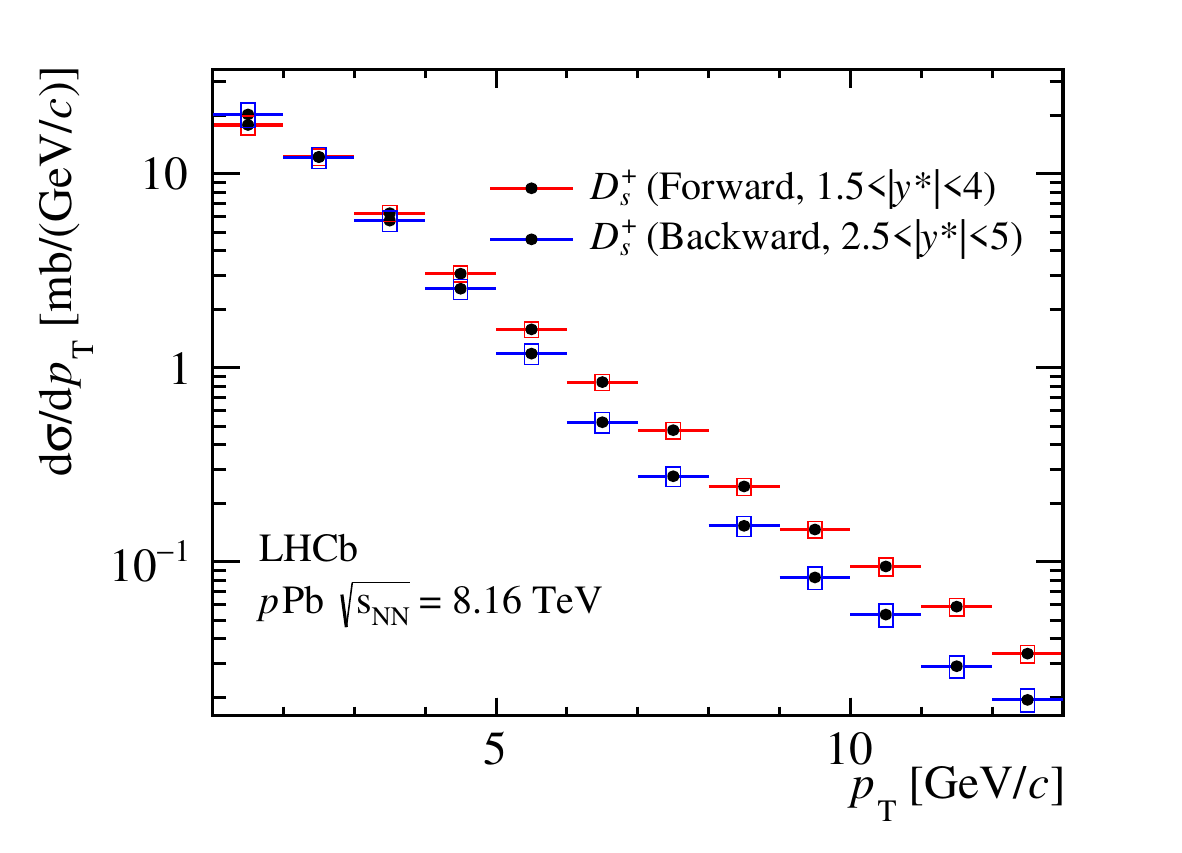}
    \includegraphics[width=0.49\linewidth]{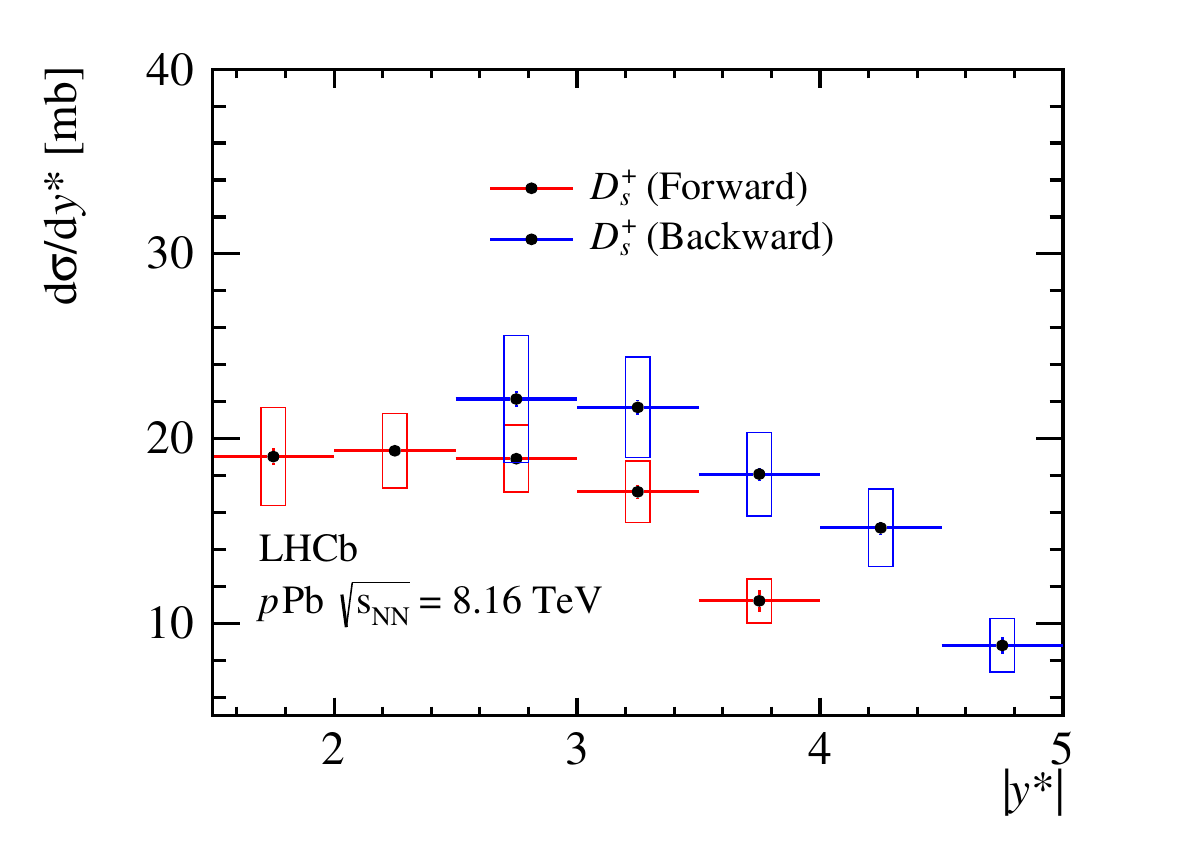}
    \vspace*{-0.5cm}
    \end{center}
    \caption{Differential cross-section of prompt $\Ds$ production in $p$Pb collisions as a function of (left) \pt and (right) $y^{*}$. The vertical error bars show the statistical uncertainties and the boxes show the systematic uncertainties.}
    \label{fig:cross_int_Ds}
\end{figure}

\begin{figure}[tb]
    \begin{center}
    \includegraphics[width=0.49\linewidth]{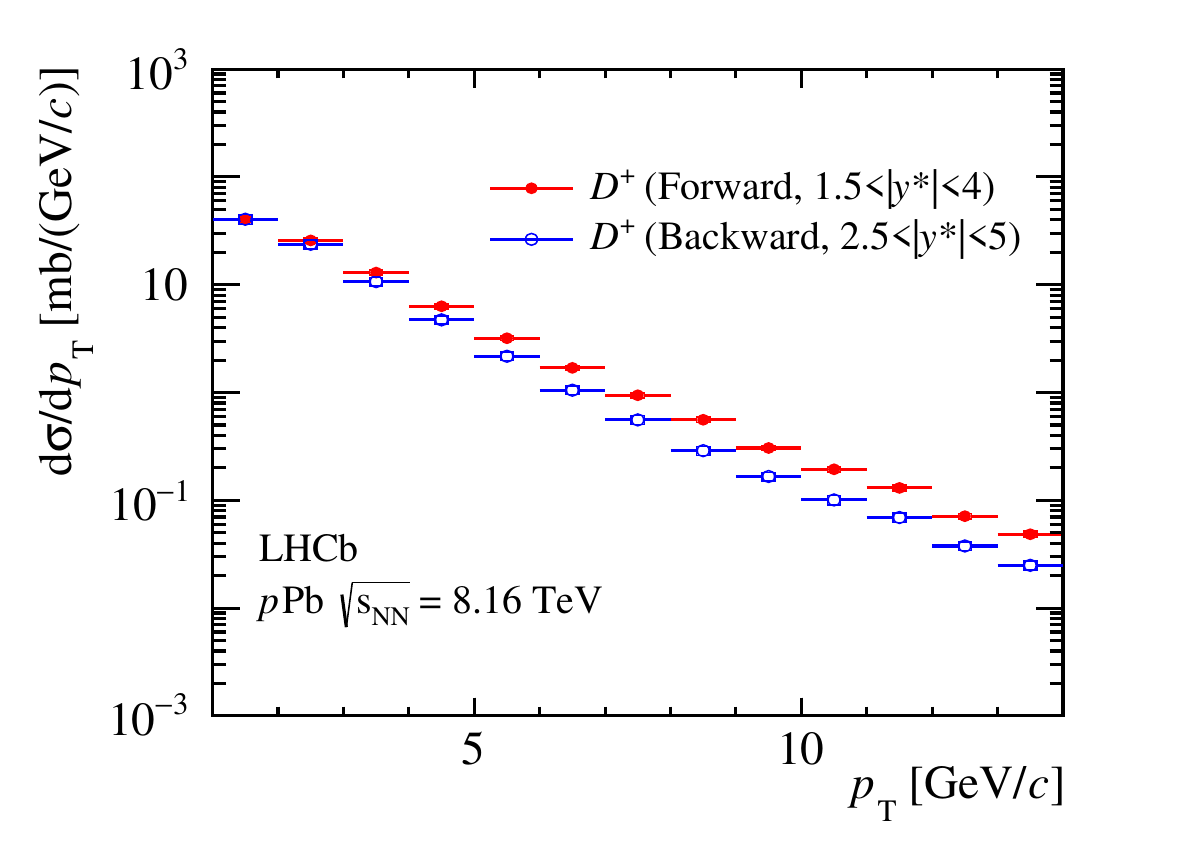}
    \includegraphics[width=0.49\linewidth]{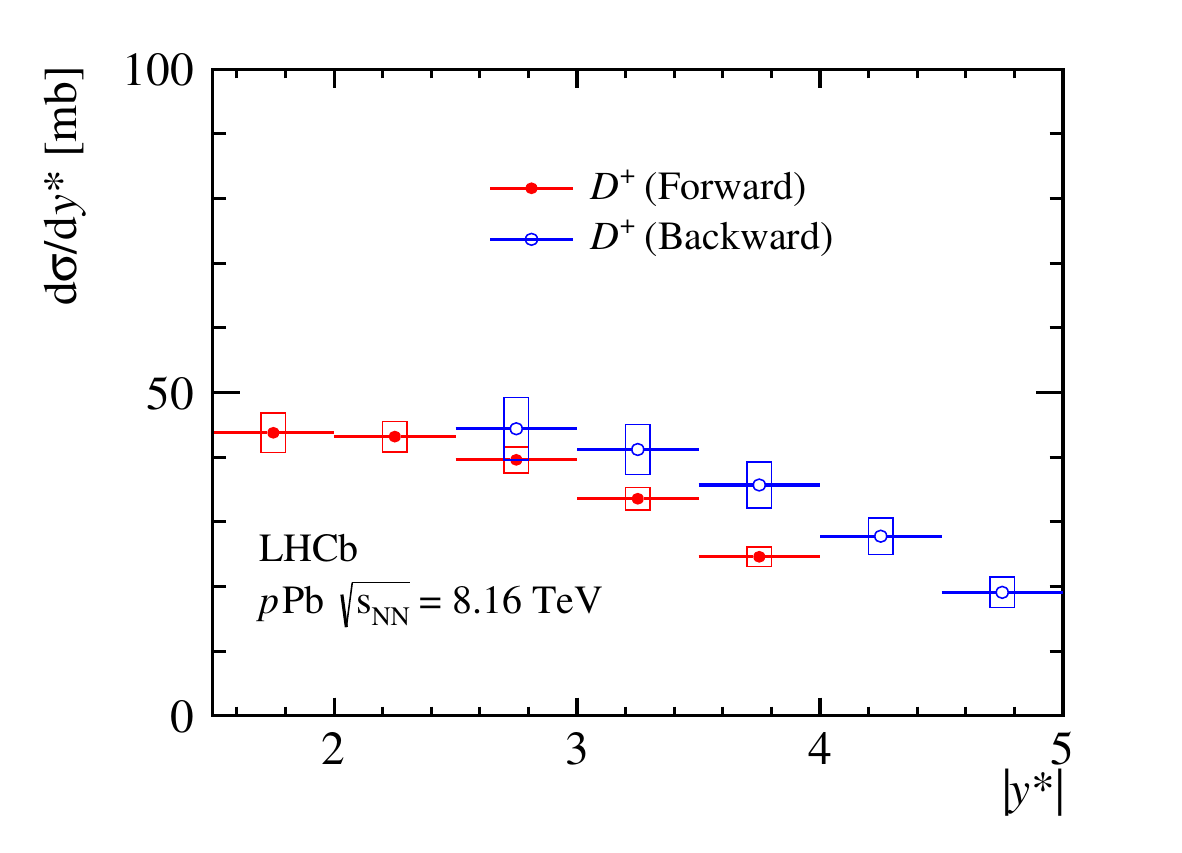}
    \vspace*{-0.5cm}
    \end{center}
    \caption{Differential cross-section of prompt $\Dp$ production in $p$Pb collisions as a function of (left) \pt and (right) $y^{*}$. The vertical error bars show the statistical uncertainties and the boxes show the systematic uncertainties.}
    \label{fig:cross_int_Dp}
\end{figure}

The differential cross-section for prompt \Ds and \Dp mesons in both forward and backward rapidities are shown in Fig.~\ref{fig:cross_section_Ds}--\ref{fig:cross_int_Dp}. The corresponding numerical values are listed in Tables~\ref{tab:cross_section_Ds}--\ref{tab:cross_y_Dp}.

\begin{sidewaystable}[htb]
    \caption{Double-differential cross-section for prompt \Ds production as a function of \pt and $y^{*}$ in $p$Pb collisions at forward and backward rapidities. The first uncertainty is statistical, the second the component of the systematic uncertainty that is uncorrelated between bins and the third the correlated systematic component.}
    \centering
    \resizebox{\textwidth}{!}{
    \scalebox{0.82}{
    \begin{tabular}{cccccc}
    \hline
    &&&$\deriv^2 \sigma /(\deriv\pt\deriv y^*)~[\mbarn/(\gevc)]$ (Forward)&& \\
    $\pt[\gevc]\backslash y^*$ &$[1.5,2]$   &$[2,2.5]$    &$[2.5,3]$     &$[3,3.5]$     &$[3.5,4]$\\
\hline
$[1,2]$&$7.006\pm0.422\pm1.613\pm0.861$&$7.658\pm0.220\pm0.745\pm0.775$&$8.021\pm0.270\pm0.441\pm0.763$&$7.770\pm0.334\pm0.389\pm0.733$&$5.197\pm0.583\pm0.378\pm0.536$\\
$[2,3]$&$5.653\pm0.156\pm0.157\pm0.625$&$5.464\pm0.049\pm0.244\pm0.520$&$5.337\pm0.048\pm0.187\pm0.485$&$4.877\pm0.073\pm0.212\pm0.456$&$3.105\pm0.093\pm0.149\pm0.296$\\
$[3,4]$&$3.027\pm0.120\pm0.051\pm0.310$&$2.961\pm0.042\pm0.094\pm0.273$&$2.694\pm0.026\pm0.034\pm0.245$&$2.314\pm0.032\pm0.064\pm0.215$&$1.521\pm0.055\pm0.074\pm0.146$\\
$[4,5]$&$1.518\pm0.029\pm0.026\pm0.147$&$1.514\pm0.020\pm0.020\pm0.137$&$1.362\pm0.015\pm0.030\pm0.123$&$1.020\pm0.024\pm0.013\pm0.095$&$0.689\pm0.028\pm0.017\pm0.066$\\
$[5,6]$&$0.792\pm0.022\pm0.017\pm0.075$&$0.755\pm0.014\pm0.012\pm0.068$&$0.686\pm0.010\pm0.014\pm0.062$&$0.543\pm0.011\pm0.016\pm0.051$&$0.376\pm0.016\pm0.014\pm0.037$\\
$[6,7]$&$0.429\pm0.012\pm0.026\pm0.041$&$0.421\pm0.005\pm0.011\pm0.038$&$0.347\pm0.006\pm0.008\pm0.031$&$0.276\pm0.011\pm0.015\pm0.026$&$0.213\pm0.016\pm0.018\pm0.022$\\
$[7,8]$&$0.235\pm0.008\pm0.012\pm0.022$&$0.245\pm0.005\pm0.007\pm0.022$&$0.197\pm0.005\pm0.004\pm0.018$&$0.150\pm0.007\pm0.008\pm0.014$&$0.126\pm0.017\pm0.011\pm0.014$\\
$[8,9]$&$0.152\pm0.011\pm0.013\pm0.014$&$0.141\pm0.004\pm0.010\pm0.013$&$0.115\pm0.003\pm0.004\pm0.010$&$0.081\pm0.004\pm0.004\pm0.008$&$-$\\
$[9,10]$&$0.093\pm0.008\pm0.004\pm0.009$&$0.083\pm0.002\pm0.005\pm0.008$&$0.069\pm0.002\pm0.003\pm0.006$&$0.048\pm0.003\pm0.003\pm0.005$&$-$\\
$[10,11]$&$0.063\pm0.003\pm0.006\pm0.006$&$0.050\pm0.002\pm0.003\pm0.005$&$0.044\pm0.002\pm0.002\pm0.004$&$0.032\pm0.003\pm0.007\pm0.003$&$-$\\
$[11,12]$&$0.041\pm0.002\pm0.002\pm0.004$&$0.033\pm0.001\pm0.002\pm0.003$&$0.025\pm0.001\pm0.002\pm0.002$&$0.018\pm0.003\pm0.002\pm0.002$&$-$\\
$[12,13]$&$0.025\pm0.001\pm0.002\pm0.002$&$0.022\pm0.002\pm0.002\pm0.002$&$0.020\pm0.002\pm0.001\pm0.002$&$-$&$-$\\
\hline
\\
    \hline
    &&&$\deriv^2 \sigma /(\deriv\pt\deriv y^*)~[\mbarn/(\gevc)]$ (Backward)&& \\
    $\pt[\gevc]\backslash y^*$ &$[-3,-2.5]$   &$[-3.5,-3]$    &$[-4,-3.5]$     &$[-4.5,-4]$     &$[-5,-4.5]$\\
\hline
$[1,2]$&$9.278\pm0.410\pm1.364\pm1.446$&$9.981\pm0.361\pm0.477\pm1.269$&$8.240\pm0.334\pm0.392\pm1.070$&$8.061\pm0.321\pm0.699\pm1.180$&$4.832\pm0.441\pm1.035\pm0.526$\\
$[2,3]$&$6.553\pm0.128\pm0.205\pm0.913$&$6.009\pm0.135\pm0.244\pm0.726$&$5.355\pm0.054\pm0.125\pm0.626$&$4.031\pm0.065\pm0.137\pm0.446$&$2.379\pm0.120\pm0.136\pm0.293$\\
$[3,4]$&$3.264\pm0.032\pm0.101\pm0.420$&$2.965\pm0.071\pm0.047\pm0.356$&$2.449\pm0.043\pm0.043\pm0.289$&$1.754\pm0.024\pm0.052\pm0.203$&$1.027\pm0.039\pm0.067\pm0.125$\\
$[4,5]$&$1.556\pm0.037\pm0.034\pm0.194$&$1.399\pm0.015\pm0.026\pm0.165$&$1.063\pm0.022\pm0.022\pm0.122$&$0.733\pm0.009\pm0.021\pm0.077$&$0.355\pm0.016\pm0.027\pm0.046$\\
$[5,6]$&$0.712\pm0.021\pm0.022\pm0.089$&$0.654\pm0.008\pm0.020\pm0.079$&$0.504\pm0.007\pm0.011\pm0.056$&$0.330\pm0.010\pm0.018\pm0.041$&$0.163\pm0.011\pm0.017\pm0.023$\\
$[6,7]$&$0.334\pm0.014\pm0.018\pm0.041$&$0.304\pm0.004\pm0.008\pm0.036$&$0.214\pm0.005\pm0.007\pm0.026$&$0.137\pm0.004\pm0.007\pm0.016$&$0.059\pm0.008\pm0.011\pm0.008$\\
$[7,8]$&$0.188\pm0.005\pm0.006\pm0.021$&$0.167\pm0.004\pm0.006\pm0.019$&$0.126\pm0.003\pm0.005\pm0.015$&$0.071\pm0.003\pm0.004\pm0.009$&$-$\\
$[8,9]$&$0.117\pm0.004\pm0.005\pm0.012$&$0.095\pm0.002\pm0.005\pm0.013$&$0.059\pm0.005\pm0.003\pm0.007$&$0.035\pm0.002\pm0.004\pm0.005$&$-$\\
$[9,10]$&$0.062\pm0.003\pm0.003\pm0.008$&$0.052\pm0.002\pm0.003\pm0.006$&$0.035\pm0.001\pm0.002\pm0.004$&$0.017\pm0.002\pm0.003\pm0.002$&$-$\\
$[10,11]$&$0.041\pm0.002\pm0.003\pm0.005$&$0.037\pm0.001\pm0.003\pm0.004$&$0.021\pm0.001\pm0.002\pm0.003$&$0.008\pm0.002\pm0.002\pm0.001$&$-$\\
$[11,12]$&$0.025\pm0.001\pm0.002\pm0.003$&$0.022\pm0.001\pm0.002\pm0.003$&$0.011\pm0.001\pm0.001\pm0.001$&$-$&$-$\\
$[12,13]$&$0.019\pm0.002\pm0.002\pm0.002$&$0.012\pm0.001\pm0.001\pm0.002$&$0.008\pm0.001\pm0.001\pm0.001$&$-$&$-$\\
\hline
\\
    \end{tabular}}
    }
    \label{tab:cross_section_Ds}
\end{sidewaystable}

\begin{sidewaystable}[htb]
    \caption{Double-differential cross-section for prompt \Dp production as a function of \pt and $y^{*}$ in $p$Pb collisions at forward and backward rapidities. The first uncertainty is statistical, the second the component of the systematic uncertainty that is uncorrelated between bins and the third the correlated systematic component.}
    \centering
    \resizebox{\textwidth}{!}{
    \scalebox{0.82}{
    \begin{tabular}{cccccc}
    \hline
    &&&$\deriv^2 \sigma /(\deriv\pt\deriv y^*)~[\mbarn/(\gevc)]$ (Forward)&& \\
    $\pt[\gevc] \backslash y^*$ &$[1.5,2]$   &$[2,2.5]$    &$[2.5,3]$     &$[3,3.5]$     &$[3.5,4]$\\
\hline
$[1,2]$&$18.276\pm0.305\pm0.884\pm1.481$&$18.390\pm0.095\pm0.563\pm1.209$&$17.369\pm0.020\pm0.607\pm1.013$&$15.329\pm0.080\pm0.439\pm0.885$&$11.032\pm0.108\pm0.632\pm0.631$\\
$[2,3]$&$12.215\pm0.059\pm0.364\pm0.886$&$12.020\pm0.024\pm0.352\pm0.701$&$11.018\pm0.083\pm0.372\pm0.597$&$9.205\pm0.026\pm0.260\pm0.506$&$7.066\pm0.035\pm0.608\pm0.400$\\
$[3,4]$&$6.286\pm0.025\pm0.171\pm0.414$&$6.172\pm0.014\pm0.174\pm0.343$&$5.410\pm0.010\pm0.141\pm0.290$&$4.552\pm0.010\pm0.231\pm0.249$&$3.498\pm0.020\pm0.245\pm0.198$\\
$[4,5]$&$3.168\pm0.014\pm0.086\pm0.192$&$3.020\pm0.005\pm0.154\pm0.161$&$2.708\pm0.009\pm0.103\pm0.145$&$2.160\pm0.008\pm0.117\pm0.118$&$1.566\pm0.014\pm0.133\pm0.092$\\
$[5,6]$&$1.664\pm0.008\pm0.047\pm0.097$&$1.543\pm0.005\pm0.046\pm0.082$&$1.350\pm0.005\pm0.062\pm0.072$&$1.062\pm0.005\pm0.062\pm0.059$&$0.757\pm0.009\pm0.065\pm0.046$\\
$[6,7]$&$0.876\pm0.018\pm0.024\pm0.050$&$0.840\pm0.004\pm0.041\pm0.045$&$0.730\pm0.004\pm0.031\pm0.039$&$0.568\pm0.004\pm0.037\pm0.032$&$0.373\pm0.015\pm0.049\pm0.024$\\
$[7,8]$&$0.491\pm0.005\pm0.014\pm0.028$&$0.482\pm0.003\pm0.017\pm0.026$&$0.405\pm0.002\pm0.019\pm0.022$&$0.308\pm0.003\pm0.021\pm0.018$&$0.200\pm0.008\pm0.026\pm0.014$\\
$[8,9]$&$0.308\pm0.000\pm0.013\pm0.018$&$0.280\pm0.002\pm0.009\pm0.015$&$0.240\pm0.002\pm0.011\pm0.013$&$0.177\pm0.003\pm0.015\pm0.011$&$0.112\pm0.010\pm0.014\pm0.009$\\
$[9,10]$&$0.184\pm0.000\pm0.007\pm0.011$&$0.176\pm0.001\pm0.007\pm0.010$&$0.141\pm0.002\pm0.007\pm0.008$&$0.110\pm0.002\pm0.010\pm0.007$&$-$\\
$[10,11]$&$0.122\pm0.002\pm0.004\pm0.007$&$0.110\pm0.002\pm0.004\pm0.006$&$0.091\pm0.001\pm0.005\pm0.005$&$0.066\pm0.002\pm0.005\pm0.004$&$-$\\
$[11,12]$&$0.090\pm0.001\pm0.004\pm0.005$&$0.070\pm0.001\pm0.004\pm0.004$&$0.061\pm0.001\pm0.004\pm0.004$&$0.039\pm0.002\pm0.004\pm0.003$&$-$\\
$[12,13]$&$0.057\pm0.001\pm0.003\pm0.003$&$0.047\pm0.001\pm0.002\pm0.003$&$0.039\pm0.001\pm0.003\pm0.002$&$-$&$-$\\
$[13,14]$&$0.037\pm0.001\pm0.003\pm0.002$&$0.032\pm0.001\pm0.002\pm0.002$&$0.029\pm0.001\pm0.002\pm0.002$&$-$&$-$\\
\hline
\\
    \hline
    &&&$\deriv^2 \sigma /(\deriv\pt\deriv y^*)~[\mbarn/(\gevc)]$ (Backward)&& \\
    $\pt[\gevc] \backslash y^*$ &$[-3,-2.5]$   &$[-3.5,-3]$    &$[-4,-3.5]$     &$[-4.5,-4]$     &$[-5,-4.5]$\\
\hline
$[1,2]$&$20.016\pm0.220\pm0.866\pm2.666$&$18.689\pm0.079\pm0.568\pm2.120$&$17.293\pm0.065\pm0.508\pm1.756$&$14.348\pm0.206\pm0.683\pm1.395$&$10.639\pm0.057\pm0.815\pm1.036$\\
$[2,3]$&$12.676\pm0.044\pm0.377\pm1.373$&$11.864\pm0.022\pm0.334\pm1.214$&$10.054\pm0.018\pm0.233\pm0.955$&$7.692\pm0.020\pm0.248\pm0.678$&$5.165\pm0.025\pm0.526\pm0.478$\\
$[3,4]$&$5.957\pm0.018\pm0.154\pm0.629$&$5.600\pm0.010\pm0.162\pm0.530$&$4.519\pm0.008\pm0.138\pm0.418$&$3.246\pm0.010\pm0.155\pm0.284$&$2.046\pm0.014\pm0.150\pm0.199$\\
$[4,5]$&$2.788\pm0.010\pm0.091\pm0.276$&$2.522\pm0.006\pm0.065\pm0.230$&$1.996\pm0.006\pm0.104\pm0.175$&$1.368\pm0.005\pm0.078\pm0.121$&$0.766\pm0.009\pm0.077\pm0.071$\\
$[5,6]$&$1.356\pm0.006\pm0.042\pm0.130$&$1.188\pm0.002\pm0.035\pm0.105$&$0.912\pm0.005\pm0.039\pm0.079$&$0.585\pm0.003\pm0.038\pm0.054$&$0.298\pm0.006\pm0.029\pm0.029$\\
$[6,7]$&$0.687\pm0.007\pm0.016\pm0.065$&$0.593\pm0.001\pm0.020\pm0.053$&$0.428\pm0.001\pm0.018\pm0.037$&$0.277\pm0.002\pm0.020\pm0.027$&$0.117\pm0.004\pm0.015\pm0.014$\\
$[7,8]$&$0.382\pm0.009\pm0.012\pm0.035$&$0.314\pm0.002\pm0.010\pm0.029$&$0.226\pm0.001\pm0.012\pm0.020$&$0.125\pm0.002\pm0.011\pm0.014$&$0.068\pm0.007\pm0.014\pm0.009$\\
$[8,9]$&$0.214\pm0.002\pm0.007\pm0.020$&$0.175\pm0.001\pm0.006\pm0.016$&$0.118\pm0.001\pm0.005\pm0.012$&$0.071\pm0.002\pm0.007\pm0.008$&$-$\\
$[9,10]$&$0.130\pm0.001\pm0.005\pm0.013$&$0.102\pm0.001\pm0.004\pm0.009$&$0.067\pm0.001\pm0.004\pm0.006$&$0.033\pm0.001\pm0.005\pm0.003$&$-$\\
$[10,11]$&$0.081\pm0.001\pm0.004\pm0.008$&$0.062\pm0.001\pm0.003\pm0.006$&$0.041\pm0.001\pm0.003\pm0.004$&$0.018\pm0.001\pm0.003\pm0.002$&$-$\\
$[11,12]$&$0.053\pm0.001\pm0.003\pm0.005$&$0.040\pm0.001\pm0.002\pm0.004$&$0.024\pm0.001\pm0.002\pm0.002$&$0.022\pm0.002\pm0.005\pm0.003$&$-$\\
$[12,13]$&$0.035\pm0.001\pm0.002\pm0.003$&$0.026\pm0.000\pm0.002\pm0.002$&$0.015\pm0.000\pm0.001\pm0.002$&$-$&$-$\\
$[13,14]$&$0.023\pm0.000\pm0.001\pm0.002$&$0.017\pm0.000\pm0.001\pm0.002$&$0.010\pm0.000\pm0.001\pm0.001$&$-$&$-$\\
\hline
\\
    \end{tabular}}
    }
    \label{tab:cross_section_Dp}
\end{sidewaystable}

\begin{table}[tbh]
        \caption{Differential cross-section for prompt \Ds production as a function of $\pt$ in $p$Pb collisions at forward and backward rapidities. The first uncertainty is statistical, the second the component of the systematic uncertainty that is uncorrelated between bins and the third the correlated systematic component.}
        \centering
        \begin{tabular}{cc}
            \hline
            $\pt[\gevc]$ & $\deriv \sigma/\deriv \pt$ [\mbarn/(\gevc)] (Forward)\\
\hline
$[1,2]$&$17.826\pm0.433\pm0.955\pm1.808$\\
$[2,3]$&$12.218\pm0.104\pm0.216\pm1.177$\\
$[3,4]$&$6.259\pm0.072\pm0.074\pm0.591$\\
$[4,5]$&$3.051\pm0.027\pm0.025\pm0.283$\\
$[5,6]$&$1.576\pm0.017\pm0.017\pm0.146$\\
$[6,7]$&$0.843\pm0.012\pm0.019\pm0.079$\\
$[7,8]$&$0.476\pm0.011\pm0.010\pm0.045$\\
$[8,9]$&$0.244\pm0.006\pm0.009\pm0.023$\\
$[9,10]$&$0.147\pm0.005\pm0.004\pm0.014$\\
$[10,11]$&$0.095\pm0.003\pm0.005\pm0.009$\\
$[11,12]$&$0.059\pm0.002\pm0.002\pm0.006$\\
$[12,13]$&$0.034\pm0.002\pm0.001\pm0.003$\\
\hline
\\
            \hline
            $\pt[\gevc]$ & $\deriv \sigma/\deriv \pt$ [\mbarn/(\gevc)] (Backward)\\
\hline
$[1,2]$&$20.196\pm0.421\pm0.975\pm2.700$\\
$[2,3]$&$12.163\pm0.119\pm0.196\pm1.490$\\
$[3,4]$&$5.729\pm0.050\pm0.073\pm0.694$\\
$[4,5]$&$2.553\pm0.025\pm0.029\pm0.300$\\
$[5,6]$&$1.182\pm0.014\pm0.020\pm0.143$\\
$[6,7]$&$0.524\pm0.009\pm0.012\pm0.063$\\
$[7,8]$&$0.276\pm0.004\pm0.005\pm0.031$\\
$[8,9]$&$0.153\pm0.004\pm0.004\pm0.018$\\
$[9,10]$&$0.083\pm0.002\pm0.003\pm0.011$\\
$[10,11]$&$0.053\pm0.002\pm0.003\pm0.007$\\
$[11,12]$&$0.029\pm0.001\pm0.001\pm0.003$\\
$[12,13]$&$0.019\pm0.001\pm0.001\pm0.002$\\
\hline
\\
        \end{tabular}
        \label{tab:cross_pt_Ds}
    \end{table}
    
\begin{table}[tbh]
        \caption{Differential cross-section for prompt \Dp production as a function of $\pt$ in $p$Pb collisions at forward and backward rapidities. The first uncertainty is statistical, the second the component of the systematic uncertainty that is uncorrelated between bins and the third the correlated systematic component.}
        \centering
        \begin{tabular}{cc}
            \hline
            $\pt[\gevc]$&$\deriv \sigma/\deriv \pt$ [\mbarn/(\gevc)] (Forward)\\
\hline
$[1,2]$&$40.198\pm0.174\pm0.717\pm2.291$\\
$[2,3]$&$25.763\pm0.057\pm0.456\pm1.326$\\
$[3,4]$&$12.959\pm0.019\pm0.219\pm0.638$\\
$[4,5]$&$6.311\pm0.012\pm0.135\pm0.300$\\
$[5,6]$&$3.188\pm0.007\pm0.064\pm0.151$\\
$[6,7]$&$1.693\pm0.012\pm0.042\pm0.081$\\
$[7,8]$&$0.943\pm0.005\pm0.022\pm0.045$\\
$[8,9]$&$0.559\pm0.005\pm0.014\pm0.028$\\
$[9,10]$&$0.306\pm0.001\pm0.008\pm0.015$\\
$[10,11]$&$0.194\pm0.002\pm0.004\pm0.010$\\
$[11,12]$&$0.130\pm0.001\pm0.004\pm0.007$\\
$[12,13]$&$0.071\pm0.001\pm0.002\pm0.004$\\
$[13,14]$&$0.048\pm0.001\pm0.002\pm0.003$\\
\hline
\\
            \hline
            $\pt[\gevc]$&$\deriv \sigma/\deriv \pt$ [\mbarn/(\gevc)] (Backward)\\
\hline
$[1,2]$&$40.492\pm0.161\pm0.785\pm4.317$\\
$[2,3]$&$23.726\pm0.031\pm0.402\pm2.241$\\
$[3,4]$&$10.684\pm0.014\pm0.170\pm0.981$\\
$[4,5]$&$4.720\pm0.008\pm0.094\pm0.414$\\
$[5,6]$&$2.170\pm0.005\pm0.041\pm0.188$\\
$[6,7]$&$1.050\pm0.004\pm0.020\pm0.093$\\
$[7,8]$&$0.557\pm0.006\pm0.013\pm0.051$\\
$[8,9]$&$0.289\pm0.002\pm0.007\pm0.026$\\
$[9,10]$&$0.166\pm0.001\pm0.004\pm0.015$\\
$[10,11]$&$0.101\pm0.001\pm0.003\pm0.010$\\
$[11,12]$&$0.069\pm0.001\pm0.003\pm0.007$\\
$[12,13]$&$0.038\pm0.000\pm0.002\pm0.003$\\
$[13,14]$&$0.025\pm0.000\pm0.001\pm0.002$\\
\hline
\\
        \end{tabular}
        \label{tab:cross_pt_Dp}
    \end{table}

\begin{table}[tbh]
        \caption{Differential cross-section for prompt \Ds production as a function of $y^{*}$ in $p$Pb collisions at forward and backward rapidities. The first uncertainty is statistical, the second the component of the systematic uncertainty that is uncorrelated between bins and the third the correlated systematic component.}
        \centering
        \begin{tabular}{cc}
            \hline
            $y^*$ & $\deriv \sigma/\deriv y^*$ [\mbarn] (Forward)\\
\hline
$[1.5,2.0]$&$19.032\pm0.467\pm1.622\pm2.098$\\
$[2.0,2.5]$&$19.347\pm0.231\pm0.790\pm1.854$\\
$[2.5,3.0]$&$18.918\pm0.276\pm0.482\pm1.749$\\
$[3.0,3.5]$&$17.129\pm0.344\pm0.449\pm1.606$\\
$[3.5,4.0]$&$11.227\pm0.594\pm0.414\pm1.113$\\
\hline
\\
            \hline
            $y^*$ & $\deriv \sigma/\deriv y^*$ [\mbarn] (Backward)\\
\hline
$[-2.5,-3.0]$&$22.148\pm0.434\pm1.383\pm3.142$\\
$[-3.0,-3.5]$&$21.695\pm0.392\pm0.539\pm2.667$\\
$[-3.5,-4.0]$&$18.086\pm0.342\pm0.414\pm2.214$\\
$[-4.0,-4.5]$&$15.176\pm0.329\pm0.714\pm1.962$\\
$[-4.5,-5.0]$&$8.814\pm0.459\pm1.047\pm1.002$\\
\hline
\\
        \end{tabular}
        \label{tab:cross_y_Ds}
\end{table}

\begin{table}[tbh]
        \caption{Differential cross-section for prompt \Dp production as a function of $y^{*}$ in $p$Pb collisions at forward and backward rapidities. The first uncertainty is statistical, the second the component of the systematic uncertainty that is uncorrelated between bins and the third the correlated systematic component.}
        \centering
        \begin{tabular}{cc}
            \hline
            $y^*$&$\deriv \sigma/\deriv y^*$ [\mbarn] (Forward)\\
\hline
$[1.5,2.0]$&$43.77\pm0.31\pm0.98\pm2.90$\\
$[2.0,2.5]$&$43.18\pm0.10\pm0.71\pm2.26$\\
$[2.5,3.0]$&$39.59\pm0.09\pm0.74\pm1.88$\\
$[3.0,3.5]$&$33.58\pm0.09\pm0.58\pm1.62$\\
$[3.5,4.0]$&$24.60\pm0.12\pm0.92\pm1.22$\\
\hline
\\
            \hline
            $y^*$&$\deriv \sigma/\deriv y^*$ [\mbarn] (Backward)\\
\hline
$[-3.0,-2.5]$&$44.40\pm0.23\pm0.96\pm4.73$\\
$[-3.5,-3.0]$&$41.19\pm0.08\pm0.68\pm3.80$\\
$[-4.0,-3.5]$&$35.70\pm0.07\pm0.59\pm3.47$\\
$[-4.5,-4.0]$&$27.78\pm0.21\pm0.75\pm2.74$\\
$[-5.0,-4.5]$&$19.10\pm0.06\pm0.99\pm2.13$\\
\hline
\\
        \end{tabular}
        \label{tab:cross_y_Dp}
\end{table}

\begin{figure}[tb]
    \begin{center}
    \includegraphics[width=0.76\linewidth]{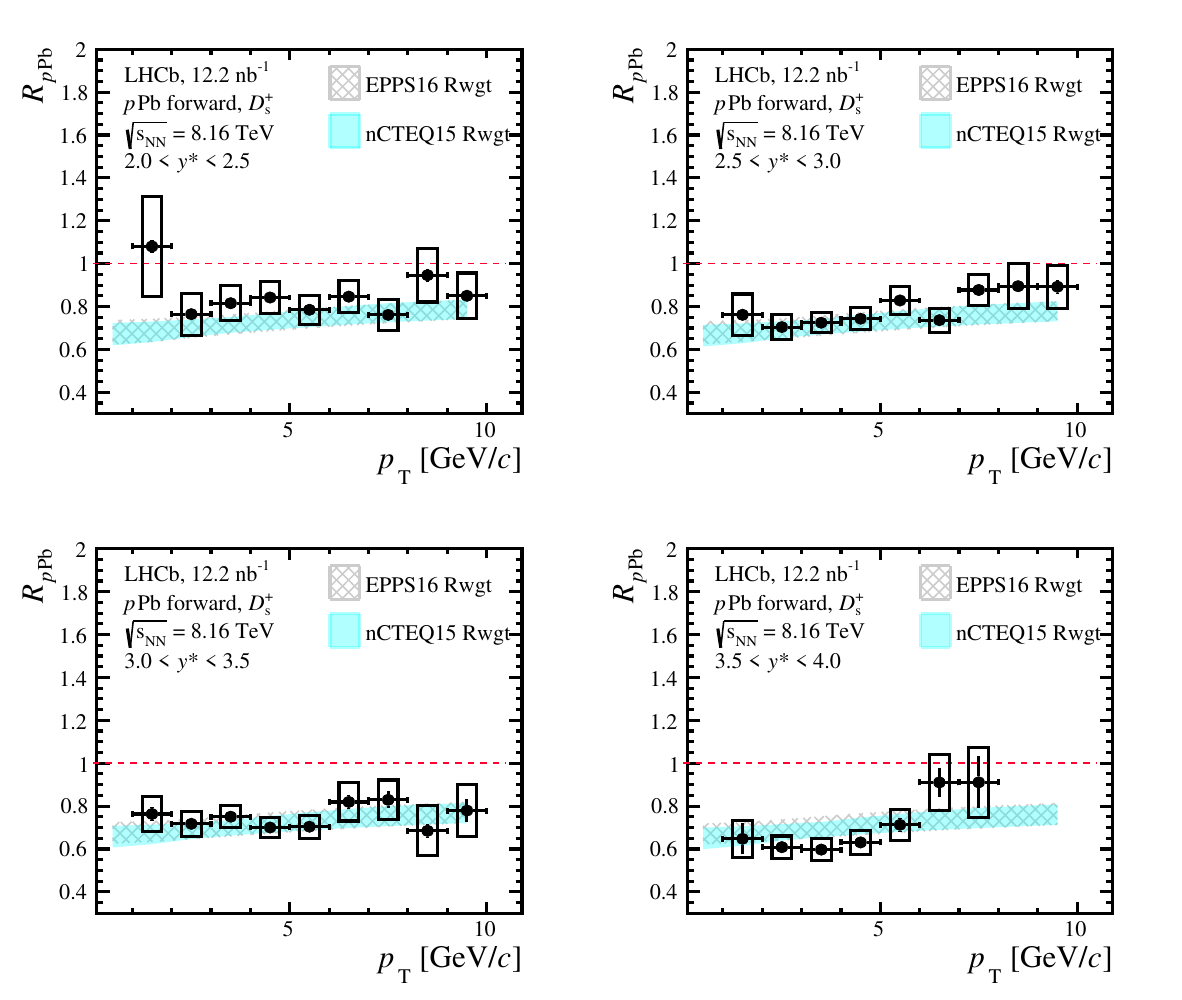}
    \includegraphics[width=0.76\linewidth]{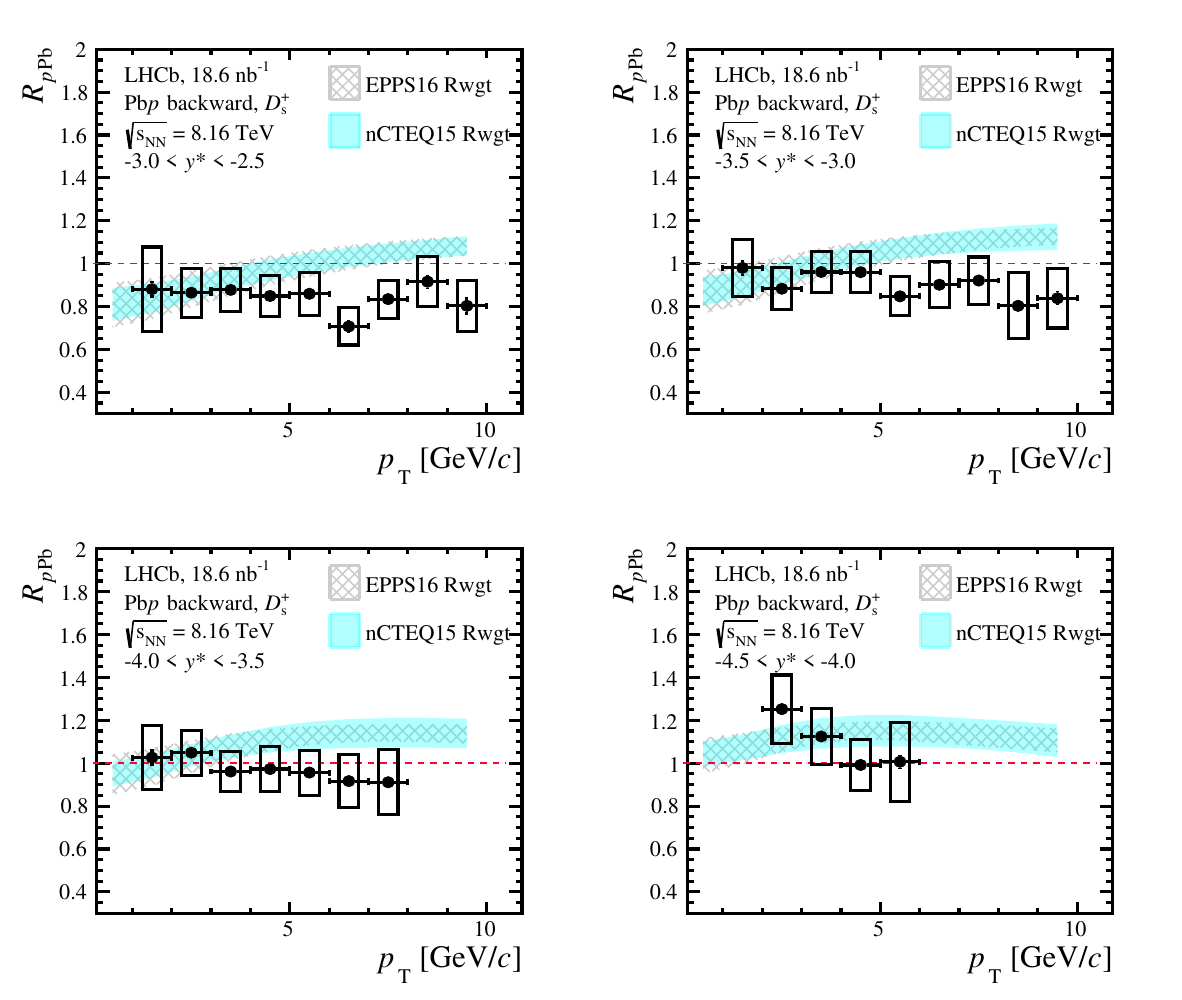}
    \vspace*{-0.5cm}
    \end{center}
    \caption{
    Nuclear modification factor $R_{p\mathrm{Pb}}$ for prompt \Ds production as a function of \pt in different $y^{*}$ intervals. The vertical error bars show the statistical uncertainties and the boxes show the systematic uncertainties. The coloured bands represent the theoretical calculations using the HELAC-Onia generator \cite{Shao:2012iz,Shao:2015vga}, incorporating nPDFs EPPS16 (grey) \cite{Eskola:2016oht} and nCTEQ15 (blue) \cite{Kovarik:2015cma}.}
    \label{fig:RpPb_2D_Ds}
\end{figure}

\begin{figure}[tb]
    \begin{center}
    \includegraphics[width=0.76\linewidth]{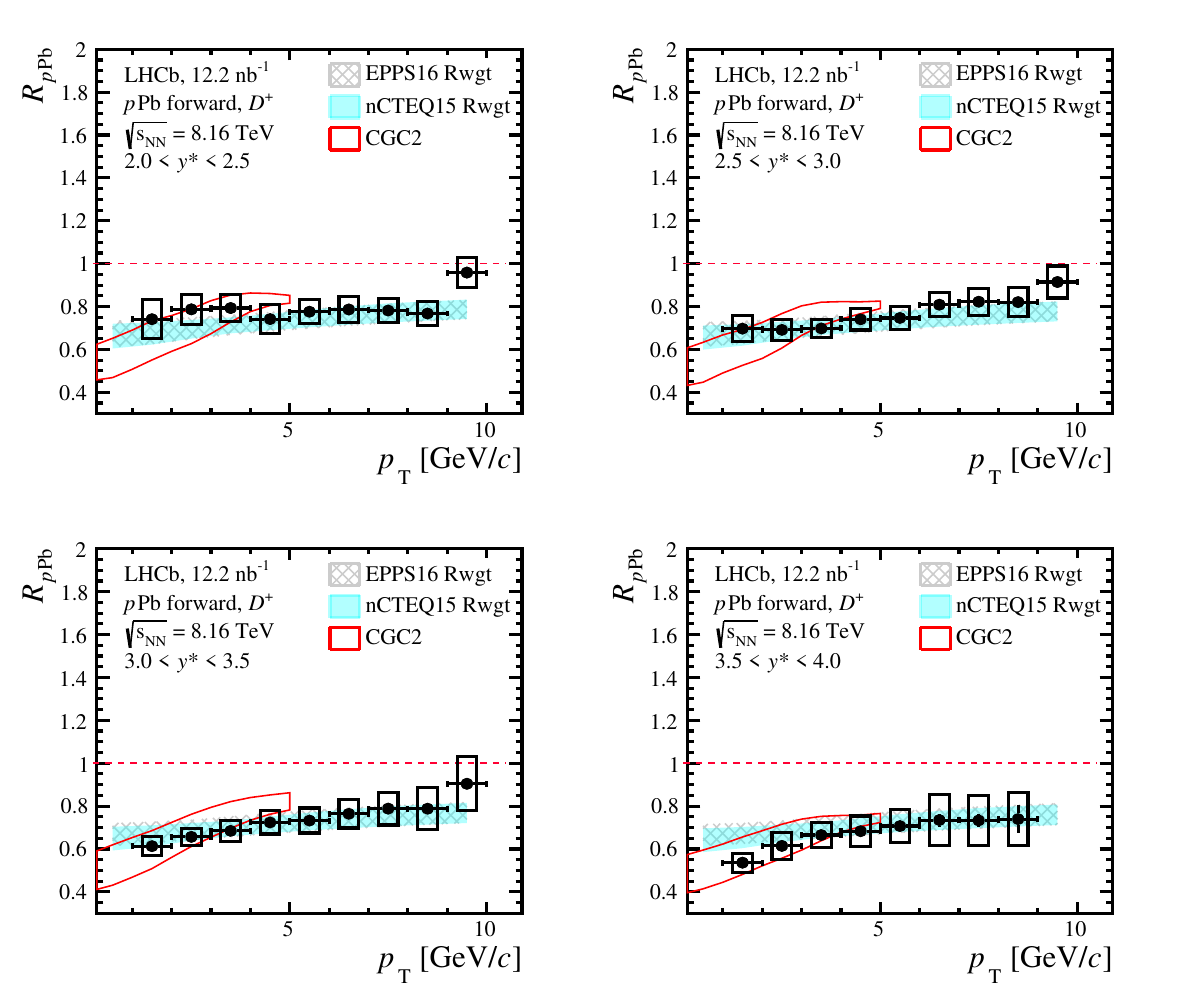}
    \includegraphics[width=0.76\linewidth]{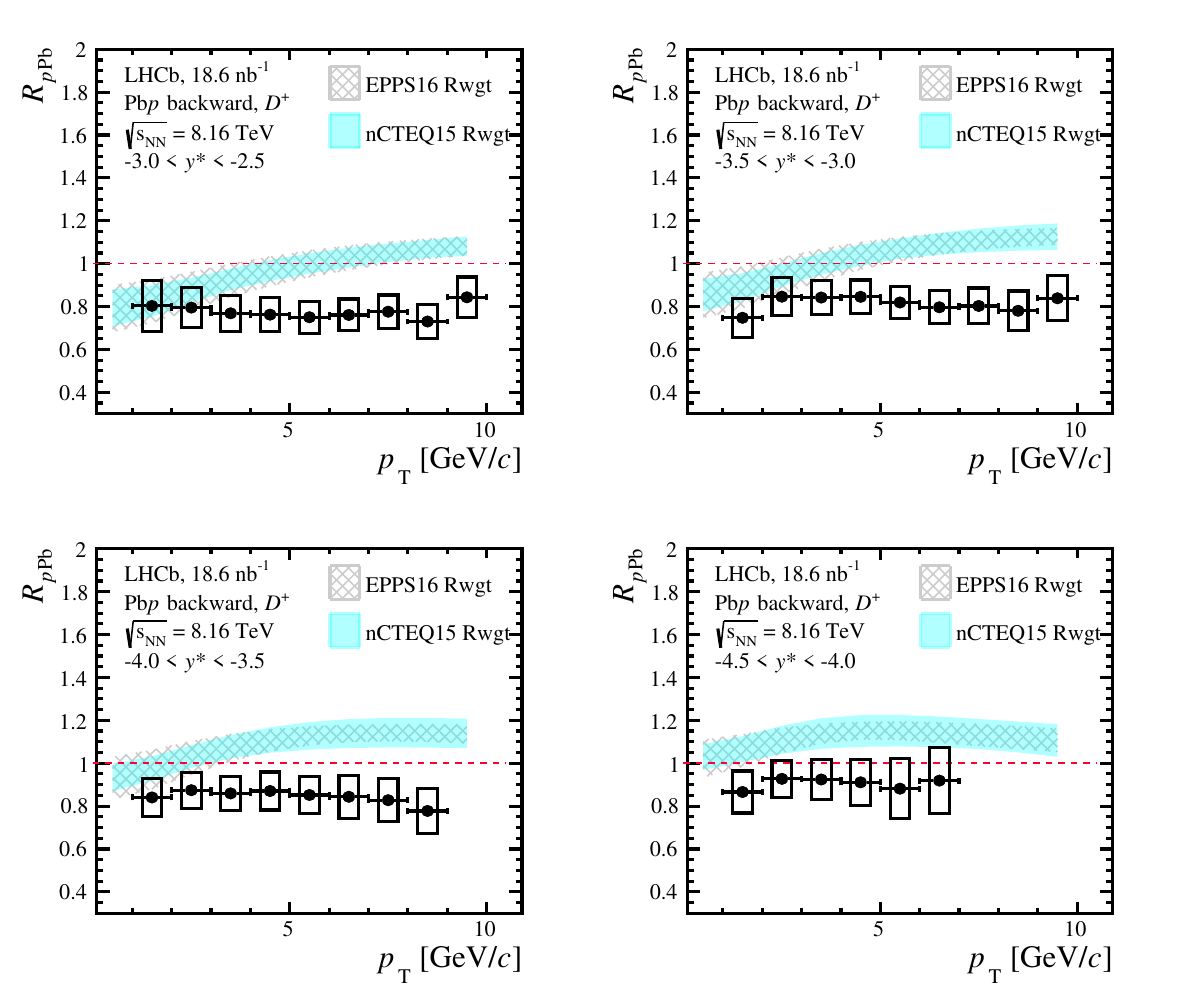}
    \vspace*{-0.5cm}
    \end{center}
    \caption{
    Nuclear modification factor $R_{p\mathrm{Pb}}$ for prompt \Dp production as a function of \pt in different $y^{*}$ intervals. The vertical error bars show the statistical uncertainties and the boxes show the systematic uncertainties. The coloured bands represent the theoretical calculations using the HELAC-Onia generator \cite{Shao:2012iz,Shao:2015vga}, incorporating nPDFs EPPS16 (grey) \cite{Eskola:2016oht} and nCTEQ15 (blue) \cite{Kovarik:2015cma}. The coloured line represent the CGC2 (red) calculations \cite{Ma:2018bax}.}
    \label{fig:RpPb_2D_Dp}
\end{figure}

\begin{figure}[tb]
    \begin{center}
    \includegraphics[width=0.8\linewidth]{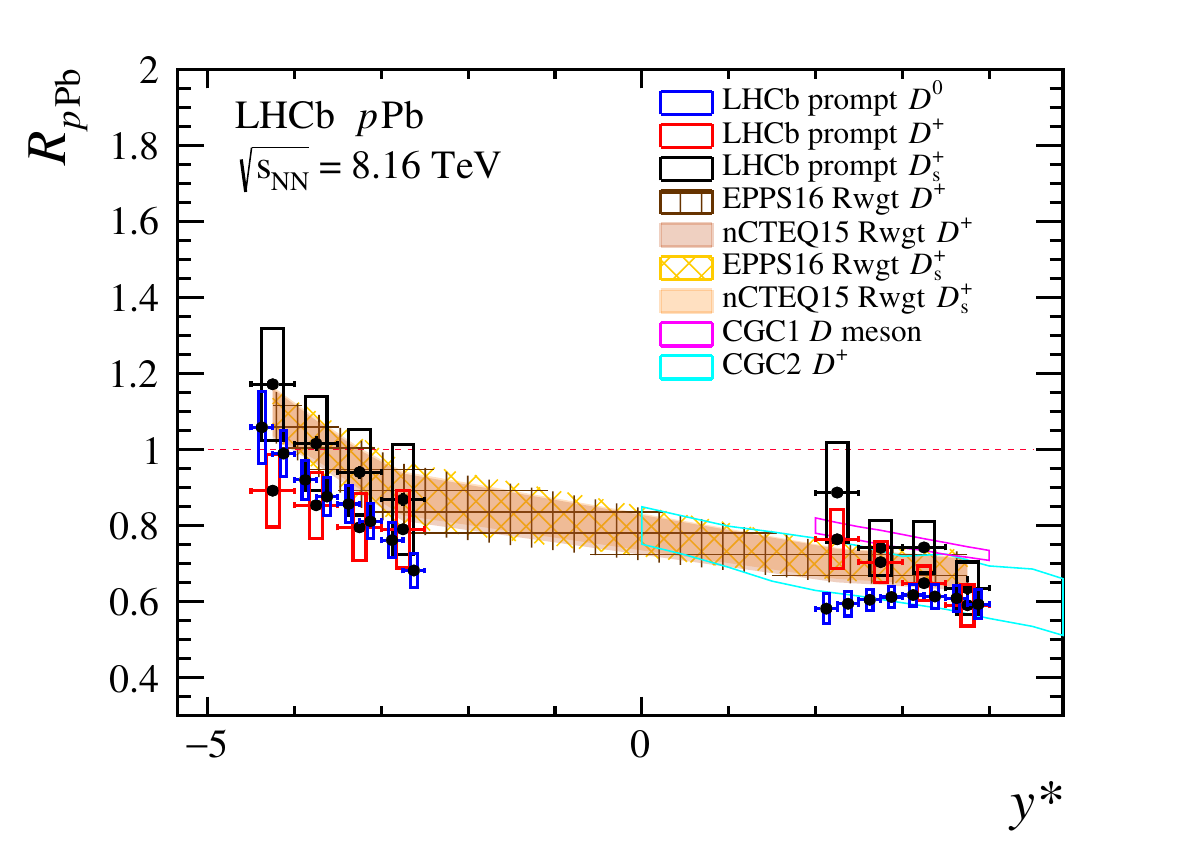}
    \vspace*{-0.5cm}
    \end{center}
    \caption{
    Nuclear modification factor as a function of $y^*$ for prompt \Dp and \Ds mesons integrated over $1 < \pt < 10$\gevc. The vertical error bars show the statistical uncertainties and the boxes show the systematic uncertainties. The LHCb \Dz results at \sqsnn = 8.16 TeV \cite{LHCb:2022dmh} and theoretical calculations at \sqsnn = 8.16 TeV are also shown \cite{Eskola:2016oht, Kovarik:2015cma, Ducloue:2015gfa, Ducloue:2016ywt, Ma:2018bax}.}
    \label{fig:RpPb_y}
\end{figure}

The nuclear modification factor $R_{p\mathrm{Pb}}$ for prompt \Ds and \Dp mesons in both forward and backward rapidities are shown in Fig.~\ref{fig:RpPb_2D_Ds}--\ref{fig:RpPb_y}. The corresponding numerical values are listed in Tables~\ref{tab:RpPb_Ds_pt}--\ref{tab:RpPb_Dp_2D}.

\begin{table}[tbh]
    \caption{
    Nuclear modification factor $R_{p\mathrm{Pb}}$ for prompt \Ds production as a function of \pt at forward (integrated over the common rapidity region of $2.0 < y^{*} < 4.0$) and backward (integrated over the common rapidity region of $-4.5 < y^{*} < -2.5$) rapidity.
    The first uncertainty is statistical, the second systematic.
    }
    \centering
    \begin{tabular}{cc}
    \hline
    \pt[\gevc]& $R_{p\mathrm{Pb}}$ (Forward)\\
\hline
$[1,2]$&$0.800\pm0.021\pm0.112$\\
$[2,3]$&$0.705\pm0.005\pm0.066$\\
$[3,4]$&$0.731\pm0.006\pm0.057$\\
$[4,5]$&$0.742\pm0.007\pm0.058$\\
$[5,6]$&$0.764\pm0.008\pm0.063$\\
$[6,7]$&$0.816\pm0.014\pm0.080$\\
$[7,8]$&$0.829\pm0.022\pm0.090$\\
$[8,9]$&$0.852\pm0.016\pm0.117$\\
$[9,10]$&$0.845\pm0.019\pm0.109$\\
\hline

    \pt[\gevc]& $R_{p\mathrm{Pb}}$ (Backward)\\
\hline
$[1,2]$&$0.957\pm0.022\pm0.160$\\
$[2,3]$&$0.967\pm0.009\pm0.111$\\
$[3,4]$&$0.956\pm0.008\pm0.101$\\
$[4,5]$&$0.928\pm0.009\pm0.099$\\
$[5,6]$&$0.896\pm0.010\pm0.107$\\
$[6,7]$&$0.817\pm0.015\pm0.100$\\
$[7,8]$&$0.883\pm0.013\pm0.110$\\
$[8,9]$&$0.862\pm0.018\pm0.136$\\
$[9,10]$&$0.819\pm0.028\pm0.127$\\
\hline

    \end{tabular}
    \label{tab:RpPb_Ds_pt}
\end{table}

\begin{table}[tbh]
    \caption{
    Nuclear modification factor $R_{p\mathrm{Pb}}$ for prompt \Ds production as a function of $y^{*}$, integrated over $1 < \pt < 10$\gevc. The first uncertainty is statistical, the second systematic.}
    \centering
    \begin{tabular}{cc}
    \hline
    $y^*$ & $R_{p\mathrm{Pb}}$\\
\hline
$[-4.5,-4.0]$&$1.172\pm0.012\pm0.147$\\
$[-4.0,-3.5]$&$1.016\pm0.019\pm0.123$\\
$[-3.5,-3.0]$&$0.941\pm0.017\pm0.112$\\
$[-3.0,-2.5]$&$0.869\pm0.017\pm0.144$\\
$[2.0,2.5]$&$0.887\pm0.011\pm0.131$\\
$[2.5,3.0]$&$0.742\pm0.011\pm0.072$\\
$[3.0,3.5]$&$0.743\pm0.015\pm0.067$\\
$[3.5,4.0]$&$0.635\pm0.034\pm0.069$\\
\hline

    \end{tabular}
    \label{tab:RpPb_Ds_y}
\end{table}    
    
\begin{sidewaystable}[htb]
    \caption{
    Nuclear modification factor $R_{p\mathrm{Pb}}$ for prompt \Ds production as a function of \pt and $y^{*}$. The first uncertainty is statistical, the second systematic.}
    \centering
    \begin{tabular}{ccccccccc}
    \hline
    &&&$R_{p\mathrm{Pb}}$ (Forward)&& \\
    $\pt[\gevc]\backslash y^*$   &$[2,2.5]$    &$[2.5,3]$     &$[3,3.5]$     &$[3.5,4]$\\
\hline
$[1,2]$&$1.080\pm0.031\pm0.232$&$0.762\pm0.026\pm0.097$&$0.763\pm0.033\pm0.081$&$0.647\pm0.073\pm0.086$\\
$[2,3]$&$0.764\pm0.007\pm0.098$&$0.704\pm0.006\pm0.057$&$0.717\pm0.011\pm0.058$&$0.608\pm0.018\pm0.052$\\
$[3,4]$&$0.816\pm0.012\pm0.081$&$0.724\pm0.007\pm0.047$&$0.750\pm0.010\pm0.051$&$0.597\pm0.022\pm0.050$\\
$[4,5]$&$0.842\pm0.011\pm0.076$&$0.743\pm0.008\pm0.052$&$0.700\pm0.017\pm0.048$&$0.630\pm0.025\pm0.056$\\
$[5,6]$&$0.785\pm0.015\pm0.067$&$0.828\pm0.012\pm0.064$&$0.704\pm0.014\pm0.053$&$0.713\pm0.031\pm0.072$\\
$[6,7]$&$0.846\pm0.010\pm0.074$&$0.737\pm0.013\pm0.056$&$0.819\pm0.034\pm0.089$&$0.910\pm0.069\pm0.131$\\
$[7,8]$&$0.761\pm0.014\pm0.073$&$0.877\pm0.021\pm0.071$&$0.829\pm0.040\pm0.093$&$0.911\pm0.121\pm0.163$\\
$[8,9]$&$0.946\pm0.029\pm0.125$&$0.895\pm0.022\pm0.105$&$0.685\pm0.030\pm0.116$&$-$\\
$[9,10]$&$0.850\pm0.019\pm0.107$&$0.893\pm0.031\pm0.100$&$0.779\pm0.055\pm0.122$&$-$\\
\hline
\\
    \hline
    &&&$R_{p\mathrm{Pb}}$ (Backward)&& \\
    $\pt[\gevc]\backslash y^*$ &$[-3,-2.5]$   &$[-3.5,-3]$    &$[-4,-3.5]$     &$[-4.5,-4]$\\
\hline
$[1,2]$&$0.881\pm0.039\pm0.197$&$0.980\pm0.035\pm0.133$&$1.026\pm0.042\pm0.148$&$-$\\
$[2,3]$&$0.865\pm0.017\pm0.114$&$0.883\pm0.020\pm0.097$&$1.049\pm0.011\pm0.105$&$1.252\pm0.020\pm0.159$\\
$[3,4]$&$0.878\pm0.009\pm0.102$&$0.961\pm0.023\pm0.096$&$0.961\pm0.017\pm0.094$&$1.125\pm0.016\pm0.130$\\
$[4,5]$&$0.849\pm0.020\pm0.094$&$0.960\pm0.011\pm0.097$&$0.973\pm0.021\pm0.105$&$0.992\pm0.012\pm0.120$\\
$[5,6]$&$0.860\pm0.026\pm0.101$&$0.848\pm0.010\pm0.091$&$0.956\pm0.012\pm0.105$&$1.007\pm0.030\pm0.184$\\
$[6,7]$&$0.707\pm0.030\pm0.087$&$0.902\pm0.012\pm0.107$&$0.916\pm0.020\pm0.124$&$-$\\
$[7,8]$&$0.835\pm0.022\pm0.088$&$0.921\pm0.023\pm0.110$&$0.911\pm0.021\pm0.151$&$-$\\
$[8,9]$&$0.916\pm0.029\pm0.118$&$0.803\pm0.020\pm0.154$&$-$&$-$\\
$[9,10]$&$0.804\pm0.042\pm0.119$&$0.839\pm0.033\pm0.139$&$-$&$-$\\
\hline
\\
    \end{tabular}
    \label{tab:RpPb_Ds_2D}
\end{sidewaystable}

\begin{table}[tbh]
    \caption{
    Nuclear modification factor $R_{p\mathrm{Pb}}$ for prompt \Dp production as a function of \pt at forward (integrated over the common rapidity region of $2.0 < y^{*} < 4.0$) and backward (integrated over the common rapidity region of $-4.5 < y^{*} < -2.5$) rapidity.
    The first uncertainty is statistical, the second systematic.
    }
    \centering
    \begin{tabular}{cc}
    \hline
    \pt[\gevc]& $R_{p\mathrm{Pb}}$ (Forward)\\
\hline
$[1,2]$&$0.652\pm0.002\pm0.058$\\
$[2,3]$&$0.693\pm0.002\pm0.053$\\
$[3,4]$&$0.715\pm0.001\pm0.051$\\
$[4,5]$&$0.727\pm0.001\pm0.059$\\
$[5,6]$&$0.746\pm0.002\pm0.058$\\
$[6,7]$&$0.779\pm0.005\pm0.070$\\
$[7,8]$&$0.787\pm0.005\pm0.071$\\
$[8,9]$&$0.783\pm0.010\pm0.078$\\
$[9,10]$&$0.929\pm0.006\pm0.087$\\
\hline

    \pt[\gevc]& $R_{p\mathrm{Pb}}$ (Backward)\\
\hline
$[1,2]$&$0.808\pm0.004\pm0.100$\\
$[2,3]$&$0.850\pm0.001\pm0.089$\\
$[3,4]$&$0.834\pm0.001\pm0.083$\\
$[4,5]$&$0.831\pm0.001\pm0.085$\\
$[5,6]$&$0.809\pm0.002\pm0.085$\\
$[6,7]$&$0.808\pm0.003\pm0.088$\\
$[7,8]$&$0.797\pm0.008\pm0.085$\\
$[8,9]$&$0.758\pm0.004\pm0.089$\\
$[9,10]$&$0.841\pm0.005\pm0.098$\\
\hline
   
    \end{tabular}
    \label{tab:RpPb_Dp_pt}
\end{table}

\begin{table}[tbh]
    \caption{
    Nuclear modification factor $R_{p\mathrm{Pb}}$ for prompt \Dp production as a function of $y^{*}$, integrated over $1 < \pt < 10$\gevc. The first uncertainty is statistical, the second systematic.}
    \centering
    \begin{tabular}{cc}
    \hline
    $y^*$ & $R_{p\mathrm{Pb}}$\\
\hline
$[-4.5,-4.0]$&$0.892\pm0.007\pm0.096$\\
$[-4.0,-3.5]$&$0.854\pm0.002\pm0.087$\\
$[-3.5,-3.0]$&$0.796\pm0.002\pm0.088$\\
$[-3.0,-2.5]$&$0.791\pm0.004\pm0.102$\\
$[2.0,2.5]$&$0.764\pm0.002\pm0.078$\\
$[2.5,3.0]$&$0.704\pm0.002\pm0.053$\\
$[3.0,3.5]$&$0.649\pm0.002\pm0.045$\\
$[3.5,4.0]$&$0.591\pm0.003\pm0.055$\\
\hline

    \end{tabular}
    \label{tab:RpPb_Dp_y}
\end{table}    
    
\begin{sidewaystable}[htb]
    \caption{
    Nuclear modification factor $R_{p\mathrm{Pb}}$ for prompt \Dp production as a function of \pt and $y^{*}$. The first uncertainty is statistical, the second systematic.}
    \centering
    \begin{tabular}{ccccccccc}
    \hline
    &&&$R_{p\mathrm{Pb}}$ (Forward)&& \\
    $\pt[\gevc]\backslash y^*$   &$[2,2.5]$    &$[2.5,3]$     &$[3,3.5]$     &$[3.5,4]$\\
\hline
$[1,2]$&$0.741\pm0.004\pm0.091$&$0.697\pm0.001\pm0.060$&$0.613\pm0.003\pm0.044$&$0.536\pm0.005\pm0.044$\\
$[2,3]$&$0.787\pm0.002\pm0.071$&$0.691\pm0.005\pm0.049$&$0.657\pm0.002\pm0.040$&$0.614\pm0.003\pm0.062$\\
$[3,4]$&$0.793\pm0.002\pm0.063$&$0.698\pm0.001\pm0.042$&$0.684\pm0.002\pm0.048$&$0.665\pm0.004\pm0.057$\\
$[4,5]$&$0.742\pm0.001\pm0.067$&$0.740\pm0.002\pm0.050$&$0.724\pm0.003\pm0.054$&$0.683\pm0.006\pm0.070$\\
$[5,6]$&$0.776\pm0.002\pm0.057$&$0.747\pm0.003\pm0.052$&$0.733\pm0.003\pm0.058$&$0.707\pm0.008\pm0.077$\\
$[6,7]$&$0.786\pm0.004\pm0.062$&$0.808\pm0.005\pm0.055$&$0.764\pm0.005\pm0.066$&$0.735\pm0.029\pm0.119$\\
$[7,8]$&$0.781\pm0.005\pm0.056$&$0.822\pm0.005\pm0.063$&$0.788\pm0.008\pm0.076$&$0.733\pm0.030\pm0.118$\\
$[8,9]$&$0.768\pm0.005\pm0.057$&$0.821\pm0.006\pm0.068$&$0.788\pm0.011\pm0.098$&$0.739\pm0.065\pm0.124$\\
$[9,10]$&$0.958\pm0.008\pm0.072$&$0.915\pm0.010\pm0.075$&$0.904\pm0.015\pm0.126$&$-$\\
\hline
\\
    \hline
    &&&$R_{p\mathrm{Pb}}$ (Backward)&& \\
    $\pt[\gevc]\backslash y^*$ &$[-3,-2.5]$   &$[-3.5,-3]$    &$[-4,-3.5]$     &$[-4.5,-4]$\\
\hline
$[1,2]$&$0.803\pm0.009\pm0.120$&$0.748\pm0.003\pm0.091$&$0.840\pm0.003\pm0.090$&$0.866\pm0.012\pm0.098$\\
$[2,3]$&$0.795\pm0.003\pm0.093$&$0.846\pm0.002\pm0.090$&$0.874\pm0.002\pm0.084$&$0.927\pm0.002\pm0.086$\\
$[3,4]$&$0.769\pm0.002\pm0.084$&$0.842\pm0.001\pm0.081$&$0.859\pm0.002\pm0.081$&$0.924\pm0.003\pm0.092$\\
$[4,5]$&$0.762\pm0.003\pm0.080$&$0.845\pm0.002\pm0.078$&$0.870\pm0.003\pm0.089$&$0.911\pm0.003\pm0.107$\\
$[5,6]$&$0.750\pm0.003\pm0.075$&$0.819\pm0.002\pm0.076$&$0.852\pm0.005\pm0.086$&$0.881\pm0.004\pm0.140$\\
$[6,7]$&$0.761\pm0.008\pm0.074$&$0.797\pm0.002\pm0.077$&$0.843\pm0.002\pm0.099$&$0.919\pm0.008\pm0.153$\\
$[7,8]$&$0.776\pm0.018\pm0.078$&$0.803\pm0.006\pm0.084$&$0.827\pm0.005\pm0.100$&$-$\\
$[8,9]$&$0.730\pm0.008\pm0.079$&$0.780\pm0.004\pm0.092$&$0.777\pm0.009\pm0.104$&$-$\\
$[9,10]$&$0.843\pm0.005\pm0.095$&$0.839\pm0.009\pm0.104$&$-$&$-$\\
\hline
\\
    \end{tabular}
    \label{tab:RpPb_Dp_2D}
\end{sidewaystable}

The numerical values for the forward and backward production ratio $R_{\mathrm{FB}}$ of prompt \Ds and \Dp mesons are given in Tables~\ref{tab:RFB_Ds} and \ref{tab:RFB_Dp}.

\begin{table}[tbh]
    \caption{
    Forward and backward production ratio $R_{\mathrm{FB}}$ for prompt \Ds mesons as a function of \pt and $y^{*}$. The first uncertainty is statistical, the second systematic. 
        }
        \centering
        \begin{tabular}{cc}
            \hline
            \pt[\gevc]& $R_{\mathrm{FB}}$\\
\hline
$[1,2]$&$0.763\pm0.032\pm0.103$\\
$[2,3]$&$0.743\pm0.011\pm0.079$\\
$[3,4]$&$0.752\pm0.011\pm0.075$\\
$[4,5]$&$0.764\pm0.013\pm0.073$\\
$[5,6]$&$0.858\pm0.016\pm0.084$\\
$[6,7]$&$0.982\pm0.030\pm0.102$\\
$[7,8]$&$0.980\pm0.030\pm0.089$\\
$[8,9]$&$0.921\pm0.028\pm0.092$\\
$[9,10]$&$1.028\pm0.051\pm0.119$\\
$[10,11]$&$0.978\pm0.057\pm0.148$\\
$[11,12]$&$1.028\pm0.074\pm0.144$\\
$[12,13]$&$1.068\pm0.144\pm0.161$\\
\hline
$|y^*|$ & $R_{\mathrm{FB}}$\\
\hline
$[2.5,3.0]$&$0.854\pm0.021\pm0.119$\\
$[3.0,3.5]$&$0.790\pm0.021\pm0.084$\\
$[3.5,4.0]$&$0.623\pm0.035\pm0.071$\\
\hline

        \end{tabular}
        \label{tab:RFB_Ds}
    \end{table}

\begin{table}[tbh]
    \caption{
    Forward and backward production ratio $R_{\mathrm{FB}}$ for prompt \Dp mesons as a function of \pt and $y^{*}$. The first uncertainty is statistical, the second systematic.}
    \centering
    \begin{tabular}{cc}
    \hline
    \pt [\gevc] & $R_{\mathrm{FB}}$ \\
\hline
$[1,2]$ &$0.775\pm 0.004 \pm 0.092$ \\
$[2,3]$ &$0.785\pm 0.003 \pm 0.082$ \\
$[3,4]$ &$0.832\pm 0.002 \pm 0.083$ \\
$[4,5]$ &$0.878\pm 0.003 \pm 0.086$ \\
$[5,6]$ &$0.913\pm 0.004 \pm 0.088$ \\
$[6,7]$ &$0.979\pm 0.010 \pm 0.097$ \\
$[7,8]$ &$0.993\pm 0.014 \pm 0.101$ \\
$[8,9]$ &$1.048\pm 0.022 \pm 0.111$ \\
$[9,10]$ &$1.081\pm 0.013 \pm 0.118$ \\
$[10,11]$ &$1.103\pm 0.022 \pm 0.127$ \\
$[11,12]$ &$1.097\pm 0.028 \pm 0.126$ \\
$[12,13]$ &$1.101\pm 0.049 \pm 0.137$ \\
$[13,14]$ &$1.272\pm 0.044 \pm 0.163$ \\
\hline
$|y^{*}|$ & $R_{\mathrm{FB}}$ \\
\hline
$[2.5,3.0]$ &$0.881\pm 0.005 \pm 0.104$ \\
$[3.0,3.5]$ &$0.814\pm 0.003 \pm 0.086$ \\
$[3.5,4.0]$ &$0.690\pm 0.004 \pm 0.072$ \\
\hline

    \end{tabular}
    \label{tab:RFB_Dp}
\end{table}

\begin{figure}[tb]
    \begin{center}
    \includegraphics[width=0.99\linewidth]{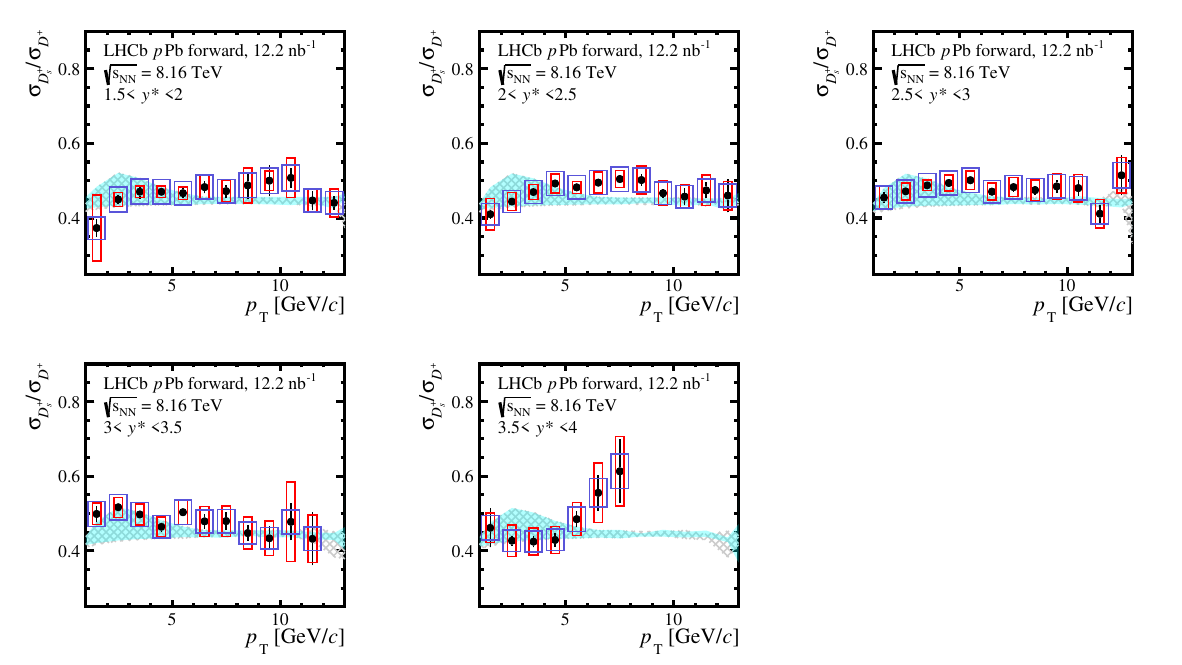}
    \includegraphics[width=0.99\linewidth]{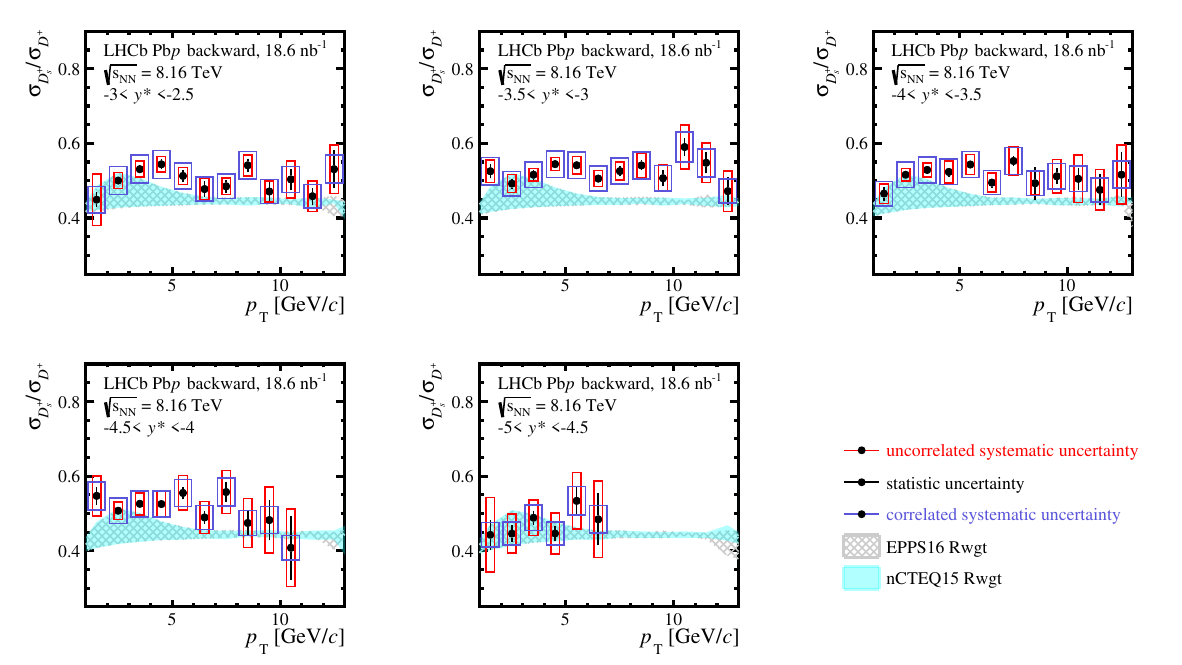}
    \vspace*{-0.5cm}
    \end{center}
    \caption{
    The production cross-section ratio $\sigma_{\Ds}/\sigma_{\Dp}$ as a function of \pt and $y^{*}$ in $p$Pb collisions. The error bars show the statistical uncertainty, the red boxes the uncorrelated systematic uncertainty and the blue boxes the correlated systematic uncertainty. The coloured bands correspond to the theoretical calculations, incorporating nPDFs EPPS16 (gray) \cite{Eskola:2016oht} and nCTEQ15 (cyan) \cite{Kovarik:2015cma}.}
    \label{fig:DsDpRatio}
\end{figure}

\begin{figure}[tb]
        \begin{center}
            \includegraphics[width=0.99\linewidth]{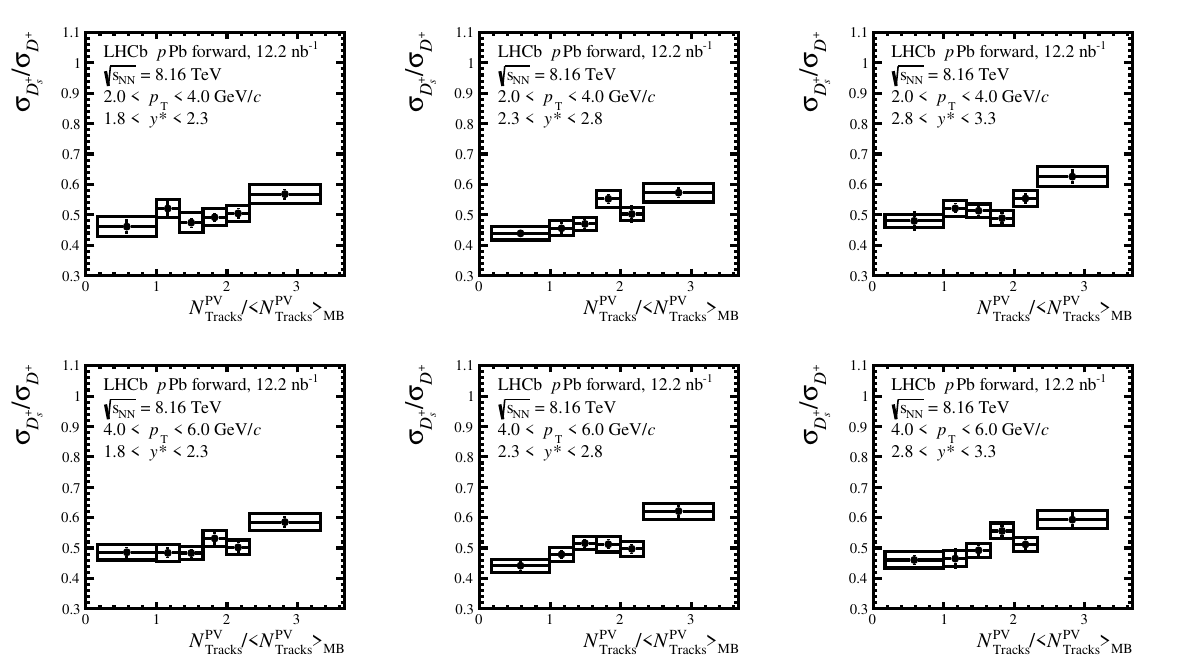}
            \includegraphics[width=0.99\linewidth]{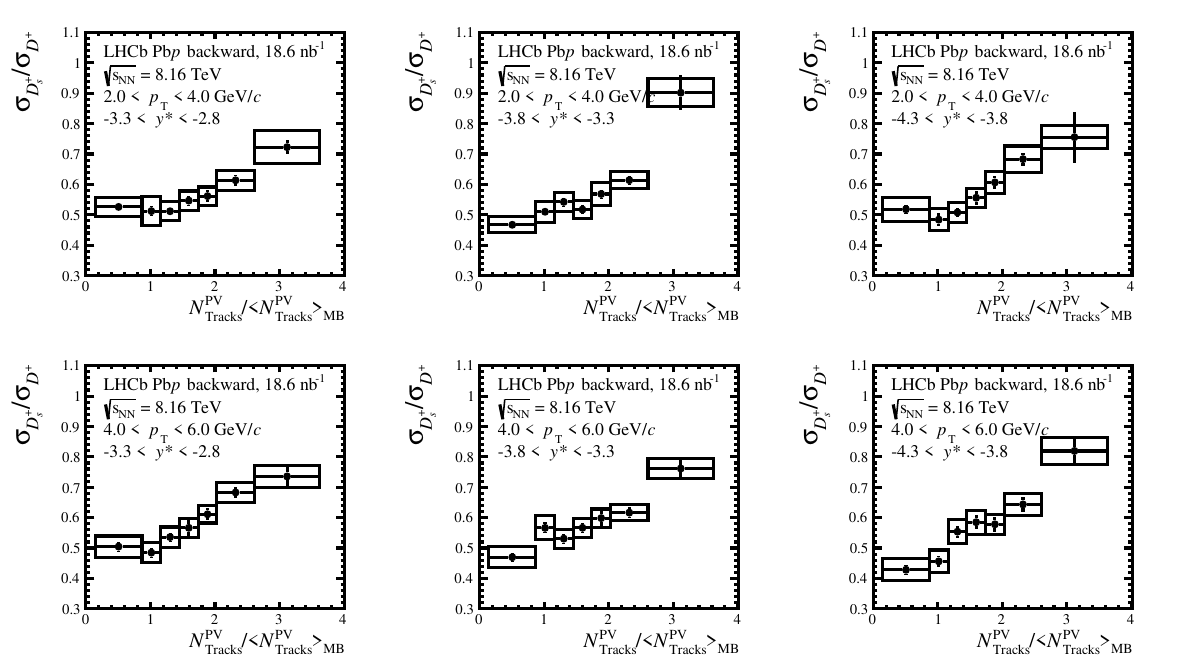}
            \vspace*{-0.5cm}
        \end{center}
        \caption{
        The production cross-section ratio, $\sigma_{\Ds}/\sigma_{\Dp}$, versus normalized event multiplicity in different $D$-meson \pt(2-6\gevc) and $y^{*}$ ranges for the (six upper plots) forward and (six lower plots) backward rapidities. The vertical error bars show the statistical uncertainty, the boxes the systematic.      
        }
        \label{fig:DsDpRatio_3D_low}
    \end{figure}

\begin{figure}[tb]
        \begin{center}
            \includegraphics[width=0.99\linewidth]{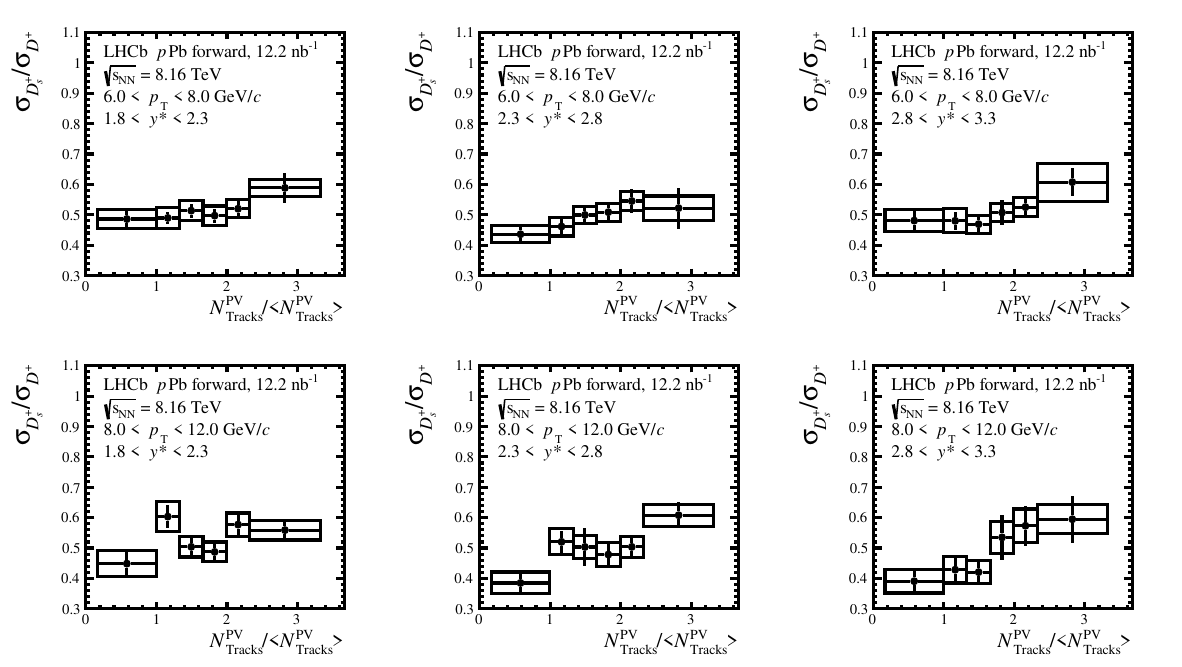}
            \includegraphics[width=0.99\linewidth]{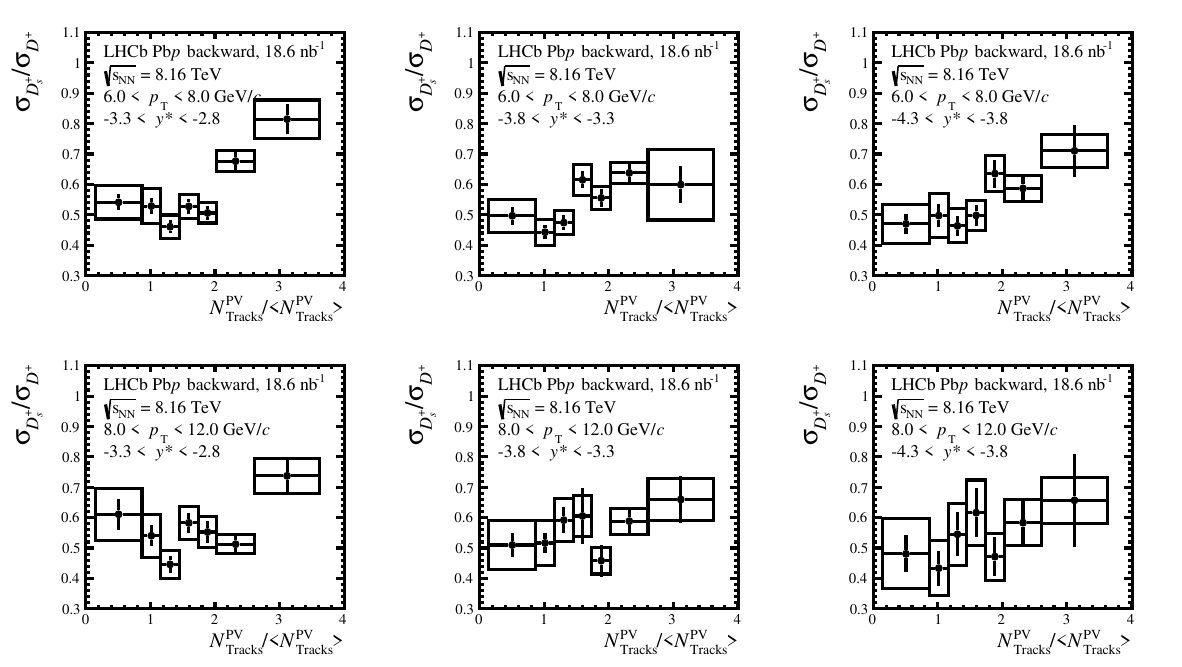}
            \vspace*{-0.5cm}
        \end{center}
        \caption{
        The production cross-section ratio, $\sigma_{\Ds}/\sigma_{\Dp}$, versus normalized event multiplicity in different $D$-meson \pt(6-12\gevc) and $y^{*}$ ranges for the (six upper plots) forward and (six lower plots) backward rapidities. The vertical error bars show the statistical uncertainty, the boxes the systematic.    
        }
        \label{fig:DsDpRatio_3D_high}
    \end{figure}

The production cross-section ratio of \Ds over \Dp mesons $\sigma_{\Ds}/\sigma_{\Dp}$ in both forward and backward rapidities are shown in Fig.~\ref{fig:DsDpRatio}--\ref{fig:DsDpRatio_3D_high}. The corresponding numerical values are listed in Tables~\ref{tab:DsDpRatio_2D} and \ref{tab:DsDpRatio}.

\begin{sidewaystable}[htb]
    \caption{
    The production cross-section ratio $\sigma_{\Ds}/\sigma_{\Dp}$ as a function of \pt and $y^{*}$ in $p$Pb collisions at (upper) forward and (lower) backward rapidities. The first uncertainty is statistical, the second the component of the systematic uncertainty that is uncorrelated between bins and the third the correlated systematic component.}
    \centering
    \scalebox{0.7}{
    \begin{tabular}{cccccc}
    \hline
    &&&$\sigma_{\Ds}/\sigma_{\Dp}$ (Forward)&& \\
    $\pt[\gevc]\backslash y^*$ &$[1.5,2]$   &$[2,2.5]$    &$[2.5,3]$     &$[3,3.5]$     &$[3.5,4]$\\
\hline
$[1,2]$&$0.373\pm0.023\pm0.088\pm0.030$&$0.410\pm0.012\pm0.042\pm0.029$&$0.455\pm0.015\pm0.030\pm0.031$&$0.499\pm0.022\pm0.029\pm0.034$&$0.461\pm0.052\pm0.039\pm0.033$\\
$[2,3]$&$0.450\pm0.013\pm0.018\pm0.034$&$0.444\pm0.004\pm0.024\pm0.030$&$0.472\pm0.006\pm0.023\pm0.031$&$0.516\pm0.008\pm0.027\pm0.034$&$0.427\pm0.013\pm0.042\pm0.029$\\
$[3,4]$&$0.471\pm0.019\pm0.015\pm0.034$&$0.470\pm0.007\pm0.020\pm0.031$&$0.488\pm0.005\pm0.014\pm0.032$&$0.497\pm0.007\pm0.029\pm0.033$&$0.425\pm0.016\pm0.036\pm0.028$\\
$[4,5]$&$0.470\pm0.009\pm0.015\pm0.032$&$0.493\pm0.007\pm0.026\pm0.032$&$0.494\pm0.006\pm0.022\pm0.032$&$0.464\pm0.011\pm0.026\pm0.030$&$0.429\pm0.018\pm0.036\pm0.029$\\
$[5,6]$&$0.466\pm0.013\pm0.017\pm0.032$&$0.482\pm0.009\pm0.016\pm0.031$&$0.501\pm0.007\pm0.025\pm0.032$&$0.503\pm0.010\pm0.033\pm0.033$&$0.485\pm0.022\pm0.044\pm0.033$\\
$[6,7]$&$0.483\pm0.017\pm0.032\pm0.033$&$0.495\pm0.006\pm0.028\pm0.032$&$0.471\pm0.009\pm0.023\pm0.030$&$0.479\pm0.020\pm0.040\pm0.031$&$0.555\pm0.048\pm0.080\pm0.038$\\
$[7,8]$&$0.472\pm0.017\pm0.028\pm0.032$&$0.504\pm0.010\pm0.023\pm0.033$&$0.483\pm0.012\pm0.025\pm0.031$&$0.480\pm0.024\pm0.041\pm0.031$&$0.613\pm0.085\pm0.094\pm0.046$\\
$[8,9]$&$0.487\pm0.036\pm0.047\pm0.033$&$0.502\pm0.016\pm0.038\pm0.033$&$0.475\pm0.012\pm0.027\pm0.031$&$0.447\pm0.020\pm0.043\pm0.029$&$-$\\
$[9,10]$&$0.500\pm0.042\pm0.027\pm0.034$&$0.467\pm0.011\pm0.033\pm0.030$&$0.485\pm0.018\pm0.034\pm0.031$&$0.434\pm0.032\pm0.046\pm0.029$&$-$\\
$[10,11]$&$0.508\pm0.026\pm0.053\pm0.035$&$0.457\pm0.023\pm0.031\pm0.030$&$0.480\pm0.022\pm0.037\pm0.031$&$0.478\pm0.049\pm0.106\pm0.032$&$-$\\
$[11,12]$&$0.447\pm0.027\pm0.033\pm0.030$&$0.474\pm0.021\pm0.041\pm0.031$&$0.412\pm0.024\pm0.039\pm0.027$&$0.432\pm0.071\pm0.064\pm0.031$&$-$\\
$[12,13]$&$0.441\pm0.019\pm0.037\pm0.030$&$0.460\pm0.045\pm0.038\pm0.030$&$0.514\pm0.053\pm0.048\pm0.034$&$-$&$-$\\
\hline
\\
    \hline
    &&&$\sigma_{\Ds}/\sigma_{\Dp}$ (Backward)&& \\
    $\pt[\gevc]\backslash y^*$ &$[-3,-2.5]$   &$[-3.5,-3]$    &$[-4,-3.5]$     &$[-4.5,-4]$     &$[-5,-4.5]$\\
\hline
$[1,2]$&$0.449\pm0.020\pm0.069\pm0.035$&$0.525\pm0.019\pm0.030\pm0.037$&$0.465\pm0.019\pm0.026\pm0.033$&$0.547\pm0.023\pm0.054\pm0.038$&$0.443\pm0.040\pm0.100\pm0.033$\\
$[2,3]$&$0.500\pm0.010\pm0.021\pm0.037$&$0.492\pm0.011\pm0.024\pm0.033$&$0.516\pm0.005\pm0.017\pm0.034$&$0.507\pm0.008\pm0.024\pm0.034$&$0.446\pm0.023\pm0.052\pm0.031$\\
$[3,4]$&$0.531\pm0.005\pm0.021\pm0.037$&$0.516\pm0.012\pm0.017\pm0.034$&$0.529\pm0.009\pm0.019\pm0.035$&$0.526\pm0.007\pm0.030\pm0.035$&$0.488\pm0.019\pm0.048\pm0.034$\\
$[4,5]$&$0.544\pm0.013\pm0.021\pm0.037$&$0.544\pm0.006\pm0.017\pm0.036$&$0.523\pm0.011\pm0.029\pm0.034$&$0.525\pm0.006\pm0.034\pm0.034$&$0.446\pm0.020\pm0.054\pm0.031$\\
$[5,6]$&$0.513\pm0.015\pm0.023\pm0.035$&$0.541\pm0.007\pm0.023\pm0.036$&$0.543\pm0.008\pm0.026\pm0.035$&$0.555\pm0.017\pm0.046\pm0.036$&$0.534\pm0.038\pm0.076\pm0.038$\\
$[6,7]$&$0.478\pm0.021\pm0.029\pm0.033$&$0.506\pm0.007\pm0.022\pm0.033$&$0.495\pm0.011\pm0.025\pm0.032$&$0.489\pm0.017\pm0.043\pm0.032$&$0.484\pm0.070\pm0.102\pm0.037$\\
$[7,8]$&$0.485\pm0.017\pm0.022\pm0.033$&$0.526\pm0.014\pm0.024\pm0.035$&$0.553\pm0.013\pm0.037\pm0.036$&$0.557\pm0.026\pm0.057\pm0.038$&$-$\\
$[8,9]$&$0.541\pm0.018\pm0.028\pm0.037$&$0.541\pm0.013\pm0.033\pm0.036$&$0.493\pm0.044\pm0.032\pm0.032$&$0.474\pm0.032\pm0.065\pm0.033$&$-$\\
$[9,10]$&$0.471\pm0.025\pm0.028\pm0.032$&$0.507\pm0.021\pm0.036\pm0.034$&$0.512\pm0.021\pm0.045\pm0.034$&$0.482\pm0.055\pm0.088\pm0.036$&$-$\\
$[10,11]$&$0.504\pm0.028\pm0.050\pm0.035$&$0.590\pm0.025\pm0.059\pm0.040$&$0.505\pm0.029\pm0.063\pm0.034$&$0.408\pm0.085\pm0.104\pm0.032$&$-$\\
$[11,12]$&$0.458\pm0.022\pm0.041\pm0.032$&$0.548\pm0.028\pm0.053\pm0.037$&$0.475\pm0.041\pm0.054\pm0.032$&$-$&$-$\\
$[12,13]$&$0.531\pm0.051\pm0.065\pm0.037$&$0.472\pm0.038\pm0.054\pm0.032$&$0.516\pm0.060\pm0.079\pm0.036$&$-$&$-$\\
\hline
\\
    \end{tabular}
    }
    \label{tab:DsDpRatio_2D}
\end{sidewaystable}

\begin{sidewaystable}[htb]
        \caption{
        The production cross-section ratio $\sigma_{\Ds}/\sigma_{\Dp}$ as a function of \pt, $y^{*}$ and  $N^{\text{PV}}_{\text{Tracks}}$ in $p$Pb collisions at (upper) forward and (lower) backward rapidities. The first uncertainty is statistical, the second the component of the systematic uncertainty that is uncorrelated between bins and the third the correlated systematic component.}
        \centering
        \resizebox{\textwidth}{!}{
        \scalebox{0.82}{
            \begin{tabular}{ccccccccc}
                \hline
                &&&$\sigma_{\Ds}/\sigma_{\Dp}$ (Forward)&& \\
                $\pt[\gevc], y^* \backslash N^{\text{PV}}_{\text{tracks}}$       &$[10,60]$       &$[60,80]$       &$[80,100]$       &$[100,120]$       &$[120,140]$       &$[140,200]$\\
\hline
$[2,4],[1.8,2.3]$&$0.46\pm0.02\pm0.02\pm0.02$&$0.52\pm0.03\pm0.02\pm0.02$&$0.47\pm0.02\pm0.03\pm0.02$&$0.49\pm0.02\pm0.02\pm0.02$&$0.50\pm0.02\pm0.02\pm0.02$&$0.57\pm0.02\pm0.02\pm0.03$\\
$[2,4],[2.3,2.8]$&$0.44\pm0.01\pm0.01\pm0.02$&$0.46\pm0.02\pm0.02\pm0.02$&$0.47\pm0.02\pm0.02\pm0.01$&$0.55\pm0.01\pm0.02\pm0.02$&$0.50\pm0.03\pm0.02\pm0.02$&$0.57\pm0.02\pm0.02\pm0.02$\\
$[2,4],[2.8,3.3]$&$0.48\pm0.03\pm0.02\pm0.02$&$0.52\pm0.02\pm0.02\pm0.02$&$0.51\pm0.02\pm0.02\pm0.02$&$0.49\pm0.02\pm0.02\pm0.02$&$0.55\pm0.02\pm0.02\pm0.02$&$0.63\pm0.02\pm0.02\pm0.02$\\
$[4,6],[1.8,2.3]$&$0.49\pm0.02\pm0.02\pm0.02$&$0.48\pm0.02\pm0.02\pm0.02$&$0.48\pm0.01\pm0.02\pm0.01$&$0.53\pm0.03\pm0.02\pm0.02$&$0.50\pm0.02\pm0.02\pm0.02$&$0.59\pm0.02\pm0.02\pm0.02$\\
$[4,6],[2.3,2.8]$&$0.44\pm0.01\pm0.01\pm0.02$&$0.48\pm0.02\pm0.02\pm0.01$&$0.52\pm0.02\pm0.02\pm0.01$&$0.51\pm0.02\pm0.02\pm0.02$&$0.50\pm0.02\pm0.02\pm0.01$&$0.62\pm0.02\pm0.02\pm0.02$\\
$[4,6],[2.8,3.3]$&$0.46\pm0.02\pm0.02\pm0.02$&$0.47\pm0.03\pm0.02\pm0.01$&$0.49\pm0.02\pm0.02\pm0.01$&$0.56\pm0.02\pm0.02\pm0.02$&$0.51\pm0.02\pm0.02\pm0.01$&$0.59\pm0.03\pm0.02\pm0.02$\\
$[6,8],[1.8,2.3]$&$0.49\pm0.03\pm0.03\pm0.02$&$0.49\pm0.02\pm0.03\pm0.02$&$0.51\pm0.02\pm0.03\pm0.01$&$0.50\pm0.02\pm0.03\pm0.02$&$0.52\pm0.03\pm0.02\pm0.02$&$0.59\pm0.05\pm0.02\pm0.02$\\
$[6,8],[2.3,2.8]$&$0.44\pm0.03\pm0.02\pm0.02$&$0.46\pm0.03\pm0.03\pm0.01$&$0.50\pm0.02\pm0.03\pm0.01$&$0.51\pm0.02\pm0.03\pm0.01$&$0.55\pm0.04\pm0.03\pm0.01$&$0.52\pm0.07\pm0.04\pm0.01$\\
$[6,8],[2.8,3.3]$&$0.48\pm0.03\pm0.03\pm0.02$&$0.48\pm0.03\pm0.04\pm0.01$&$0.47\pm0.03\pm0.03\pm0.01$&$0.51\pm0.04\pm0.03\pm0.01$&$0.53\pm0.04\pm0.03\pm0.01$&$0.61\pm0.05\pm0.06\pm0.02$\\
$[8,12],[1.8,2.3]$&$0.45\pm0.04\pm0.04\pm0.02$&$0.60\pm0.04\pm0.04\pm0.02$&$0.50\pm0.03\pm0.03\pm0.01$&$0.49\pm0.03\pm0.03\pm0.01$&$0.58\pm0.04\pm0.03\pm0.02$&$0.56\pm0.04\pm0.03\pm0.02$\\
$[8,12],[2.3,2.8]$&$0.39\pm0.03\pm0.03\pm0.01$&$0.52\pm0.04\pm0.04\pm0.02$&$0.50\pm0.06\pm0.04\pm0.01$&$0.48\pm0.04\pm0.04\pm0.01$&$0.50\pm0.04\pm0.03\pm0.01$&$0.61\pm0.04\pm0.03\pm0.02$\\
$[8,12],[2.8,3.3]$&$0.39\pm0.04\pm0.04\pm0.01$&$0.43\pm0.05\pm0.04\pm0.01$&$0.42\pm0.04\pm0.04\pm0.01$&$0.54\pm0.07\pm0.05\pm0.02$&$0.57\pm0.07\pm0.05\pm0.02$&$0.59\pm0.08\pm0.04\pm0.02$\\
\hline
\\
                \hline
                &&&$\sigma_{\Ds}/\sigma_{\Dp}$ (Backward)&& \\
                $\pt[\gevc], y^* \backslash N^{\text{PV}}_{\text{tracks}}$       &$[10,60]$       &$[60,80]$       &$[80,100]$       &$[100,120]$       &$[120,140]$       &$[140,180]$       &$[180,250]$\\
\hline
$[2,4],[-3.3,-2.8]$&$0.53\pm0.01\pm0.03\pm0.02$&$0.51\pm0.02\pm0.04\pm0.03$&$0.51\pm0.01\pm0.02\pm0.02$&$0.55\pm0.02\pm0.02\pm0.02$&$0.56\pm0.02\pm0.02\pm0.02$&$0.61\pm0.02\pm0.02\pm0.02$&$0.72\pm0.02\pm0.04\pm0.04$\\
$[2,4],[-3.8,-3.3]$&$0.47\pm0.01\pm0.02\pm0.01$&$0.51\pm0.01\pm0.03\pm0.02$&$0.54\pm0.01\pm0.03\pm0.02$&$0.52\pm0.02\pm0.02\pm0.02$&$0.57\pm0.01\pm0.03\pm0.02$&$0.61\pm0.02\pm0.02\pm0.02$&$0.90\pm0.06\pm0.03\pm0.04$\\
$[2,4],[-4.3,-3.8]$&$0.52\pm0.01\pm0.03\pm0.02$&$0.48\pm0.02\pm0.03\pm0.02$&$0.51\pm0.01\pm0.03\pm0.02$&$0.56\pm0.02\pm0.03\pm0.02$&$0.61\pm0.02\pm0.03\pm0.02$&$0.68\pm0.02\pm0.04\pm0.03$&$0.76\pm0.08\pm0.03\pm0.03$\\
$[4,6],[-3.3,-2.8]$&$0.50\pm0.01\pm0.03\pm0.02$&$0.48\pm0.02\pm0.03\pm0.02$&$0.54\pm0.01\pm0.03\pm0.02$&$0.57\pm0.03\pm0.02\pm0.02$&$0.61\pm0.02\pm0.02\pm0.02$&$0.68\pm0.02\pm0.02\pm0.02$&$0.74\pm0.03\pm0.02\pm0.03$\\
$[4,6],[-3.8,-3.3]$&$0.47\pm0.01\pm0.03\pm0.02$&$0.57\pm0.02\pm0.04\pm0.02$&$0.53\pm0.01\pm0.03\pm0.01$&$0.57\pm0.02\pm0.03\pm0.02$&$0.60\pm0.03\pm0.03\pm0.02$&$0.62\pm0.02\pm0.02\pm0.02$&$0.76\pm0.04\pm0.02\pm0.03$\\
$[4,6],[-4.3,-3.8]$&$0.43\pm0.02\pm0.03\pm0.02$&$0.46\pm0.02\pm0.03\pm0.02$&$0.55\pm0.02\pm0.04\pm0.02$&$0.58\pm0.02\pm0.04\pm0.02$&$0.58\pm0.02\pm0.03\pm0.02$&$0.64\pm0.03\pm0.03\pm0.02$&$0.82\pm0.05\pm0.03\pm0.03$\\
$[6,8],[-3.3,-2.8]$&$0.54\pm0.03\pm0.05\pm0.02$&$0.53\pm0.03\pm0.05\pm0.02$&$0.46\pm0.02\pm0.04\pm0.01$&$0.53\pm0.03\pm0.03\pm0.02$&$0.51\pm0.02\pm0.03\pm0.02$&$0.68\pm0.03\pm0.03\pm0.02$&$0.81\pm0.05\pm0.06\pm0.03$\\
$[6,8],[-3.8,-3.3]$&$0.50\pm0.03\pm0.05\pm0.02$&$0.44\pm0.02\pm0.04\pm0.01$&$0.47\pm0.02\pm0.04\pm0.01$&$0.62\pm0.03\pm0.05\pm0.02$&$0.56\pm0.03\pm0.03\pm0.02$&$0.64\pm0.03\pm0.03\pm0.02$&$0.60\pm0.06\pm0.11\pm0.02$\\
$[6,8],[-4.3,-3.8]$&$0.47\pm0.03\pm0.06\pm0.02$&$0.50\pm0.04\pm0.07\pm0.02$&$0.46\pm0.03\pm0.05\pm0.01$&$0.50\pm0.04\pm0.05\pm0.01$&$0.64\pm0.05\pm0.06\pm0.02$&$0.59\pm0.04\pm0.04\pm0.02$&$0.71\pm0.09\pm0.05\pm0.02$\\
$[8,12],[-3.3,-2.8]$&$0.61\pm0.05\pm0.08\pm0.02$&$0.54\pm0.03\pm0.07\pm0.02$&$0.45\pm0.03\pm0.04\pm0.02$&$0.58\pm0.03\pm0.05\pm0.02$&$0.55\pm0.04\pm0.05\pm0.02$&$0.51\pm0.03\pm0.03\pm0.02$&$0.74\pm0.06\pm0.05\pm0.02$\\
$[8,12],[-3.8,-3.3]$&$0.51\pm0.04\pm0.08\pm0.02$&$0.52\pm0.03\pm0.07\pm0.02$&$0.59\pm0.04\pm0.07\pm0.02$&$0.61\pm0.09\pm0.06\pm0.02$&$0.46\pm0.05\pm0.04\pm0.01$&$0.59\pm0.04\pm0.04\pm0.02$&$0.66\pm0.08\pm0.07\pm0.02$\\
$[8,12],[-4.3,-3.8]$&$0.48\pm0.06\pm0.11\pm0.02$&$0.43\pm0.06\pm0.09\pm0.02$&$0.55\pm0.07\pm0.10\pm0.02$&$0.62\pm0.08\pm0.11\pm0.02$&$0.47\pm0.06\pm0.08\pm0.02$&$0.58\pm0.08\pm0.07\pm0.02$&$0.66\pm0.15\pm0.07\pm0.02$\\
\hline
\\
            \end{tabular}}
        }
        \label{tab:DsDpRatio}
    \end{sidewaystable}

\clearpage


\addcontentsline{toc}{section}{References}
\bibliographystyle{LHCb}
\bibliography{main,standard,LHCb-PAPER,LHCb-CONF,LHCb-DP,LHCb-TDR}

\ifx\mcitethebibliography\mciteundefinedmacro
\PackageError{LHCb.bst}{mciteplus.sty has not been loaded}
{This bibstyle requires the use of the mciteplus package.}\fi
\providecommand{\href}[2]{#2}
\begin{mcitethebibliography}{10}
\mciteSetBstSublistMode{n}
\mciteSetBstMaxWidthForm{subitem}{\alph{mcitesubitemcount})}
\mciteSetBstSublistLabelBeginEnd{\mcitemaxwidthsubitemform\space}
{\relax}{\relax}

\bibitem{Webber:1983if}
B.~R. Webber, \ifthenelse{\boolean{articletitles}}{\emph{{A QCD model for jet
  fragmentation including soft gluon interference}},
  }{}\href{https://doi.org/10.1016/0550-3213(84)90333-X}{Nucl.\ Phys.\
  \textbf{B238} (1984) 492}\relax
\mciteBstWouldAddEndPuncttrue
\mciteSetBstMidEndSepPunct{\mcitedefaultmidpunct}
{\mcitedefaultendpunct}{\mcitedefaultseppunct}\relax
\EndOfBibitem
\bibitem{Andersson:1983ia}
B.~Andersson, G.~Gustafson, G.~Ingelman, and T.~Sjostrand,
  \ifthenelse{\boolean{articletitles}}{\emph{{Parton fragmentation and string
  dynamics}}, }{}\href{https://doi.org/10.1016/0370-1573(83)90080-7}{Phys.\
  Rept.\  \textbf{97} (1983) 31}\relax
\mciteBstWouldAddEndPuncttrue
\mciteSetBstMidEndSepPunct{\mcitedefaultmidpunct}
{\mcitedefaultendpunct}{\mcitedefaultseppunct}\relax
\EndOfBibitem
\bibitem{Hirai:2007sx}
M.~Hirai, S.~Kumano, and T.-H. Nagai,
  \ifthenelse{\boolean{articletitles}}{\emph{{Determination of nuclear parton
  distribution functions and their uncertainties in next-to-leading order}},
  }{}\href{https://doi.org/10.1103/PhysRevC.76.065207}{Phys.\ Rev.\
  \textbf{C76} (2007) 065207},
  \href{http://arxiv.org/abs/0709.3038}{{\normalfont\ttfamily
  arXiv:0709.3038}}\relax
\mciteBstWouldAddEndPuncttrue
\mciteSetBstMidEndSepPunct{\mcitedefaultmidpunct}
{\mcitedefaultendpunct}{\mcitedefaultseppunct}\relax
\EndOfBibitem
\bibitem{Eskola:2021nhw}
K.~J. Eskola, P.~Paakkinen, H.~Paukkunen, and C.~A. Salgado,
  \ifthenelse{\boolean{articletitles}}{\emph{{EPPS21: a global QCD analysis of
  nuclear PDFs}},
  }{}\href{https://doi.org/10.1140/epjc/s10052-022-10359-0}{Eur.\ Phys.\ J.\
  \textbf{C82} (2022) 413},
  \href{http://arxiv.org/abs/2112.12462}{{\normalfont\ttfamily
  arXiv:2112.12462}}\relax
\mciteBstWouldAddEndPuncttrue
\mciteSetBstMidEndSepPunct{\mcitedefaultmidpunct}
{\mcitedefaultendpunct}{\mcitedefaultseppunct}\relax
\EndOfBibitem
\bibitem{Gelis:2012ri}
F.~Gelis, \ifthenelse{\boolean{articletitles}}{\emph{{Color Glass Condensate
  and Glasma}}, }{}\href{https://doi.org/10.1142/S0217751X13300019}{Int.\ J.\
  Mod.\ Phys.\  \textbf{A28} (2013) 1330001},
  \href{http://arxiv.org/abs/1211.3327}{{\normalfont\ttfamily
  arXiv:1211.3327}}\relax
\mciteBstWouldAddEndPuncttrue
\mciteSetBstMidEndSepPunct{\mcitedefaultmidpunct}
{\mcitedefaultendpunct}{\mcitedefaultseppunct}\relax
\EndOfBibitem
\bibitem{Fujii:2013yja}
H.~Fujii and K.~Watanabe, \ifthenelse{\boolean{articletitles}}{\emph{{Heavy
  quark pair production in high energy pA collisions: Open heavy flavors}},
  }{}\href{https://doi.org/10.1016/j.nuclphysa.2013.10.006}{Nucl.\ Phys.\
  \textbf{A920} (2013) 78},
  \href{http://arxiv.org/abs/1308.1258}{{\normalfont\ttfamily
  arXiv:1308.1258}}\relax
\mciteBstWouldAddEndPuncttrue
\mciteSetBstMidEndSepPunct{\mcitedefaultmidpunct}
{\mcitedefaultendpunct}{\mcitedefaultseppunct}\relax
\EndOfBibitem
\bibitem{LHCb:2022dmh}
LHCb collaboration, I.~Bezshyiko {\em et~al.},
  \ifthenelse{\boolean{articletitles}}{\emph{{Measurement of the Prompt D0
  Nuclear Modification Factor in p-Pb Collisions at sNN=8.16\,\,TeV}},
  }{}\href{https://doi.org/10.1103/PhysRevLett.131.102301}{Phys.\ Rev.\ Lett.\
  \textbf{131} (2023) 102301},
  \href{http://arxiv.org/abs/2205.03936}{{\normalfont\ttfamily
  arXiv:2205.03936}}\relax
\mciteBstWouldAddEndPuncttrue
\mciteSetBstMidEndSepPunct{\mcitedefaultmidpunct}
{\mcitedefaultendpunct}{\mcitedefaultseppunct}\relax
\EndOfBibitem
\bibitem{Braaten:1994bz}
E.~Braaten, K.-m. Cheung, S.~Fleming, and T.~C. Yuan,
  \ifthenelse{\boolean{articletitles}}{\emph{{Perturbative QCD fragmentation
  functions as a model for heavy quark fragmentation}},
  }{}\href{https://doi.org/10.1103/PhysRevD.51.4819}{Phys.\ Rev.\  \textbf{D51}
  (1995) 4819},
  \href{http://arxiv.org/abs/hep-ph/9409316}{{\normalfont\ttfamily
  arXiv:hep-ph/9409316}}\relax
\mciteBstWouldAddEndPuncttrue
\mciteSetBstMidEndSepPunct{\mcitedefaultmidpunct}
{\mcitedefaultendpunct}{\mcitedefaultseppunct}\relax
\EndOfBibitem
\bibitem{ALICE:2021dhb}
ALICE collaboration, S.~Acharya {\em et~al.},
  \ifthenelse{\boolean{articletitles}}{\emph{{Charm-quark fragmentation
  fractions and production cross section at midrapidity in pp collisions at the
  LHC}}, }{}\href{https://doi.org/10.1103/PhysRevD.105.L011103}{Phys.\ Rev.\
  \textbf{D105} (2022) L011103},
  \href{http://arxiv.org/abs/2105.06335}{{\normalfont\ttfamily
  arXiv:2105.06335}}\relax
\mciteBstWouldAddEndPuncttrue
\mciteSetBstMidEndSepPunct{\mcitedefaultmidpunct}
{\mcitedefaultendpunct}{\mcitedefaultseppunct}\relax
\EndOfBibitem
\bibitem{ALICE:2023sgl}
ALICE collaboration, S.~Acharya {\em et~al.},
  \ifthenelse{\boolean{articletitles}}{\emph{{Charm production and
  fragmentation fractions at midrapidity in pp collisions at $\sqrt{s} = 13$
  TeV}}, }{}\href{http://arxiv.org/abs/2308.04877}{{\normalfont\ttfamily
  arXiv:2308.04877}}\relax
\mciteBstWouldAddEndPuncttrue
\mciteSetBstMidEndSepPunct{\mcitedefaultmidpunct}
{\mcitedefaultendpunct}{\mcitedefaultseppunct}\relax
\EndOfBibitem
\bibitem{Oh:2009zj}
Y.~Oh, C.~M. Ko, S.~H. Lee, and S.~Yasui,
  \ifthenelse{\boolean{articletitles}}{\emph{{Heavy baryon/meson ratios in
  relativistic heavy ion collisions}},
  }{}\href{https://doi.org/10.1103/PhysRevC.79.044905}{Phys.\ Rev.\
  \textbf{C79} (2009) 044905},
  \href{http://arxiv.org/abs/0901.1382}{{\normalfont\ttfamily
  arXiv:0901.1382}}\relax
\mciteBstWouldAddEndPuncttrue
\mciteSetBstMidEndSepPunct{\mcitedefaultmidpunct}
{\mcitedefaultendpunct}{\mcitedefaultseppunct}\relax
\EndOfBibitem
\bibitem{He:2019vgs}
M.~He and R.~Rapp, \ifthenelse{\boolean{articletitles}}{\emph{{Hadronization
  and charm-hadron ratios in heavy-ion collisions}},
  }{}\href{https://doi.org/10.1103/PhysRevLett.124.042301}{Phys.\ Rev.\ Lett.\
  \textbf{124} (2020) 042301},
  \href{http://arxiv.org/abs/1905.09216}{{\normalfont\ttfamily
  arXiv:1905.09216}}\relax
\mciteBstWouldAddEndPuncttrue
\mciteSetBstMidEndSepPunct{\mcitedefaultmidpunct}
{\mcitedefaultendpunct}{\mcitedefaultseppunct}\relax
\EndOfBibitem
\bibitem{Minissale:2020bif}
V.~Minissale, S.~Plumari, and V.~Greco,
  \ifthenelse{\boolean{articletitles}}{\emph{{Charm hadrons in pp collisions at
  LHC energy within a coalescence plus fragmentation approach}},
  }{}\href{https://doi.org/10.1016/j.physletb.2021.136622}{Phys.\ Lett.\
  \textbf{B821} (2021) 136622},
  \href{http://arxiv.org/abs/2012.12001}{{\normalfont\ttfamily
  arXiv:2012.12001}}\relax
\mciteBstWouldAddEndPuncttrue
\mciteSetBstMidEndSepPunct{\mcitedefaultmidpunct}
{\mcitedefaultendpunct}{\mcitedefaultseppunct}\relax
\EndOfBibitem
\bibitem{CMS:2018loe}
CMS collaboration, A.~M. Sirunyan {\em et~al.},
  \ifthenelse{\boolean{articletitles}}{\emph{{Elliptic flow of charm and
  strange hadrons in high-multiplicity pPb collisions at
  $\sqrt{s_{_\mathrm{NN}}} =$ 8.16 TeV}},
  }{}\href{https://doi.org/10.1103/PhysRevLett.121.082301}{Phys.\ Rev.\ Lett.\
  \textbf{121} (2018) 082301},
  \href{http://arxiv.org/abs/1804.09767}{{\normalfont\ttfamily
  arXiv:1804.09767}}\relax
\mciteBstWouldAddEndPuncttrue
\mciteSetBstMidEndSepPunct{\mcitedefaultmidpunct}
{\mcitedefaultendpunct}{\mcitedefaultseppunct}\relax
\EndOfBibitem
\bibitem{STAR:2005gfr}
STAR collaboration, J.~Adams {\em et~al.},
  \ifthenelse{\boolean{articletitles}}{\emph{{Experimental and theoretical
  challenges in the search for the quark gluon plasma: The STAR Collaboration's
  critical assessment of the evidence from RHIC collisions}},
  }{}\href{https://doi.org/10.1016/j.nuclphysa.2005.03.085}{Nucl.\ Phys.\
  \textbf{A757} (2005) 102},
  \href{http://arxiv.org/abs/nucl-ex/0501009}{{\normalfont\ttfamily
  arXiv:nucl-ex/0501009}}\relax
\mciteBstWouldAddEndPuncttrue
\mciteSetBstMidEndSepPunct{\mcitedefaultmidpunct}
{\mcitedefaultendpunct}{\mcitedefaultseppunct}\relax
\EndOfBibitem
\bibitem{PHENIX:2004vcz}
PHENIX collaboration, K.~Adcox {\em et~al.},
  \ifthenelse{\boolean{articletitles}}{\emph{{Formation of dense partonic
  matter in relativistic nucleus-nucleus collisions at RHIC: Experimental
  evaluation by the PHENIX collaboration}},
  }{}\href{https://doi.org/10.1016/j.nuclphysa.2005.03.086}{Nucl.\ Phys.\
  \textbf{A757} (2005) 184},
  \href{http://arxiv.org/abs/nucl-ex/0410003}{{\normalfont\ttfamily
  arXiv:nucl-ex/0410003}}\relax
\mciteBstWouldAddEndPuncttrue
\mciteSetBstMidEndSepPunct{\mcitedefaultmidpunct}
{\mcitedefaultendpunct}{\mcitedefaultseppunct}\relax
\EndOfBibitem
\bibitem{PhysRevLett.48.1066}
J.~Rafelski and B.~M\"uller,
  \ifthenelse{\boolean{articletitles}}{\emph{Strangeness production in the
  quark-gluon plasma},
  }{}\href{https://doi.org/10.1103/PhysRevLett.48.1066}{Phys.\ Rev.\ Lett.\
  \textbf{48} (1982) 1066}\relax
\mciteBstWouldAddEndPuncttrue
\mciteSetBstMidEndSepPunct{\mcitedefaultmidpunct}
{\mcitedefaultendpunct}{\mcitedefaultseppunct}\relax
\EndOfBibitem
\bibitem{STAR:2011fbd}
STAR collaboration, G.~Agakishiev {\em et~al.},
  \ifthenelse{\boolean{articletitles}}{\emph{{Strangeness enhancement in Cu+Cu
  and Au+Au collisions at $\sqrt{s_{NN}} = 200$ GeV}},
  }{}\href{https://doi.org/10.1103/PhysRevLett.108.072301}{Phys.\ Rev.\ Lett.\
  \textbf{108} (2012) 072301},
  \href{http://arxiv.org/abs/1107.2955}{{\normalfont\ttfamily
  arXiv:1107.2955}}\relax
\mciteBstWouldAddEndPuncttrue
\mciteSetBstMidEndSepPunct{\mcitedefaultmidpunct}
{\mcitedefaultendpunct}{\mcitedefaultseppunct}\relax
\EndOfBibitem
\bibitem{ALICE:2013xmt}
ALICE collaboration, B.~B. Abelev {\em et~al.},
  \ifthenelse{\boolean{articletitles}}{\emph{{Multi-strange baryon production
  at mid-rapidity in Pb-Pb collisions at $\sqrt{s_{NN}}$ = 2.76 TeV}},
  }{}\href{https://doi.org/doi.org/10.1016/j.physletb.2013.11.048}{Phys.\
  Lett.\  \textbf{B728} (2014) 216}, Erratum
  \href{https://doi.org/doi.org/10.1016/j.physletb.2014.05.052}{ibid.\
  \textbf{734} (2014) 409},
  \href{http://arxiv.org/abs/1307.5543}{{\normalfont\ttfamily
  arXiv:1307.5543}}\relax
\mciteBstWouldAddEndPuncttrue
\mciteSetBstMidEndSepPunct{\mcitedefaultmidpunct}
{\mcitedefaultendpunct}{\mcitedefaultseppunct}\relax
\EndOfBibitem
\bibitem{STAR:2021tte}
STAR collaboration, J.~Adam {\em et~al.},
  \ifthenelse{\boolean{articletitles}}{\emph{{Observation of $D_{s}^{\pm}/D^0$
  enhancement in Au+Au collisions at $\sqrt{s_{_{NN}}}$ = 200 GeV}},
  }{}\href{https://doi.org/10.1103/PhysRevLett.127.092301}{Phys.\ Rev.\ Lett.\
  \textbf{127} (2021) 092301},
  \href{http://arxiv.org/abs/2101.11793}{{\normalfont\ttfamily
  arXiv:2101.11793}}\relax
\mciteBstWouldAddEndPuncttrue
\mciteSetBstMidEndSepPunct{\mcitedefaultmidpunct}
{\mcitedefaultendpunct}{\mcitedefaultseppunct}\relax
\EndOfBibitem
\bibitem{ALICE:2021kfc}
ALICE collaboration, S.~Acharya {\em et~al.},
  \ifthenelse{\boolean{articletitles}}{\emph{{Measurement of prompt
  $D_s^+$-meson production and azimuthal anisotropy in Pb\textendash{}Pb
  collisions at $\sqrt {s_{NN}}$=5.02TeV}},
  }{}\href{https://doi.org/10.1016/j.physletb.2022.136986}{Phys.\ Lett.\
  \textbf{B827} (2022) 136986},
  \href{http://arxiv.org/abs/2110.10006}{{\normalfont\ttfamily
  arXiv:2110.10006}}\relax
\mciteBstWouldAddEndPuncttrue
\mciteSetBstMidEndSepPunct{\mcitedefaultmidpunct}
{\mcitedefaultendpunct}{\mcitedefaultseppunct}\relax
\EndOfBibitem
\bibitem{ALICE:2016fzo}
ALICE collaboration, J.~Adam {\em et~al.},
  \ifthenelse{\boolean{articletitles}}{\emph{{Enhanced production of
  multi-strange hadrons in high-multiplicity proton-proton collisions}},
  }{}\href{https://doi.org/10.1038/nphys4111}{Nature Phys.\  \textbf{13} (2017)
  535}, \href{http://arxiv.org/abs/1606.07424}{{\normalfont\ttfamily
  arXiv:1606.07424}}\relax
\mciteBstWouldAddEndPuncttrue
\mciteSetBstMidEndSepPunct{\mcitedefaultmidpunct}
{\mcitedefaultendpunct}{\mcitedefaultseppunct}\relax
\EndOfBibitem
\bibitem{ALICE:2013wgn}
ALICE collaboration, B.~B. Abelev {\em et~al.},
  \ifthenelse{\boolean{articletitles}}{\emph{{Multiplicity dependence of pion,
  kaon, proton and lambda production in p-Pb collisions at $\sqrt{s_{NN}}$ =
  5.02 TeV}}, }{}\href{https://doi.org/10.1016/j.physletb.2013.11.020}{Phys.\
  Lett.\  \textbf{B728} (2014) 25},
  \href{http://arxiv.org/abs/1307.6796}{{\normalfont\ttfamily
  arXiv:1307.6796}}\relax
\mciteBstWouldAddEndPuncttrue
\mciteSetBstMidEndSepPunct{\mcitedefaultmidpunct}
{\mcitedefaultendpunct}{\mcitedefaultseppunct}\relax
\EndOfBibitem
\bibitem{ALICE:2015mpp}
ALICE collaboration, J.~Adam {\em et~al.},
  \ifthenelse{\boolean{articletitles}}{\emph{{Multi-strange baryon production
  in p-Pb collisions at $\sqrt{s_\mathbf{NN}}=5.02$ TeV}},
  }{}\href{https://doi.org/10.1016/j.physletb.2016.05.027}{Phys.\ Lett.\
  \textbf{B758} (2016) 389},
  \href{http://arxiv.org/abs/1512.07227}{{\normalfont\ttfamily
  arXiv:1512.07227}}\relax
\mciteBstWouldAddEndPuncttrue
\mciteSetBstMidEndSepPunct{\mcitedefaultmidpunct}
{\mcitedefaultendpunct}{\mcitedefaultseppunct}\relax
\EndOfBibitem
\bibitem{Kanakubo:2019ogh}
Y.~Kanakubo, Y.~Tachibana, and T.~Hirano,
  \ifthenelse{\boolean{articletitles}}{\emph{{Unified description of hadron
  yield ratios from dynamical core-corona initialization}},
  }{}\href{https://doi.org/10.1103/PhysRevC.101.024912}{Phys.\ Rev.\
  \textbf{C101} (2020) 024912},
  \href{http://arxiv.org/abs/1910.10556}{{\normalfont\ttfamily
  arXiv:1910.10556}}\relax
\mciteBstWouldAddEndPuncttrue
\mciteSetBstMidEndSepPunct{\mcitedefaultmidpunct}
{\mcitedefaultendpunct}{\mcitedefaultseppunct}\relax
\EndOfBibitem
\bibitem{Bierlich:2022ned}
C.~Bierlich, S.~Chakraborty, G.~Gustafson, and L.~L\"onnblad,
  \ifthenelse{\boolean{articletitles}}{\emph{{Strangeness enhancement across
  collision systems without a plasma}},
  }{}\href{https://doi.org/10.1016/j.physletb.2022.137571}{Phys.\ Lett.\
  \textbf{B835} (2022) 137571},
  \href{http://arxiv.org/abs/2205.11170}{{\normalfont\ttfamily
  arXiv:2205.11170}}\relax
\mciteBstWouldAddEndPuncttrue
\mciteSetBstMidEndSepPunct{\mcitedefaultmidpunct}
{\mcitedefaultendpunct}{\mcitedefaultseppunct}\relax
\EndOfBibitem
\bibitem{LHCb-DP-2008-001}
LHCb collaboration, A.~A. Alves~Jr.\ {\em et~al.},
  \ifthenelse{\boolean{articletitles}}{\emph{{The \lhcb detector at the LHC}},
  }{}\href{https://doi.org/10.1088/1748-0221/3/08/S08005}{JINST \textbf{3}
  (2008) S08005}\relax
\mciteBstWouldAddEndPuncttrue
\mciteSetBstMidEndSepPunct{\mcitedefaultmidpunct}
{\mcitedefaultendpunct}{\mcitedefaultseppunct}\relax
\EndOfBibitem
\bibitem{LHCb-DP-2014-002}
LHCb collaboration, R.~Aaij {\em et~al.},
  \ifthenelse{\boolean{articletitles}}{\emph{{LHCb detector performance}},
  }{}\href{https://doi.org/10.1142/S0217751X15300227}{Int.\ J.\ Mod.\ Phys.\
  \textbf{A30} (2015) 1530022},
  \href{http://arxiv.org/abs/1412.6352}{{\normalfont\ttfamily
  arXiv:1412.6352}}\relax
\mciteBstWouldAddEndPuncttrue
\mciteSetBstMidEndSepPunct{\mcitedefaultmidpunct}
{\mcitedefaultendpunct}{\mcitedefaultseppunct}\relax
\EndOfBibitem
\bibitem{Sjostrand:2007gs}
T.~Sj\"{o}strand, S.~Mrenna, and P.~Skands,
  \ifthenelse{\boolean{articletitles}}{\emph{{A brief introduction to PYTHIA
  8.1}}, }{}\href{https://doi.org/10.1016/j.cpc.2008.01.036}{Comput.\ Phys.\
  Commun.\  \textbf{178} (2008) 852},
  \href{http://arxiv.org/abs/0710.3820}{{\normalfont\ttfamily
  arXiv:0710.3820}}\relax
\mciteBstWouldAddEndPuncttrue
\mciteSetBstMidEndSepPunct{\mcitedefaultmidpunct}
{\mcitedefaultendpunct}{\mcitedefaultseppunct}\relax
\EndOfBibitem
\bibitem{Sjostrand:2006za}
T.~Sj\"{o}strand, S.~Mrenna, and P.~Skands,
  \ifthenelse{\boolean{articletitles}}{\emph{{PYTHIA 6.4 physics and manual}},
  }{}\href{https://doi.org/10.1088/1126-6708/2006/05/026}{JHEP \textbf{05}
  (2006) 026}, \href{http://arxiv.org/abs/hep-ph/0603175}{{\normalfont\ttfamily
  arXiv:hep-ph/0603175}}\relax
\mciteBstWouldAddEndPuncttrue
\mciteSetBstMidEndSepPunct{\mcitedefaultmidpunct}
{\mcitedefaultendpunct}{\mcitedefaultseppunct}\relax
\EndOfBibitem
\bibitem{Pierog:2013ria}
T.~Pierog {\em et~al.}, \ifthenelse{\boolean{articletitles}}{\emph{{EPOS LHC:
  Test of collective hadronization with data measured at the CERN Large Hadron
  Collider}}, }{}\href{https://doi.org/10.1103/PhysRevC.92.034906}{Phys.\ Rev.\
   \textbf{C92} (2015) 034906},
  \href{http://arxiv.org/abs/1306.0121}{{\normalfont\ttfamily
  arXiv:1306.0121}}\relax
\mciteBstWouldAddEndPuncttrue
\mciteSetBstMidEndSepPunct{\mcitedefaultmidpunct}
{\mcitedefaultendpunct}{\mcitedefaultseppunct}\relax
\EndOfBibitem
\bibitem{LHCb:2011dpk}
LHCb collaboration, I.~Belyaev {\em et~al.},
  \ifthenelse{\boolean{articletitles}}{\emph{{Handling of the generation of
  primary events in Gauss, the LHCb simulation framework}},
  }{}\href{https://doi.org/10.1088/1742-6596/331/3/032047}{J.\ Phys.\ Conf.\
  Ser.\  \textbf{331} (2011) 032047}\relax
\mciteBstWouldAddEndPuncttrue
\mciteSetBstMidEndSepPunct{\mcitedefaultmidpunct}
{\mcitedefaultendpunct}{\mcitedefaultseppunct}\relax
\EndOfBibitem
\bibitem{Lange:2001uf}
D.~J. Lange, \ifthenelse{\boolean{articletitles}}{\emph{{The EvtGen particle
  decay simulation package}},
  }{}\href{https://doi.org/10.1016/S0168-9002(01)00089-4}{Nucl.\ Instrum.\
  Meth.\  \textbf{A462} (2001) 152}\relax
\mciteBstWouldAddEndPuncttrue
\mciteSetBstMidEndSepPunct{\mcitedefaultmidpunct}
{\mcitedefaultendpunct}{\mcitedefaultseppunct}\relax
\EndOfBibitem
\bibitem{Golonka:2005pn}
P.~Golonka and Z.~Was, \ifthenelse{\boolean{articletitles}}{\emph{{PHOTOS Monte
  Carlo: A precision tool for QED corrections in $Z$ and $W$ decays}},
  }{}\href{https://doi.org/10.1140/epjc/s2005-02396-4}{Eur.\ Phys.\ J.\
  \textbf{C45} (2006) 97},
  \href{http://arxiv.org/abs/hep-ph/0506026}{{\normalfont\ttfamily
  arXiv:hep-ph/0506026}}\relax
\mciteBstWouldAddEndPuncttrue
\mciteSetBstMidEndSepPunct{\mcitedefaultmidpunct}
{\mcitedefaultendpunct}{\mcitedefaultseppunct}\relax
\EndOfBibitem
\bibitem{GEANT4:2002zbu}
GEANT4 collaboration, S.~Agostinelli {\em et~al.},
  \ifthenelse{\boolean{articletitles}}{\emph{{GEANT4--a simulation toolkit}},
  }{}\href{https://doi.org/10.1016/S0168-9002(03)01368-8}{Nucl.\ Instrum.\
  Meth.\  \textbf{A506} (2003) 250}\relax
\mciteBstWouldAddEndPuncttrue
\mciteSetBstMidEndSepPunct{\mcitedefaultmidpunct}
{\mcitedefaultendpunct}{\mcitedefaultseppunct}\relax
\EndOfBibitem
\bibitem{Clemencic:2011zza}
LHCb collaboration, M.~Clemencic {\em et~al.},
  \ifthenelse{\boolean{articletitles}}{\emph{{The LHCb simulation application,
  Gauss: Design, evolution and experience}},
  }{}\href{https://doi.org/10.1088/1742-6596/331/3/032023}{J.\ Phys.\ Conf.\
  Ser.\  \textbf{331} (2011) 032023}\relax
\mciteBstWouldAddEndPuncttrue
\mciteSetBstMidEndSepPunct{\mcitedefaultmidpunct}
{\mcitedefaultendpunct}{\mcitedefaultseppunct}\relax
\EndOfBibitem
\bibitem{Pivk:2004ty}
M.~Pivk and F.~R. Le~Diberder,
  \ifthenelse{\boolean{articletitles}}{\emph{{sPlot: A statistical tool to
  unfold data distributions}},
  }{}\href{https://doi.org/10.1016/j.nima.2005.08.106}{Nucl.\ Instrum.\ Meth.\
  \textbf{A555} (2005) 356},
  \href{http://arxiv.org/abs/physics/0402083}{{\normalfont\ttfamily
  arXiv:physics/0402083}}\relax
\mciteBstWouldAddEndPuncttrue
\mciteSetBstMidEndSepPunct{\mcitedefaultmidpunct}
{\mcitedefaultendpunct}{\mcitedefaultseppunct}\relax
\EndOfBibitem
\bibitem{CLEO:2008hzo}
CLEO collaboration, J.~P. Alexander {\em et~al.},
  \ifthenelse{\boolean{articletitles}}{\emph{{Absolute measurement of hadronic
  branching fractions of the \Dsp meson}},
  }{}\href{https://doi.org/10.1103/PhysRevLett.100.161804}{Phys.\ Rev.\ Lett.\
  \textbf{100} (2008) 161804},
  \href{http://arxiv.org/abs/0801.0680}{{\normalfont\ttfamily
  arXiv:0801.0680}}\relax
\mciteBstWouldAddEndPuncttrue
\mciteSetBstMidEndSepPunct{\mcitedefaultmidpunct}
{\mcitedefaultendpunct}{\mcitedefaultseppunct}\relax
\EndOfBibitem
\bibitem{ParticleDataGroup:2022pth}
Particle Data Group, R.~L. Workman {\em et~al.},
  \ifthenelse{\boolean{articletitles}}{\emph{{Review of Particle Physics}},
  }{}\href{https://doi.org/10.1093/ptep/ptac097}{PTEP \textbf{2022} (2022)
  083C01}\relax
\mciteBstWouldAddEndPuncttrue
\mciteSetBstMidEndSepPunct{\mcitedefaultmidpunct}
{\mcitedefaultendpunct}{\mcitedefaultseppunct}\relax
\EndOfBibitem
\bibitem{Skwarnicki:1986xj}
T.~Skwarnicki, {\em {A study of the radiative cascade transitions between the
  Upsilon-prime and Upsilon resonances}}, PhD thesis, Institute of Nuclear
  Physics, Krakow, 1986,
  {\href{http://inspirehep.net/record/230779/}{DESY-F31-86-02}}\relax
\mciteBstWouldAddEndPuncttrue
\mciteSetBstMidEndSepPunct{\mcitedefaultmidpunct}
{\mcitedefaultendpunct}{\mcitedefaultseppunct}\relax
\EndOfBibitem
\bibitem{Bukin:2007}
A.~D. Bukin, \ifthenelse{\boolean{articletitles}}{\emph{{Fitting function for
  asymmetric peaks}},
  }{}\href{http://arxiv.org/abs/0711.4449}{{\normalfont\ttfamily
  arXiv:0711.4449}}\relax
\mciteBstWouldAddEndPuncttrue
\mciteSetBstMidEndSepPunct{\mcitedefaultmidpunct}
{\mcitedefaultendpunct}{\mcitedefaultseppunct}\relax
\EndOfBibitem
\bibitem{Supplemental:2023}
R.~Aaij {\em et~al.}, \ifthenelse{\boolean{articletitles}}{\emph{{See
  Supplemental Material at [URL to be added] for further details}}, }{}\relax
\mciteBstWouldAddEndPuncttrue
\mciteSetBstMidEndSepPunct{\mcitedefaultmidpunct}
{\mcitedefaultendpunct}{\mcitedefaultseppunct}\relax
\EndOfBibitem
\bibitem{LHCb-DP-2013-002}
LHCb collaboration, R.~Aaij {\em et~al.},
  \ifthenelse{\boolean{articletitles}}{\emph{{Measurement of the track
  reconstruction efficiency at LHCb}},
  }{}\href{https://doi.org/10.1088/1748-0221/10/02/P02007}{JINST \textbf{10}
  (2015) P02007}, \href{http://arxiv.org/abs/1408.1251}{{\normalfont\ttfamily
  arXiv:1408.1251}}\relax
\mciteBstWouldAddEndPuncttrue
\mciteSetBstMidEndSepPunct{\mcitedefaultmidpunct}
{\mcitedefaultendpunct}{\mcitedefaultseppunct}\relax
\EndOfBibitem
\bibitem{LHCb-PUB-2016-021}
L.~Anderlini {\em et~al.}, \ifthenelse{\boolean{articletitles}}{\emph{{The
  PIDCalib package}}, }{}
  \href{http://cdsweb.cern.ch/search?p=LHCb-PUB-2016-021&f=reportnumber&action_search=Search&c=LHCb+Notes}
  {LHCb-PUB-2016-021}, 2016\relax
\mciteBstWouldAddEndPuncttrue
\mciteSetBstMidEndSepPunct{\mcitedefaultmidpunct}
{\mcitedefaultendpunct}{\mcitedefaultseppunct}\relax
\EndOfBibitem
\bibitem{LHCb-DP-2018-001}
R.~Aaij {\em et~al.}, \ifthenelse{\boolean{articletitles}}{\emph{{Selection and
  processing of calibration samples to measure the particle identification
  performance of the LHCb experiment in Run 2}},
  }{}\href{https://doi.org/10.1140/epjti/s40485-019-0050-z}{Eur.\ Phys.\ J.\
  Tech.\ Instr.\  \textbf{6} (2019) 1},
  \href{http://arxiv.org/abs/1803.00824}{{\normalfont\ttfamily
  arXiv:1803.00824}}\relax
\mciteBstWouldAddEndPuncttrue
\mciteSetBstMidEndSepPunct{\mcitedefaultmidpunct}
{\mcitedefaultendpunct}{\mcitedefaultseppunct}\relax
\EndOfBibitem
\bibitem{LHCb:2023kqs}
LHCb collaboration, I.~Bezshyiko {\em et~al.},
  \ifthenelse{\boolean{articletitles}}{\emph{{Measurement of prompt D$^{+}$ and
  $ {D}_s^{+} $ production in pPb collisions at $ \sqrt{s_{NN}} $ = 5.02 TeV}},
  }{}\href{https://doi.org/10.1007/JHEP01(2024)070}{JHEP \textbf{01} (2024)
  070}, \href{http://arxiv.org/abs/2309.14206}{{\normalfont\ttfamily
  arXiv:2309.14206}}\relax
\mciteBstWouldAddEndPuncttrue
\mciteSetBstMidEndSepPunct{\mcitedefaultmidpunct}
{\mcitedefaultendpunct}{\mcitedefaultseppunct}\relax
\EndOfBibitem
\bibitem{LHCb-PUB-2014-039}
S.~Tolk, J.~Albrecht, F.~Dettori, and A.~Pellegrino,
  \ifthenelse{\boolean{articletitles}}{\emph{{Data driven trigger efficiency
  determination at LHCb}}, }{}
  \href{http://cdsweb.cern.ch/search?p=LHCb-PUB-2014-039&f=reportnumber&action_search=Search&c=LHCb+Notes}
  {LHCb-PUB-2014-039}, 2014\relax
\mciteBstWouldAddEndPuncttrue
\mciteSetBstMidEndSepPunct{\mcitedefaultmidpunct}
{\mcitedefaultendpunct}{\mcitedefaultseppunct}\relax
\EndOfBibitem
\bibitem{LHCb-PAPER-2016-042}
LHCb collaboration, R.~Aaij {\em et~al.},
  \ifthenelse{\boolean{articletitles}}{\emph{{Measurements of prompt charm
  production cross-sections in \proton\proton collisions at $\sqs = $5\tev}},
  }{}\href{https://doi.org/10.1007/JHEP06(2017)147}{JHEP \textbf{06} (2017)
  147}, \href{http://arxiv.org/abs/1610.02230}{{\normalfont\ttfamily
  arXiv:1610.02230}}\relax
\mciteBstWouldAddEndPuncttrue
\mciteSetBstMidEndSepPunct{\mcitedefaultmidpunct}
{\mcitedefaultendpunct}{\mcitedefaultseppunct}\relax
\EndOfBibitem
\bibitem{LHCb-PAPER-2015-041}
LHCb collaboration, R.~Aaij {\em et~al.},
  \ifthenelse{\boolean{articletitles}}{\emph{{Measurements of prompt charm
  production cross-sections in \proton\proton collisions at $\sqs = $13\tev}},
  }{}\href{https://doi.org/10.1007/JHEP03(2016)159}{JHEP \textbf{03} (2016)
  159}, Erratum \href{https://doi.org/10.1007/JHEP09(2016)013}{ibid.\
  \textbf{09} (2016) 013}, Erratum
  \href{https://doi.org/10.1007/JHEP05(2017)074}{ibid.\   \textbf{05} (2017)
  074}, \href{http://arxiv.org/abs/1510.01707}{{\normalfont\ttfamily
  arXiv:1510.01707}}\relax
\mciteBstWouldAddEndPuncttrue
\mciteSetBstMidEndSepPunct{\mcitedefaultmidpunct}
{\mcitedefaultendpunct}{\mcitedefaultseppunct}\relax
\EndOfBibitem
\bibitem{Shao:2012iz}
H.-S. Shao, \ifthenelse{\boolean{articletitles}}{\emph{{HELAC-Onia: An
  automatic matrix element generator for heavy quarkonium physics}},
  }{}\href{https://doi.org/10.1016/j.cpc.2013.05.023}{Comput.\ Phys.\ Commun.\
  \textbf{184} (2013) 2562},
  \href{http://arxiv.org/abs/1212.5293}{{\normalfont\ttfamily
  arXiv:1212.5293}}\relax
\mciteBstWouldAddEndPuncttrue
\mciteSetBstMidEndSepPunct{\mcitedefaultmidpunct}
{\mcitedefaultendpunct}{\mcitedefaultseppunct}\relax
\EndOfBibitem
\bibitem{Shao:2015vga}
H.-S. Shao, \ifthenelse{\boolean{articletitles}}{\emph{{HELAC-Onia 2.0: an
  upgraded matrix-element and event generator for heavy quarkonium physics}},
  }{}\href{https://doi.org/10.1016/j.cpc.2015.09.011}{Comput.\ Phys.\ Commun.\
  \textbf{198} (2016) 238},
  \href{http://arxiv.org/abs/1507.03435}{{\normalfont\ttfamily
  arXiv:1507.03435}}\relax
\mciteBstWouldAddEndPuncttrue
\mciteSetBstMidEndSepPunct{\mcitedefaultmidpunct}
{\mcitedefaultendpunct}{\mcitedefaultseppunct}\relax
\EndOfBibitem
\bibitem{Lansberg:2016deg}
J.-P. Lansberg and H.-S. Shao,
  \ifthenelse{\boolean{articletitles}}{\emph{{Towards an automated tool to
  evaluate the impact of the nuclear modification of the gluon density on
  quarkonium, D and B meson production in proton\textendash{}nucleus
  collisions}}, }{}\href{https://doi.org/10.1140/epjc/s10052-016-4575-x}{Eur.\
  Phys.\ J.\  \textbf{C77} (2017) 1},
  \href{http://arxiv.org/abs/1610.05382}{{\normalfont\ttfamily
  arXiv:1610.05382}}\relax
\mciteBstWouldAddEndPuncttrue
\mciteSetBstMidEndSepPunct{\mcitedefaultmidpunct}
{\mcitedefaultendpunct}{\mcitedefaultseppunct}\relax
\EndOfBibitem
\bibitem{Eskola:2016oht}
K.~J. Eskola, P.~Paakkinen, H.~Paukkunen, and C.~A. Salgado,
  \ifthenelse{\boolean{articletitles}}{\emph{{EPPS16: Nuclear parton
  distributions with LHC data}},
  }{}\href{https://doi.org/10.1140/epjc/s10052-017-4725-9}{Eur.\ Phys.\ J.\
  \textbf{C77} (2017) 163},
  \href{http://arxiv.org/abs/1612.05741}{{\normalfont\ttfamily
  arXiv:1612.05741}}\relax
\mciteBstWouldAddEndPuncttrue
\mciteSetBstMidEndSepPunct{\mcitedefaultmidpunct}
{\mcitedefaultendpunct}{\mcitedefaultseppunct}\relax
\EndOfBibitem
\bibitem{Kovarik:2015cma}
K.~Kovarik {\em et~al.}, \ifthenelse{\boolean{articletitles}}{\emph{{nCTEQ15 -
  Global analysis of nuclear parton distributions with uncertainties in the
  CTEQ framework}}, }{}\href{https://doi.org/10.1103/PhysRevD.93.085037}{Phys.\
  Rev.\  \textbf{D93} (2016) 085037},
  \href{http://arxiv.org/abs/1509.00792}{{\normalfont\ttfamily
  arXiv:1509.00792}}\relax
\mciteBstWouldAddEndPuncttrue
\mciteSetBstMidEndSepPunct{\mcitedefaultmidpunct}
{\mcitedefaultendpunct}{\mcitedefaultseppunct}\relax
\EndOfBibitem
\bibitem{LHCb-PAPER-2017-015}
LHCb collaboration, R.~Aaij {\em et~al.},
  \ifthenelse{\boolean{articletitles}}{\emph{{Study of prompt \Dz meson
  production in \proton{}Pb collisions at $\sqsnn = $5\tev}},
  }{}\href{https://doi.org/10.1007/JHEP10(2017)090}{JHEP \textbf{10} (2017)
  090}, \href{http://arxiv.org/abs/1707.02750}{{\normalfont\ttfamily
  arXiv:1707.02750}}\relax
\mciteBstWouldAddEndPuncttrue
\mciteSetBstMidEndSepPunct{\mcitedefaultmidpunct}
{\mcitedefaultendpunct}{\mcitedefaultseppunct}\relax
\EndOfBibitem
\bibitem{ALICE:2014xjz}
ALICE collaboration, B.~B. Abelev {\em et~al.},
  \ifthenelse{\boolean{articletitles}}{\emph{{Measurement of prompt $D$-meson
  production in $p-Pb$ collisions at $\sqrt{s_{NN}}$ = 5.02 TeV}},
  }{}\href{https://doi.org/10.1103/PhysRevLett.113.232301}{Phys.\ Rev.\ Lett.\
  \textbf{113} (2014) 232301},
  \href{http://arxiv.org/abs/1405.3452}{{\normalfont\ttfamily
  arXiv:1405.3452}}\relax
\mciteBstWouldAddEndPuncttrue
\mciteSetBstMidEndSepPunct{\mcitedefaultmidpunct}
{\mcitedefaultendpunct}{\mcitedefaultseppunct}\relax
\EndOfBibitem
\bibitem{ALICE:2016cpm}
ALICE collaboration, J.~Adam {\em et~al.},
  \ifthenelse{\boolean{articletitles}}{\emph{{Measurement of D-meson production
  versus multiplicity in p-Pb collisions at $
  \sqrt{{\mathrm{s}}_{\mathrm{NN}}}=5.02 $ TeV}},
  }{}\href{https://doi.org/10.1007/JHEP08(2016)078}{JHEP \textbf{08} (2016)
  078}, \href{http://arxiv.org/abs/1602.07240}{{\normalfont\ttfamily
  arXiv:1602.07240}}\relax
\mciteBstWouldAddEndPuncttrue
\mciteSetBstMidEndSepPunct{\mcitedefaultmidpunct}
{\mcitedefaultendpunct}{\mcitedefaultseppunct}\relax
\EndOfBibitem
\bibitem{ALICE:2016yta}
ALICE collaboration, J.~Adam {\em et~al.},
  \ifthenelse{\boolean{articletitles}}{\emph{{$D$-meson production in $p$-Pb
  collisions at $\sqrt{s_{\rm NN}}=$5.02 TeV and in pp collisions at
  $\sqrt{s}=$7 TeV}},
  }{}\href{https://doi.org/10.1103/PhysRevC.94.054908}{Phys.\ Rev.\
  \textbf{C94} (2016) 054908},
  \href{http://arxiv.org/abs/1605.07569}{{\normalfont\ttfamily
  arXiv:1605.07569}}\relax
\mciteBstWouldAddEndPuncttrue
\mciteSetBstMidEndSepPunct{\mcitedefaultmidpunct}
{\mcitedefaultendpunct}{\mcitedefaultseppunct}\relax
\EndOfBibitem
\bibitem{Kusina:2017gkz}
A.~Kusina, J.-P. Lansberg, I.~Schienbein, and H.-S. Shao,
  \ifthenelse{\boolean{articletitles}}{\emph{{Gluon shadowing in heavy-flavor
  production at the LHC}},
  }{}\href{https://doi.org/10.1103/PhysRevLett.121.052004}{Phys.\ Rev.\ Lett.\
  \textbf{121} (2018) 052004},
  \href{http://arxiv.org/abs/1712.07024}{{\normalfont\ttfamily
  arXiv:1712.07024}}\relax
\mciteBstWouldAddEndPuncttrue
\mciteSetBstMidEndSepPunct{\mcitedefaultmidpunct}
{\mcitedefaultendpunct}{\mcitedefaultseppunct}\relax
\EndOfBibitem
\bibitem{Ducloue:2015gfa}
B.~Duclou\'e, T.~Lappi, and H.~M\"antysaari,
  \ifthenelse{\boolean{articletitles}}{\emph{{Forward $J/\psi$ production in
  proton-nucleus collisions at high energy}},
  }{}\href{https://doi.org/10.1103/PhysRevD.91.114005}{Phys.\ Rev.\
  \textbf{D91} (2015) 114005},
  \href{http://arxiv.org/abs/1503.02789}{{\normalfont\ttfamily
  arXiv:1503.02789}}\relax
\mciteBstWouldAddEndPuncttrue
\mciteSetBstMidEndSepPunct{\mcitedefaultmidpunct}
{\mcitedefaultendpunct}{\mcitedefaultseppunct}\relax
\EndOfBibitem
\bibitem{Ducloue:2016ywt}
B.~Duclou\'e, T.~Lappi, and H.~M\"antysaari,
  \ifthenelse{\boolean{articletitles}}{\emph{{Forward $J/\psi$ and $D$ meson
  nuclear suppression at the LHC}},
  }{}\href{https://doi.org/10.1016/j.nuclphysbps.2017.05.071}{Nucl.\ Part.\
  Phys.\ Proc.\  \textbf{289-290} (2017) 309},
  \href{http://arxiv.org/abs/1612.04585}{{\normalfont\ttfamily
  arXiv:1612.04585}}\relax
\mciteBstWouldAddEndPuncttrue
\mciteSetBstMidEndSepPunct{\mcitedefaultmidpunct}
{\mcitedefaultendpunct}{\mcitedefaultseppunct}\relax
\EndOfBibitem
\bibitem{Ma:2018bax}
Y.-Q. Ma, P.~Tribedy, R.~Venugopalan, and K.~Watanabe,
  \ifthenelse{\boolean{articletitles}}{\emph{{Event engineering studies for
  heavy flavor production and hadronization in high multiplicity hadron-hadron
  and hadron-nucleus collisions}},
  }{}\href{https://doi.org/10.1103/PhysRevD.98.074025}{Phys.\ Rev.\
  \textbf{D98} (2018) 074025},
  \href{http://arxiv.org/abs/1803.11093}{{\normalfont\ttfamily
  arXiv:1803.11093}}\relax
\mciteBstWouldAddEndPuncttrue
\mciteSetBstMidEndSepPunct{\mcitedefaultmidpunct}
{\mcitedefaultendpunct}{\mcitedefaultseppunct}\relax
\EndOfBibitem
\bibitem{Lisovyi:2015uqa}
M.~Lisovyi, A.~Verbytskyi, and O.~Zenaiev,
  \ifthenelse{\boolean{articletitles}}{\emph{{Combined analysis of charm-quark
  fragmentation-fraction measurements}},
  }{}\href{https://doi.org/10.1140/epjc/s10052-016-4246-y}{Eur.\ Phys.\ J.\ C
  \textbf{76} (2016) 397},
  \href{http://arxiv.org/abs/1509.01061}{{\normalfont\ttfamily
  arXiv:1509.01061}}\relax
\mciteBstWouldAddEndPuncttrue
\mciteSetBstMidEndSepPunct{\mcitedefaultmidpunct}
{\mcitedefaultendpunct}{\mcitedefaultseppunct}\relax
\EndOfBibitem
\bibitem{LHCb:2016ikn}
LHCb collaboration, R.~Aaij {\em et~al.},
  \ifthenelse{\boolean{articletitles}}{\emph{{Measurements of prompt charm
  production cross-sections in pp collisions at $ \sqrt{s}=5 $ TeV}},
  }{}\href{https://doi.org/10.1007/JHEP06(2017)147}{JHEP \textbf{06} (2017)
  147}, \href{http://arxiv.org/abs/1610.02230}{{\normalfont\ttfamily
  arXiv:1610.02230}}\relax
\mciteBstWouldAddEndPuncttrue
\mciteSetBstMidEndSepPunct{\mcitedefaultmidpunct}
{\mcitedefaultendpunct}{\mcitedefaultseppunct}\relax
\EndOfBibitem
\bibitem{ALICE:2019fhe}
ALICE collaboration, S.~Acharya {\em et~al.},
  \ifthenelse{\boolean{articletitles}}{\emph{{Measurement of prompt D$^{0}$,
  D$^{+}$, D$^{*+}$, and $ {\mathrm{D}}_{\mathrm{S}}^{+} $ production in
  p\textendash{}Pb collisions at $ \sqrt{{\mathrm{s}}_{\mathrm{NN}}} $ = 5.02
  TeV}}, }{}\href{https://doi.org/10.1007/JHEP12(2019)092}{JHEP \textbf{12}
  (2019) 092}, \href{http://arxiv.org/abs/1906.03425}{{\normalfont\ttfamily
  arXiv:1906.03425}}\relax
\mciteBstWouldAddEndPuncttrue
\mciteSetBstMidEndSepPunct{\mcitedefaultmidpunct}
{\mcitedefaultendpunct}{\mcitedefaultseppunct}\relax
\EndOfBibitem
\bibitem{ALICE:2018lyv}
ALICE collaboration, S.~Acharya {\em et~al.},
  \ifthenelse{\boolean{articletitles}}{\emph{{Measurement of D$^{0}$, D$^{+}$,
  D$^{*+}$ and D$_{s}^{+}$ production in Pb-Pb collisions at $
  \sqrt{{\mathrm{s}}_{\mathrm{NN}}}=5.02 $ TeV}},
  }{}\href{https://doi.org/10.1007/JHEP10(2018)174}{JHEP \textbf{10} (2018)
  174}, \href{http://arxiv.org/abs/1804.09083}{{\normalfont\ttfamily
  arXiv:1804.09083}}\relax
\mciteBstWouldAddEndPuncttrue
\mciteSetBstMidEndSepPunct{\mcitedefaultmidpunct}
{\mcitedefaultendpunct}{\mcitedefaultseppunct}\relax
\EndOfBibitem
\bibitem{Skands:2014pea}
P.~Skands, S.~Carrazza, and J.~Rojo,
  \ifthenelse{\boolean{articletitles}}{\emph{{Tuning PYTHIA 8.1: the Monash
  2013 Tune}}, }{}\href{https://doi.org/10.1140/epjc/s10052-014-3024-y}{Eur.\
  Phys.\ J.\ C \textbf{74} (2014) 3024},
  \href{http://arxiv.org/abs/1404.5630}{{\normalfont\ttfamily
  arXiv:1404.5630}}\relax
\mciteBstWouldAddEndPuncttrue
\mciteSetBstMidEndSepPunct{\mcitedefaultmidpunct}
{\mcitedefaultendpunct}{\mcitedefaultseppunct}\relax
\EndOfBibitem
\bibitem{Christiansen:2015yqa}
J.~R. Christiansen and P.~Z. Skands,
  \ifthenelse{\boolean{articletitles}}{\emph{{String Formation Beyond Leading
  Colour}}, }{}\href{https://doi.org/10.1007/JHEP08(2015)003}{JHEP \textbf{08}
  (2015) 003}, \href{http://arxiv.org/abs/1505.01681}{{\normalfont\ttfamily
  arXiv:1505.01681}}\relax
\mciteBstWouldAddEndPuncttrue
\mciteSetBstMidEndSepPunct{\mcitedefaultmidpunct}
{\mcitedefaultendpunct}{\mcitedefaultseppunct}\relax
\EndOfBibitem
\bibitem{Zhao:2023ucp}
J.~Zhao, J.~Aichelin, P.~B. Gossiaux, and K.~Werner,
  \ifthenelse{\boolean{articletitles}}{\emph{{Heavy flavor as a probe of hot
  QCD matter produced in proton-proton collisions}},
  }{}\href{http://arxiv.org/abs/2310.08684}{{\normalfont\ttfamily
  arXiv:2310.08684}}\relax
\mciteBstWouldAddEndPuncttrue
\mciteSetBstMidEndSepPunct{\mcitedefaultmidpunct}
{\mcitedefaultendpunct}{\mcitedefaultseppunct}\relax
\EndOfBibitem
\bibitem{Zhao:2024ecc}
J.~Zhao, J.~Aichelin, P.~B. Gossiaux, and K.~Werner,
  \ifthenelse{\boolean{articletitles}}{\emph{{Heavy flavour hadron production
  in relativistic heavy ion collisions at RHIC and LHC in EPOS4HQ}},
  }{}\href{http://arxiv.org/abs/2401.17096}{{\normalfont\ttfamily
  arXiv:2401.17096}}\relax
\mciteBstWouldAddEndPuncttrue
\mciteSetBstMidEndSepPunct{\mcitedefaultmidpunct}
{\mcitedefaultendpunct}{\mcitedefaultseppunct}\relax
\EndOfBibitem
\end{mcitethebibliography}

\newpage
\centerline
{\large\bf LHCb collaboration}
\begin
{flushleft}
\small
R.~Aaij$^{35}$\lhcborcid{0000-0003-0533-1952},
A.S.W.~Abdelmotteleb$^{54}$\lhcborcid{0000-0001-7905-0542},
C.~Abellan~Beteta$^{48}$,
F.~Abudin{\'e}n$^{54}$\lhcborcid{0000-0002-6737-3528},
T.~Ackernley$^{58}$\lhcborcid{0000-0002-5951-3498},
B.~Adeva$^{44}$\lhcborcid{0000-0001-9756-3712},
M.~Adinolfi$^{52}$\lhcborcid{0000-0002-1326-1264},
P.~Adlarson$^{78}$\lhcborcid{0000-0001-6280-3851},
H.~Afsharnia$^{11}$,
C.~Agapopoulou$^{46}$\lhcborcid{0000-0002-2368-0147}
% ...

{\footnotesize \it

% ...
$^{11}$Universit{\'e} Clermont Auvergne, CNRS/IN2P3, LPC, Clermont-Ferrand, France\\
% ...
$^{35}$Nikhef National Institute for Subatomic Physics, Amsterdam, Netherlands\\
% ...
$^{44}$Instituto Galego de F{\'\i}sica de Altas Enerx{\'\i}as (IGFAE), Universidade de Santiago de Compostela, Santiago de Compostela, Spain\\
% ...
$^{46}$European Organization for Nuclear Research (CERN), Geneva, Switzerland\\
% ...
$^{48}$Physik-Institut, Universit{\"a}t Z{\"u}rich, Z{\"u}rich, Switzerland\\
% ...
$^{52}$H.H. Wills Physics Laboratory, University of Bristol, Bristol, United Kingdom\\
% ...
$^{54}$Department of Physics, University of Warwick, Coventry, United Kingdom\\
% ...
$^{57}$School of Physics and Astronomy, University of Glasgow, Glasgow, United Kingdom\\
$^{58}$Oliver Lodge Laboratory, University of Liverpool, Liverpool, United Kingdom\\
% ...
$^{78}$Department of Physics and Astronomy, Uppsala University, Uppsala, Sweden, associated to $^{57}$\\
% ...
}
\end{flushleft}

\end{document}